\documentclass[12pt, journal, onecolumn]{IEEEtran}

\usepackage{xr-hyper}

\usepackage[T1]{fontenc}
\usepackage[utf8]{inputenc}
\usepackage{amsmath,amssymb,amsfonts,mathrsfs,bm}
\usepackage{mathtools}
\usepackage{amsthm}
\usepackage{scalerel}
\usepackage{nicefrac}
\usepackage[shortlabels]{enumitem}
\usepackage{graphicx}
\usepackage{epstopdf}
\DeclareGraphicsExtensions{.eps,.png,.jpg,.pdf}

\usepackage{url}
\usepackage{colortbl}
\usepackage{booktabs}
\usepackage{multirow}
\usepackage[table,dvipsnames]{xcolor}
\usepackage[normalem]{ulem}
\usepackage{xparse}
\usepackage{calc}
\usepackage{etoolbox}

\makeatletter
\@ifpackageloaded{natbib}{
	\relax
}{
	\usepackage{cite}
}
\makeatother

\usepackage{array}
\newcolumntype{L}[1]{>{\raggedright\let\newline\\\arraybackslash\hspace{0pt}}m{#1}}
\newcolumntype{C}[1]{>{\centering\let\newline\\\arraybackslash\hspace{0pt}}m{#1}}
\newcolumntype{R}[1]{>{\raggedleft\let\newline\\\arraybackslash\hspace{0pt}}m{#1}}

\makeatletter
\let\MYcaption\@makecaption
\makeatother
\usepackage[font=footnotesize]{subcaption}
\makeatletter
\let\@makecaption\MYcaption
\makeatother

\usepackage{glossaries}
\makeatletter
\let\oldgls\gls
\let\oldglspl\glspl

\newcommand\fussy@ifnextchar[3]{%
	\let\reserved@d=#1%
	\def\reserved@a{#2}%
	\def\reserved@b{#3}%
	\futurelet\@let@token\fussy@ifnch}
\def\fussy@ifnch{%
	\ifx\@let@token\reserved@d
		\let\reserved@c\reserved@a
	\else
		\let\reserved@c\reserved@b
	\fi
	\reserved@c}

\renewcommand{\gls}[1]{%
\oldgls{#1}\fussy@ifnextchar.{\@checkperiod}{\@}}
\renewcommand{\glspl}[1]{%
\oldglspl{#1}\fussy@ifnextchar.{\@checkperiod}{\@}}

\newcommand{\@checkperiod}[1]{%
	\ifnum\sfcode`\.=\spacefactor\else#1\fi
}
\makeatother

\newacronym{wrt}{w.r.t.}{with respect to}
\newacronym{RHS}{R.H.S.}{right-hand side}
\newacronym{LHS}{L.H.S.}{left-hand side}
\newacronym{iid}{i.i.d.}{independent and identically distributed}
\newacronym{SVD}{SVD}{singular value decomposition}
\newacronym{5G}{5G}{fifth generation wireless}
\newacronym{3GPP}{3GPP}{third generation partnership project}
\newacronym{OFDM}{OFDM}{orthogonal frequency-division multiplexing}

\usepackage{float}

\ifx\notloadhyperref\undefined
	\ifx\loadbibentry\undefined
		\usepackage[hidelinks,hypertexnames=false]{hyperref}
	\else
		\usepackage{bibentry}
		\makeatletter\let\saved@bibitem\@bibitem\makeatother
		\usepackage[hidelinks,hypertexnames=false]{hyperref}
		\makeatletter\let\@bibitem\saved@bibitem\makeatother
	\fi
\else
	\ifx\loadbibentry\undefined
		\relax
	\else
		\usepackage{bibentry}
	\fi
\fi

\usepackage[capitalize]{cleveref}
\crefname{equation}{}{}
\Crefname{equation}{}{}
\crefname{claim}{claim}{claims}
\crefname{step}{step}{steps}
\crefname{line}{line}{lines}
\crefname{condition}{condition}{conditions}
\crefname{dmath}{}{}
\crefname{dseries}{}{}
\crefname{dgroup}{}{}

\crefname{Problem}{Problem}{Problems}
\crefformat{Problem}{Problem~(#2#1#3)}
\crefrangeformat{Problem}{Problems~(#3#1#4) to~(#5#2#6)}

\crefname{Theorem}{Theorem}{Theorems}
\crefname{Corollary}{Corollary}{Corollaries}
\crefname{Proposition}{Proposition}{Propositions}
\crefname{Lemma}{Lemma}{Lemmas}
\crefname{Definition}{Definition}{Definitions}
\crefname{Example}{Example}{Examples}
\crefname{Assumption}{Assumption}{Assumptions}
\crefname{Remark}{Remark}{Remarks}
\crefname{Rem}{Remark}{Remarks}
\crefname{remarks}{Remarks}{Remarks}
\crefname{Appendix}{Appendix}{Appendices}
\crefname{Supplement}{Supplement}{Supplements}
\crefname{Exercise}{Exercise}{Exercises}
\crefname{Theorem_A}{Theorem}{Theorems}
\crefname{Corollary_A}{Corollary}{Corollaries}
\crefname{Proposition_A}{Proposition}{Propositions}
\crefname{Lemma_A}{Lemma}{Lemmas}
\crefname{Definition_A}{Definition}{Definitions}

\usepackage{crossreftools}
\ifx\notloadhyperref\undefined
\else
	\relax
\fi

\usepackage{algorithm,algorithmic}

\ifx\loadbreqn\undefined
	\relax
\else
	\usepackage{breqn}
\fi

\makeatletter
\def\cleartheorem#1{%
    \expandafter\let\csname#1\endcsname\relax
    \expandafter\let\csname c@#1\endcsname\relax
}
\def\clearthms#1{ \@for\tname:=#1\do{\cleartheorem\tname} }
\makeatother

\ifx\renewtheorem\undefined
	\ifx\useTheoremCounter\undefined
		\newtheorem{Theorem}{Theorem}
		\newtheorem{Corollary}{Corollary}
		\newtheorem{Proposition}{Proposition}
		\newtheorem{Lemma}{Lemma}
	\else
		\newtheorem{Theorem}{Theorem}
		
		\newtheorem{Proposition}[Theorem]{Proposition}
	\fi

	\newtheorem{Remark}{Remark}

\fi

\theoremstyle{remark}

\theoremstyle{plain}

\newcommand{\qednew}{\nobreak \ifvmode \relax \else
		\ifdim\lastskip<1.5em \hskip-\lastskip
			\hskip1.5em plus0em minus0.5em \fi \nobreak
		\vrule height0.75em width0.5em depth0.25em\fi}



\NewDocumentCommand{\movedownsub}{e{^_}}{%
	\IfNoValueTF{#1}{%
		\IfNoValueF{#2}{^{}}
	}{%
		^{#1}
	}%
	\IfNoValueF{#2}{_{#2}}
}

\let\latexchi\chi
\RenewDocumentCommand{\chi}{}{\latexchi\movedownsub}

\newcommand{\iu}{\mathfrak{i}\mkern1mu}

\newcommand{\calC}{\mathcal{C}}

\newcommand{\calN}{\mathcal{N}}

\newcommand{\ba}{\mathbf{a}}
\newcommand{\bA}{\mathbf{A}}
\newcommand{\bb}{\mathbf{b}}
\newcommand{\bB}{\mathbf{B}}

\newcommand{\bC}{\mathbf{C}}

\newcommand{\bD}{\mathbf{D}}

\newcommand{\bE}{\mathbf{E}}
\newcommand{\boldf}{\mathbf{f}}
\newcommand{\bF}{\mathbf{F}}

\newcommand{\bh}{\mathbf{h}}
\newcommand{\bH}{\mathbf{H}}

\newcommand{\bI}{\mathbf{I}}

\newcommand{\bJ}{\mathbf{J}}

\newcommand{\bn}{\mathbf{n}}
\newcommand{\bN}{\mathbf{N}}

\newcommand{\bO}{\mathbf{O}}
\newcommand{\bp}{\mathbf{p}}
\newcommand{\bP}{\mathbf{P}}

\newcommand{\bQ}{\mathbf{Q}}
\newcommand{\br}{\mathbf{r}}
\newcommand{\bR}{\mathbf{R}}
\newcommand{\bs}{\mathbf{s}}

\newcommand{\bw}{\mathbf{w}}
\newcommand{\bW}{\mathbf{W}}
\newcommand{\bx}{\mathbf{x}}
\newcommand{\bX}{\mathbf{X}}
\newcommand{\by}{\mathbf{y}}

\newcommand{\bz}{\mathbf{z}}
\newcommand{\bZ}{\mathbf{Z}}

\DeclareSymbolFont{bsfletters}{OT1}{cmss}{bx}{n}
\DeclareSymbolFont{ssfletters}{OT1}{cmss}{m}{n}
\DeclareMathSymbol{\bsfGamma}{0}{bsfletters}{'000}
\DeclareMathSymbol{\ssfGamma}{0}{ssfletters}{'000}
\DeclareMathSymbol{\bsfDelta}{0}{bsfletters}{'001}
\DeclareMathSymbol{\ssfDelta}{0}{ssfletters}{'001}
\DeclareMathSymbol{\bsfTheta}{0}{bsfletters}{'002}
\DeclareMathSymbol{\ssfTheta}{0}{ssfletters}{'002}
\DeclareMathSymbol{\bsfLambda}{0}{bsfletters}{'003}
\DeclareMathSymbol{\ssfLambda}{0}{ssfletters}{'003}
\DeclareMathSymbol{\bsfXi}{0}{bsfletters}{'004}
\DeclareMathSymbol{\ssfXi}{0}{ssfletters}{'004}
\DeclareMathSymbol{\bsfPi}{0}{bsfletters}{'005}
\DeclareMathSymbol{\ssfPi}{0}{ssfletters}{'005}
\DeclareMathSymbol{\bsfSigma}{0}{bsfletters}{'006}
\DeclareMathSymbol{\ssfSigma}{0}{ssfletters}{'006}
\DeclareMathSymbol{\bsfUpsilon}{0}{bsfletters}{'007}
\DeclareMathSymbol{\ssfUpsilon}{0}{ssfletters}{'007}
\DeclareMathSymbol{\bsfPhi}{0}{bsfletters}{'010}
\DeclareMathSymbol{\ssfPhi}{0}{ssfletters}{'010}
\DeclareMathSymbol{\bsfPsi}{0}{bsfletters}{'011}
\DeclareMathSymbol{\ssfPsi}{0}{ssfletters}{'011}
\DeclareMathSymbol{\bsfOmega}{0}{bsfletters}{'012}
\DeclareMathSymbol{\ssfOmega}{0}{ssfletters}{'012}

\newcommand{\bgamma}{\bm{\gamma}}

\newcommand{\btheta}{\bm{\theta}}

\newcommand{\bmeta}{\bm{\eta}}
\newcommand{\bkappa}{\bm{\kappa}}

\newcommand{\bxi}{\bm{\xi}}

\newcommand{\bGamma}{\bm{\Gamma}}

\newcommand{\bTheta}{\bm{\Theta}}

\makeatletter
\newcommand*\rel@kern[1]{\kern#1\dimexpr\macc@kerna}
\newcommand*\widebar[1]{%
  \begingroup
  \def\mathaccent##1##2{%
    \rel@kern{0.8}%
    \overline{\rel@kern{-0.8}\macc@nucleus\rel@kern{0.2}}%
    \rel@kern{-0.2}%
  }%
  \macc@depth\@ne
  \let\math@bgroup\@empty \let\math@egroup\macc@set@skewchar
  \mathsurround\z@ \frozen@everymath{\mathgroup\macc@group\relax}%
  \macc@set@skewchar\relax
  \let\mathaccentV\macc@nested@a
  \macc@nested@a\relax111{#1}%
  \endgroup
}
\makeatother

\DeclareMathOperator*{\argmax}{arg\,max}
\DeclareMathOperator*{\argmin}{arg\,min}

\DeclareMathOperator{\st}{s.t.\ }

\DeclareMathOperator{\diag}{diag}

\DeclareMathOperator{\tr}{tr}

\DeclareMathOperator{\vect}{vec}

\newcommand{\ifbcdot}[1]{\ifblank{#1}{\cdot}{#1}}

\DeclarePairedDelimiterX\abs[1]{\lvert}{\rvert}{\ifbcdot{#1}}
\DeclarePairedDelimiterX\parens[1]{(}{)}{\ifbcdot{#1}}
\DeclarePairedDelimiterX\brk[1]{[}{]}{\ifbcdot{#1}}
\DeclarePairedDelimiterX\braces[1]{\{}{\}}{\ifbcdot{#1}}
\DeclarePairedDelimiterX\angles[1]{\langle}{\rangle}{\ifblank{#1}{\cdot,\cdot}{#1}}
\DeclarePairedDelimiterX\ip[2]{\langle}{\rangle}{\ifbcdot{#1},\ifbcdot{#2}}
\DeclarePairedDelimiterX\norm[1]{\lVert}{\rVert}{\ifbcdot{#1}}
\DeclarePairedDelimiterX\ceil[1]{\lceil}{\rceil}{\ifbcdot{#1}}
\DeclarePairedDelimiterX\floor[1]{\lfloor}{\rfloor}{\ifbcdot{#1}}

\DeclarePairedDelimiterXPP\trace[1]{\operatorname{Tr}}{(}{)}{}{\ifbcdot{#1}} 
\DeclarePairedDelimiterXPP\col[1]{\operatorname{col}}{\{}{\}}{}{\ifbcdot{#1}} 
\DeclarePairedDelimiterXPP\row[1]{\operatorname{row}}{\{}{\}}{}{\ifbcdot{#1}} 
\DeclarePairedDelimiterXPP\erf[1]{\operatorname{erf}}{(}{)}{}{\ifbcdot{#1}}
\DeclarePairedDelimiterXPP\erfc[1]{\operatorname{erfc}}{(}{)}{}{\ifbcdot{#1}}
\DeclarePairedDelimiterXPP\KLD[2]{D}{(}{)}{}{\ifbcdot{#1}\, \delimsize\|\, \ifbcdot{#2}} 
\DeclarePairedDelimiterXPP\op[2]{\operatorname{#1}}{(}{)}{}{#2} 

\newcommand{\T}{^{\mkern-1.5mu\mathop\intercal}}
\newcommand{\He}{^{\mkern-1.5mu\mathsf{H}}}

\DeclarePairedDelimiterXPP\indicate[1]{{\bf 1}}{\{}{\}}{}{\ifbcdot{#1}}

\providecommand\given{}

\DeclarePairedDelimiterX\Set[2]\{\}{%
\renewcommand\given{\SetSymbol[\delimsize]{#1}}
#2
}
\DeclarePairedDelimiterX\Setc[1]\{\}{%
\renewcommand\given{\SetSymbol{:}}
#1
}

\NewDocumentCommand\set{s o m}{%
	\IfBooleanTF#1%
	{\IfValueTF{#2}{\Set*{#2}{#3}}{\Setc*{#3}}}%
	{\IfValueTF{#2}{\Set{#2}{#3}}{\Setc{#3}}}%
}

\NewDocumentCommand{\evalat}{ s O{\big} m e{_^} }{%
\IfBooleanTF{#1}%
{\left. #3 \right|}{#3#2|}%
\IfValueT{#4}{_{#4}}%
\IfValueT{#5}{^{#5}}%
}

\NewDocumentCommand \ifcondp {m m} {%
	#1%
	\IfValueT{#2}{\given #2}%
}

\providecommand\given{}
\DeclarePairedDelimiterXPP\cprob[1]{}(){}{
\renewcommand\given{\nonscript\,\delimsize\vert\allowbreak\nonscript\,\mathopen{}}
\ifcondp#1
}
\DeclarePairedDelimiterXPP\cexp[1]{}[]{}{
\renewcommand\given{\nonscript\,\delimsize\vert\allowbreak\nonscript\,\mathopen{}}
\ifcondp#1
}

\DeclareDocumentCommand \P { s e{_^} >{\SplitArgument{ 1 }{ @| }}d() g } {%
	\mathbb{P}%
	\IfBooleanTF{#1}%
		{
			\IfValueT{#2}{_{#2}}%
			\IfValueT{#3}{^{#3}}%
			\IfValueTF{#5}%
				{\cprob{#4 \given #5}}%
				{\IfValueT{#4}{\cprob{#4}}}%
		}%
		{
			\IfValueT{#2}{_{#2}}%
			\IfValueT{#3}{^{#3}}%
			\IfValueTF{#5}%
				{\cprob*{#4 \given #5}}%
				{\IfValueT{#4}{\cprob*{#4}}}%
		}%
}

\DeclareDocumentCommand \E { s e{_^} >{\SplitArgument{ 1 }{ @| }}d[] g } {%
	\mathbb{E}%
	\IfBooleanTF{#1}%
		{
			\IfValueT{#2}{_{#2}}%
			\IfValueT{#3}{^{#3}}%
			\IfValueTF{#5}%
				{\cexp{#4 \given #5}}%
				{\IfValueT{#4}{\cexp{#4}}}%
		}%
		{
			\IfValueT{#2}{_{#2}}%
			\IfValueT{#3}{^{#3}}%
			\IfValueTF{#5}%
				{\cexp*{#4 \given #5}}%
				{\IfValueT{#4}{\cexp*{#4}}}%
		}%
}

\ExplSyntaxOn
\NewDocumentCommand \dist {m o o} {%
\mathrm{#1}\left(%
	\IfValueT{#3}{%
		\tl_if_blank:nTF{ #3 }{\cdot\, \middle|\, }{#3\, \middle|\, }%
	}
	\IfValueT{#2}{#2}%
\right)%
}
\ExplSyntaxOff

\NewDocumentCommand {\cbrace} {t+ D[]{black} D(){\widthof{#5}} m m } {%
	\begingroup%
		\color{#2}
		\IfBooleanTF{#1}{%
			\overbrace{#4}^%
		}{
			\underbrace{#4}_%
		}%
		{\parbox[c]{#3}{\centering\footnotesize{#5}}}%
	\endgroup%
}

\let\oldforall\forall
\renewcommand{\forall}{\oldforall \, }

\let\oldexist\exists
\renewcommand{\exists}{\oldexist \, }

\graphicspath{{./Figures/}{./figures/}}
\DeclareDocumentCommand{\includeCroppedPdf}{ o O{./Figures/} m }{
	\IfFileExists{#2#3-crop.pdf}{}{%
		\immediate\write18{pdfcrop #2#3.pdf #2#3-crop.pdf}}%
	\includegraphics[#1]{#2#3-crop.pdf}
}

\makeatletter
\newcommand*{\addFileDependency}[1]{
  \typeout{(#1)}
  \@addtofilelist{#1}
  \IfFileExists{#1}{}{\typeout{No file #1.}}
}
\makeatother

\newcommand*{\myexternaldocument}[1]{%
    \externaldocument{#1}%
    \addFileDependency{#1.tex}%
    \addFileDependency{#1.aux}%
}

\definecolor{gray90}{gray}{0.9}

\ifx\nohighlights\undefined

	\newcommand{\msout}[1]{\text{\color{green} \sout{\ensuremath{#1}}}}
	\newcommand{\del}[1]{{\color{green}\ifmmode \msout{#1}\else\sout{#1}\fi}}
\else

	\newcommand{\msout}[1]{#1}
	\newcommand{\del}[1]{#1}
\fi

\newcommand{\hhide}[1]{}

\newcommand{\txp}[2]{\texorpdfstring{#1}{#2}}

\ifx\diagnoselabel\undefined
	\relax
\else
	\makeatletter
	\def\@testdef #1#2#3{%
		\def\reserved@a{#3}\expandafter \ifx \csname #1@#2\endcsname
			\reserved@a  \else
			\typeout{^^Jlabel #2 changed:^^J%
				\meaning\reserved@a^^J%
				\expandafter\meaning\csname #1@#2\endcsname^^J}%
			\@tempswatrue \fi}
	\makeatother
\fi

  \def\R{{\mathbb{R}}} \def\C{{\mathbb{C}}}   
\def\Eb{{\mathbb{E}}}

\newcommand{\beq}{\begin{eqnarray}}
\newcommand{\eeq}{\end{eqnarray}}

  \def\cC{{\mathcal{C}}}

 \def\cN{{\mathcal{N}}}

           \def\lA{\left\|}     \def\rA{\right\|}

\newcommand{\opRe}{\operatorname{Re}}
\newcommand{\opIm}{\operatorname{Im}}
\newcommand{\Col}{\operatorname{Col}}

\newacronym{MLE}{MLE}{maximum likelihood estimate}
\newacronym{BLUE}{BLUE}{best linear unbiased estimator}
\newacronym{WLS}{WLS}{weighted least squares}

\usepackage{setspace}
\usepackage{autobreak}
\usepackage{pdfpages}

\myexternaldocument{supp}

\begin{document}


\title{Approximate Maximum-Likelihood \protect\\ RIS-Aided Positioning}

\author{Wei Zhang, \IEEEmembership{Member,~IEEE}, Zhenni Wang, and
Wee Peng Tay, \IEEEmembership{Senior Member,~IEEE}
\thanks{

{Wei Zhang was with the School of Electrical and Electronic Engineering, Nanyang Technological University, Singapore 639798. He is
now with the School of Electronics and Information Engineering,
Harbin Institute of Technology, Shenzhen, Shenzhen 518055, China
(e-mail: zhangwei.sz@hit.edu.cn).}

{Zhenni Wang is with the Department of Electrical Engineering, City University of Hong Kong, Hong Kong SAR, China (e-mail: zhenni126@126.com).}

{Wee Peng Tay is with the School of Electrical and Electronic Engineering, Nanyang Technological University, Singapore 639798 (e-mail: wptay@ntu.edu.sg).}

}
}

\maketitle
\begin{abstract}
A reconfigurable intelligent surface (RIS) allows a reflection transmission path between a base station (BS) and user equipment (UE). In wireless localization, this reflection path aids in positioning accuracy, especially when the line-of-sight (LOS) path is subject to severe blockage and fading. In this paper, we develop a RIS-aided positioning framework to locate a UE in environments where the LOS path may or may not be available. We first estimate the RIS-aided channel parameters from the received signals at the UE. To infer the UE position and clock bias from the estimated channel parameters, we propose a fusion method consisting of weighted least squares over the estimates of the LOS and reflection paths. We show that this approximates the maximum likelihood estimator under the large-sample regime and when the estimates from different paths are independent. We then optimize the RIS phase shifts to improve the positioning accuracy and extend the proposed approach to the case with multiple BSs and UEs. We derive Cram\'er–Rao bound (CRB) and demonstrate numerically that our proposed positioning method approaches the CRB.
\end{abstract}
\vspace{-0.4cm}

\section{Introduction} \label{se:intro}
Localization or positioning is an important task in wireless communication systems \cite{Yassin2017Localization, WEN2019Survey}. In \gls{5G} systems, the positioning of the user equipment (UE) has diverse applications, including industrial use cases, smart mobility, and location-based services. The use of \gls{5G} millimeter wave (mmWave) has the potential to provide better positioning accuracy compared to global navigation satellite systems (GNSS) \cite{Wymeersch2017mmWave}. As such, the \gls{3GPP} Release 16 \cite{3gpp.21.916} has incorporated standards for location management in the \gls{5G} new radio (NR) framework.
By using sufficient measurements, the position of a UE can be estimated by measuring the received
signal strength indicator (RSSI) \cite{Vari2014mmV, Lin2018indoor}.
Apart from RSSI-based methods, some work \cite{Zhou2017low, Wen2021Pos} obtains the position of a UE in mmWave systems by estimating the channel parameters, i.e., time of arrival (ToA) and angle of arrival (AoA) or angle of departure (AoD). However, because of the high frequency of mmWave signals \cite{Heath16, ZhangSequ}, their transmission is easily blocked by obstacles, which makes positioning using mmWave systems challenging.
The NLOS paths can be leveraged to perform the positioning \cite{xu2015distributed, Shah2018Position, Wen2021Pos}, and from a theoretical perspective, incorporating NLOS paths into the position estimation process should result in improved accuracy. However, the degree of improvement in accuracy
may be limited by factors such as the signal strength of the NLOS paths relative to the LOS path. Utilizing them to perform accurate positioning also requires estimating or knowing additional information like the orientation or positions of their corresponding scatterers \cite{xu2015distributed, Shah2018Position, Wen2021Pos}, which make the positioning algorithms complex.

The reconfigurable intelligent surface (RIS) has been proposed to aid wireless communication systems.
A RIS consists of many low-cost meta elements \cite{DongSur2020, Basar2019WirelessAccess}, through which the performance of existing wireless communication systems can be improved without high additional hardware costs.
Unlike a relay, a RIS passively reflects the received signal, only changing its phase shift before transmission to a UE \cite{RenzoSur2020, EmilBeat2020,Zhang2021Cost}.
Compared to traditional wireless communication with transmit beamforming, the phase shifts of a RIS can be configured to achieve passive beamforming for RIS-aided systems \cite{ChenSum2019, ZhaoRIS2021}.
With properly designed passive beamforming, much work in the literature has shown that the RIS can improve various system performance metrics, such as spectral efficiency \cite{ChenSum2019, ZhaoRIS2021,Peng2022Peformance},  received signal-to-noise ratio (SNR) \cite{Atapattu2020RIS,Sun2022Energy} and bit error rate \cite{Ferreira2020Bit}. With the help of a RIS whose position is known a priori and whose phase shifts are controllable (unlike the non-RIS NLOS scenario), a reflected transmission path can be established if the LOS transmission path is blocked, which makes the RIS potentially useful and promising for urban or indoor positioning \cite{Liang2019Large,Hu2018Beyond}.

\begin{table*}
  \footnotesize
  \centering
  \caption{Summary of RIS-Aided Positioning Methods}\label{tab:referencetable}
  \begin{tabular}{C{1cm}L{1.3cm}L{0.8cm}L{0.8cm} L{0.8cm}L{1.0cm} L{1.4cm} L{1.5cm} L{1.9cm} L{3.6cm}  }\hline
    Ref.   & Multi-RIS  & 3D/2D& Multi-BS& Multi-UE&System & Rely on LOS path& Clock bias estimation & Parameters for positioning & Inference method    \\ \hline
    \cite{Zhang2021Meta}    & No   & 3D   & No  & Yes    & SISO                                        & No& No  & RSSI & MLE based on RSSI.\\ \hline
    \cite{abu2021near}         & No & 3D& No  & No  & MISO  &No & No & AoA, distance &  Exactly-determined using AoA and distance. \\ \hline
    \cite{wang2021joint}     & Yes& 3D& No  & No     & MIMO                         & Yes& No &AoA
 & LS based on AoA \\ \hline
    \cite{Key2021SISO}     & No& 2D & No  & No     &  SISO         &Yes & Yes &AoA, ToA & Exactly-determined using AoD and ToA.   \\ \hline
    \cite{Lin2021Channel}     & Yes & 3D& No  & No  & MIMO          & No  & No  &  AoA, ToA&Positioning from a single path using AoA and ToA. \\ \hline

    \cite{alexandropoulos2022localization} &Yes& 3D& No  & No  &   MIMO                      & No& No   &AoA &Positioning from AoA. \\ \hline
Ours &Yes, simultaneous& 3D& Yes  & Yes  &   MIMO       & Adaptive & Yes &AoA, AoD,  ToA& Approximate MLE based on all the estimated parameters.\\ \hline
  \end{tabular}
\end{table*}

Since a RIS creates a reflection path between a BS and a UE, the UE can utilize the measurements from this reflection path as additional information for positioning. The Cram\'er-Rao bound (CRB) of the positioning accuracy is analyzed in \cite{Hu2018Beyond,  Ji2020large, elzanaty2021reconfigurable}, which reveals the potential of using the RIS to improve positioning accuracy.
Meanwhile, some work focuses on explicit positioning algorithms in RIS-aided systems
\cite{Zhang2021Meta, abu2021near, wang2021joint, Key2021SISO, Lin2021Channel,alexandropoulos2022localization,keykhosravi2022ris}. A positioning framework typically consists of two steps: the first step is the estimation of channel parameters, and the second is the inference of UE position from the channel parameters.
For example, in \cite{Zhang2021Meta}, an indoor positioning using the RSSI is investigated, which estimates the position of a UE using the probability distribution of the RSSI.
Apart from RSSI, positioning methods based on AoA/AoD and ToA in RIS-aided systems are studied in \cite{abu2021near, wang2021joint, Key2021SISO, keykhosravi2022ris, Lin2021Channel,alexandropoulos2022localization} due to the advantages of massive multi-input multi-output (MIMO) systems.
In \cite{abu2021near}, a near-field positioning technique with RIS infers the UE position from the estimated AoAs and the distance between the RIS and UE, where the distance is estimated through near-field channel phases of  RIS elements.
In \cite{wang2021joint,alexandropoulos2022localization}, the authors consider the scenario with multiple RISs, and the UE position is calculated through the estimated AoAs.
Different from the assumption in \cite{abu2021near,wang2021joint,alexandropoulos2022localization} that UE and BS are synchronized, localization and clock bias estimation are investigated in \cite{Key2021SISO,keykhosravi2022ris} with both single-antenna UE and BS, and the position of a UE is obtained from the estimated AoD and ToA.
When both UE and BS are equipped with multiple antennas,  \cite{Lin2021Channel} studies the channel estimation and positioning under the twin-RIS scenario, and the UE position is obtained through the AoA and ToA of a single path.
However, the existing work has not addressed the following challenges (see \cref{tab:referencetable} for a summary):
\begin{itemize}
  \item When there are multiple RISs for positioning, the methods in the existing work \cite{wang2021joint, Lin2021Channel,alexandropoulos2022localization} activate each RIS one by one to estimate the parameters associated with the channel of each RIS. It requires that the BS or the controller manipulates the state of each RIS. To the best of our knowledge, no existing work has utilized multiple RISs simultaneously in positioning a UE.

  \item Different from the single RIS scenario \cite{abu2021near, Key2021SISO, keykhosravi2022ris, Zhang2021Meta} or the case where only one RIS is activated each time \cite{wang2021joint, Lin2021Channel,alexandropoulos2022localization}, matching of channel parameters with each path when using multiple RISs simultaneously is an important issue. When the matching is not accurate, the positioning accuracy is adversely affected.

  \item The theoretical CRB indicates that positioning accuracy can be improved when more parameters are utilized to infer the UE position. Instead of utilizing the parameters of all the paths, the references \cite{wang2021joint, alexandropoulos2022localization} only utilize partial information such as AoA as shown in \cref{tab:referencetable}. The reference \cite{Lin2021Channel} uses AoA and ToA but does not perform clock bias estimation, which is critical in ToA estimation. The clock bias between the local oscillator at the UE and that at the BS needs to be accounted for \cite{Key2021SISO}. Otherwise, the obtained ToA may be highly inaccurate.
\end{itemize}

\begin{itemize}
  \item The existing inference methods for UE positioning based on estimated channel parameters (AoA, AoD, and ToA) in the works \cite{wang2021joint, Lin2021Channel, abu2021near, Key2021SISO,keykhosravi2022ris} do not consider the accuracy of the LOS and reflection path parameters.
  These methods are not necessarily optimal from a CRB perspective. For instance, the channel parameters of strong signal paths can be estimated more accurately than those of weak signal paths, which should be given more weight during the UE positioning inference. However, the approach in \cite{wang2021joint} uses the LS criterion equally for all paths, and \cite{Lin2021Channel} focuses on single-path positioning, which can result in suboptimal positioning results.
  The work in \cite{alexandropoulos2022localization} considers parameter accuracy but lacks theoretical optimality analysis.
\end{itemize}

In this paper, we develop a three-dimensional positioning and clock bias estimation framework for {RIS-aided} systems using channel estimation techniques.
In our framework, multiple RISs work simultaneously, and the BS and UE are equipped with multiple antennas.
In addition, different from the LS method or exactly-determined solution in \cite{abu2021near, wang2021joint, Key2021SISO, Lin2021Channel}, our proposed inference model considers the estimation accuracy of the channel parameters of different paths.
Through simulations and theoretical analysis, we show that the proposed inference approach yields a UE position estimation error close to the CRB. \cref{tab:referencetable} summarizes the differences between this paper and the literature. The main contributions of this paper are:
\begin{itemize}
  \item We consider the downlink MIMO \gls{OFDM} setup in this work, where multiple RISs are used simultaneously. We utilize the two-step positioning framework to estimate the UE position and clock bias from the received signals. In the first step, we estimate the channel parameters of the LOS and reflection paths. Since multiple RISs are used simultaneously, which generates multiple paths, we propose an energy-based method to match the channel parameters with each path. In the second step, we infer the UE position from all estimated channel parameters. We also derive the CRB of the UE position estimate.
  \item  To infer the UE position and clock bias from the estimated channel parameters, we use a \gls{WLS} optimization with the positioning estimates of the LOS and reflection paths, which can be treated as an information fusion of these paths. In particular, the proposed \gls{WLS} information fusion method adaptively relies less on those paths with low SNRs.
  The weights in the formulated problem depend only on the covariance of the estimates. The proposed \gls{WLS} method utilizes the approximate covariance based on the Fisher information matrix (FIM).
  We show that when the channel estimates from different paths are independent, the proposed WLS solution is approximately the \gls{MLE} of UE position in the large-sample regime.
  \item To optimize the positioning framework, we propose a \gls{SVD}-based approach for designing the RIS phase shifts. Specifically, assuming each RIS serves the UEs in a certain range, the phase shift design problem maximizes the expectation of the reflection path gain, which can then be solved using \gls{SVD}. Our proposed RIS-aided positioning framework is also readily extended to the multi-UE and multi-BS scenarios, and the proposed \gls{SVD}-based design for phase shifts of the RISs can also be applied in these scenarios.

\end{itemize}

The rest of this paper is organized as follows. In \cref{sect:signal model}, the signal and channel model and our system assumptions are introduced.
In \cref{sec:CRB}, we derive the CRB of the UE positioning error under the signal and channel model.
The proposed RIS-aided channel parameter estimation approach is discussed in \cref{sec:estimation}. In \cref{sec:fusion}, we present our fusion method to infer the UE position from the estimated channel parameters. In \cref{sec:discussions}, we propose a method to optimize the RIS phase shifts and discuss the extension of our positioning framework to the multi-UE and multi-BS scenarios.
We present numerical results in \cref{sec:numerical}. Finally, we conclude in \cref{sec:conclusions}.

\emph{Notations:} A bold lower case letter $\mathbf{a}$ is a vector and a bold capital letter $\mathbf{A}$ represents a matrix. $\bA\T$, ${{\mathbf{A}}^{\He}}$, ${{\mathbf{A}}^{-1}}$, $\mathrm{tr}(\mathbf{A})$, $\left| \mathbf{A} \right|$,  ${{\left\| \mathbf{A} \right\|}_{\text{F}}}$ and ${{\left\| \mathbf{a} \right\|}_{2}}$ are, respectively, the transpose, Hermitian, inverse, trace, determinant, Frobenius norm of $\bA$, and the $2$-norm of $\mathbf{a}$.
${{[\mathbf{A}]}_{:.i}}$, ${{[\mathbf{A}]}_{i,:}}$, ${{[\mathbf{A}]}_{i, j}}$, and $[\ba]_i$ are, respectively, the $i$th column, $i$th row, $i$th row and $j$th column entry of $\mathbf{A}$, and the $i$th entry of vector $\mathbf{a}$. The operation $\mathrm{\mathop{vec}}(\mathbf{A})$ stacks the columns of $\mathbf{A}$ to form a column vector.
$\Col(\bA)$ is the column space of matrix $\bA$.
We use $\diag(\ba)$ to represent a diagonal matrix with the vector $\ba$ on the main diagonal.
The circular symmetric complex Gaussian distribution with mean $\mu$ and variance $\sigma^2$ is given by $\dist{\calC\calN}[\mu, \sigma^2]$. The uniform distribution over $[a,b]$ is denoted by $U[a,b]$. We let $\otimes$ denote the Kronecker product.

For convenience, we list some commonly-used notations in \cref{tab:notation definition}. The symbols in \cref{tab:notation definition} are defined formally where they first appear in the paper.

\begin{table}[!htb]
\renewcommand{\arraystretch}{1.2}
  \footnotesize
  \centering
  \caption{Commonly-used Notations}\label{tab:notation definition}

  \def\defcol{4.5cm}
  \def\symcol{2cm}

\begin{subtable}[t]{0.45\textwidth}
  \begin{tabular}{C{\symcol}ll}\hline
    \textbf{Symbol}                            & \textbf{Definition}                &                      \\ \hline
    $\bp_B$, $\bp_U$, $\bp_{R,q}$              & \multicolumn{2}{L{\defcol}}{position of BS, UE, $q$-th RIS}  \\ \hline
    $\tau_{BR,q}$, $\tau_{RU,q}$, $\tau_{BU}$  & \multicolumn{2}{L{\defcol}}{propagation delay from BS to $q$-th RIS, $q$-th RIS to UE,  BS to UE}                     \\ \hline
        $c$              & \multicolumn{2}{L{\defcol}}{speed of light, $3\times 10^8 \text{m/s}$}  \\ \hline
                                               & \multicolumn{2}{L{\defcol}}{elevation and azimuth AoDs for} \\ \hline
    ${\theta}_{BR,q}$, ${\phi}_{BR,q}$ &                                    & BS-RIS link          \\ \hline
    $\theta_{RU,q}$, $\phi_{RU,q}$ &                                    & RIS-UE link          \\ \hline
    $\theta_{BU}$, $\phi_{BU}$     &                                    & BS-UE link           \\ \hline
                                               & \multicolumn{2}{L{\defcol}}{elevation and azimuth AoAs for} \\ \hline
    $\bar{\theta}_{BR,q}$, $\bar{\phi}_{BR,q}$ &                                    & BS-RIS link          \\ \hline

      $\Delta_{\text{clock}} $                   & \multicolumn{2}{L{\defcol}}{clock bias between BS and UE}   \\ \hline
     $\bO$              & \multicolumn{2}{L{\defcol}}{rotation matrix of UE}  \\ \hline
        $\bmeta,\bmeta_B,\bmeta_{R,q}$                  & \multicolumn{2}{L{\defcol}}{channel parameters in \cref{eq:bmeta}}   \\ \hline
            $\bxi,\bxi_B,\bxi_{R,q}$                  & \multicolumn{2}{L{\defcol}}{UE position parameters in \cref{eq:bxi}}   \\ \hline
                                 $\bF_{\bmeta},\bF_{\bxi}$, $\bF_{\bmeta_B},\bF_{\bmeta_{R,q}}$                  & \multicolumn{2}{L{\defcol}} {FIM of $\bmeta,\bxi,\bmeta_B,\bmeta_{R,q}$ }   \\ \hline
                                    $\bJ, \bJ_B, \bJ_{R,q} $                  & \multicolumn{2}{L{\defcol}}{Jacobian matrix $\frac{\partial \bmeta}{\partial\bxi\T},\frac{\partial \bmeta_B}{\partial\bxi_B\T},\frac{\partial \bmeta_{R,q}}{\partial\bxi_{R,q}\T}$ }   \\ \hline
  $F,F_B, F_{R,q}$                  & \multicolumn{2}{L{\defcol}} {function of $\bxi\rightarrow \bmeta$, $\bxi_B\rightarrow \bmeta_B$, $\bxi_{R,q}\rightarrow \bmeta_{R,q}$}   \\ \hline

                $N$, $D$              & \multicolumn{2}{L{\defcol}}{number of antennas of BS, UE}  \\ \hline

  \end{tabular}
\end{subtable}
\begin{subtable}[t]{0.45\textwidth}
  \begin{tabular}{C{\symcol}ll}\hline
    \textbf{Symbol}                            & \textbf{Definition}                &                      \\ \hline
  $M$              & \multicolumn{2}{L{\defcol}}{number of elements in one RIS}  \\ \hline
$Q$              & \multicolumn{2}{L{\defcol}}{number of RISs}  \\ \hline

           $K$              & \multicolumn{2}{L{\defcol}}{number of \gls{OFDM} subcarriers}  \\ \hline
       $T$              & \multicolumn{2}{L{\defcol}}{number of transmit slots}  \\ \hline
           $K_{BU}, K_{RU}$              & \multicolumn{2}{L{\defcol}}{Rician factor of BS-UE channel, RIS-UE channel}  \\ \hline
                $\bZ_{BU,k}$              & \multicolumn{2}{L{\defcol}}{small scale channel fading  of BS-UE link}  \\ \hline
       $\bR_k$              & \multicolumn{2}{L{\defcol}}{received signal at UE for $k$-th subcarrier}  \\ \hline
        $\tilde{\bR}_k$              & \multicolumn{2}{L{\defcol}}{defined in \cref{eq:R observations}}  \\ \hline
           $h_{BR,q}$, $h_{RU,q}$, $h_{BU}$              &   \multicolumn{2}{L{\defcol}}{complex path gain of BS-RIS link, RIS-UE link,  BS-UE link}  \\ \hline
            $\beta_{BR,q}$, $\beta_{RU,q}$, $\beta_{BU}$              &   \multicolumn{2}{L{\defcol}}{large scale path gain of BS-RIS link, RIS-UE link,  BS-UE link}  \\ \hline
             $\alpha_{BR,q}$, $\alpha_{RU,q}$, $\alpha_{BU}$              &   \multicolumn{2}{L{\defcol}}{complex coefficient of BS-RIS link, RIS-UE link, BS-UE link}  \\ \hline
             ${\bp}_U^{(B)},{\bp}_U^{(R,q)}$ &   \multicolumn{2}{L{\defcol}}{defined in \cref{eq:sub clock}}  \\ \hline
               $ \bC_{{\bp}_{U}}^{(B)},\bC_{{\bp}_{U}}^{(R,q)}$              &   \multicolumn{2}{L{\defcol}}{error covariance matrix of ${\bp}_U^{(B)},{\bp}_U^{(R,q)}$  defined in \cref{eq:cov_direct}, \cref{eq:cov_reflect}}  \\ \hline
$ \br^{(B)},\br^{(R,q)}$              &   \multicolumn{2}{L{\defcol}}{unit directional vector of the LOS path, reflection path } \\ \hline

  \end{tabular}
\end{subtable}

\end{table}

\section{System Model} \label{sect:signal model}

In this section, we present our system model and assumptions. We first discuss the channel model, which contains BS-RIS, RIS-UE, and BS-UE links. We then present the received signal model at UE. Throughout this paper, we let the abbreviations $B$, $R$, and $U$ in symbol notations denote quantities related to the BS, a RIS, and the UE, respectively. The BS to the $q$-th RIS channel, the $q$-th RIS to the UE channel, and the BS to the UE channel are thus denoted as $(BR, q)$, $(RU, q)$ and $BU$, respectively, in symbol notations.

\begin{figure}[!htbp]
\centering
\begin{minipage}[t]{0.48\textwidth}
\centering
\includegraphics[width=0.95\textwidth]{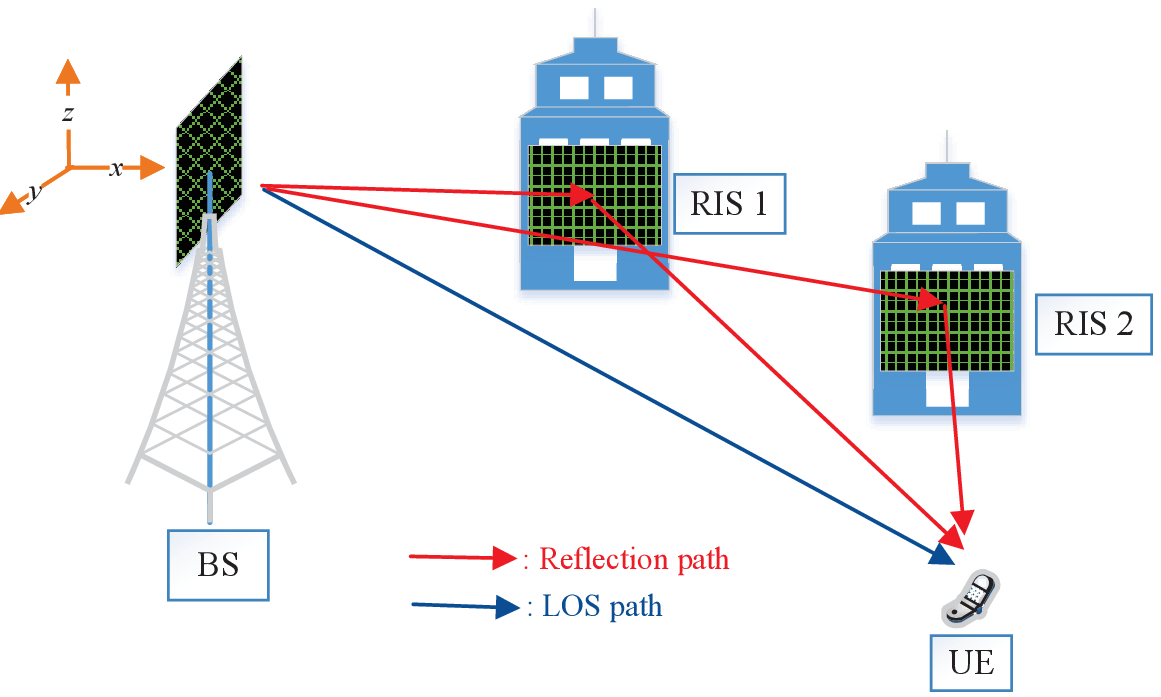}
\caption{Positioning of a UE with the aid of two RISs.} \label{fig:system}
\end{minipage}
\begin{minipage}[t]{0.45\textwidth}
\centering
\includegraphics[width=0.60\textwidth]{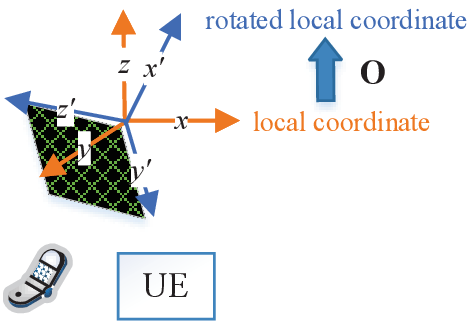}
\caption{Orientation of the UE URA w.r.t.\ its local coordinate system.} \label{fig:rotation}
\end{minipage}
\end{figure}

\subsection{Channel Model}\label{sect:channel model}
Suppose there is a single BS with $Q$ RISs working simultaneously. The BS controls the RISs via reliable links.
In this work, we assume the UE is stationary during the positioning time period.\footnote{For the mobile case, the speed of the UE can be estimated through the Doppler-induced phase rotations of the received signal, which has been investigated for the SISO scenario \cite{keykhosravi2022ris}. The positioning for mobile UEs in the MIMO scenario is an interesting future work.}

We adopt a global coordinate system with the BS at its origin, i.e., $\bp_B=[x_B, y_B, z_B]\T=[0,0,0]\T$.
We denote the position of the UE as $\bp_U=[x_U, y_U, z_U]\T$, and the position of the $q$-th RIS as $\bp_{R, q}=[x_{R, q},y_{R, q},z_{R, q}]\T$. The local coordinate system at each RIS or UE is obtained through the translation of the global coordinate by $\bp_{R, q}$ or $\bp_U$. Without loss of generality, we assume that the URA of the BS is in the $y-z$ plane of the global coordinate (see \cref{fig:system} for an illustration). Each RIS' URA is assumed to be contained in the $x-z$ plane of the local coordinate of the RIS, which is perpendicular to the $y-z$ plane of the BS URA.
The UE URA lies in a plane characterized by a rotation matrix $\bO \in \R^{3 \times 3}$ \gls{wrt} its local coordinate system (see \cref{fig:rotation}).
The rotation matrix $\bO$ is determined by three Euler angles under the 3D scenario. In practice, $\bO$ can be obtained from an inertial measurement unit (IMU), which is a common sensor on UEs like smartphones.

We suppose the communication system uses \gls{OFDM} with $K$ subcarriers.
We assume the BS has a uniform rectangular array (URA) with $N=N_r \times N_c$ antennas, where $N_r$ and $N_c$ are the numbers of rows and columns of the URA, respectively. Similarly, we assume that each RIS is equipped with a URA with $M=M_r \times M_c$ elements, and the UE is equipped with a URA with $D=D_r \times D_c$ antennas.
We also assume that each integer $N_r, N_c, D_r, D_c$ is larger than $Q+1$, and $K >Q+1$ so that different AoDs, AoAs, and delays can be estimated \cite{shakeri2012direction}.
For the $k$-th subcarrier, the channel from BS to the $q$-th RIS is denoted as $\bH_{BR,k, q}\in \C^{M \times N}$, the channel from the $q$-th RIS to the UE is $\bH_{RU, k,q} \in \C^{D\times M}$, and the channel from BS to UE is $\bH_{BU, k} \in \C^{D\times M}$.

In the following, we present explicit characterizations of these channels.

To define the channel model for each link, we first define the URA response vector at the BS, RIS, and UE. Let $n \in \{ N, M,D \}$ denote the number of antennas or elements with $n_r$ and $n_c$ being the corresponding number of rows and columns, respectively.
Thus, the URA response vector $\ba_n(x,y) \in \C^{n\times 1}$ at the BS, RIS, or UE is given by
\begin{align}
\ba_n(x, y) &=  \tilde\ba_{n_r}(x) \otimes \tilde\ba_{n_c}(y), \label{eq:URA response}
\end{align}
where $\tilde{\ba}_{m}(x) = \frac{1}{\sqrt{m}}\left[ 1, e^{\iu \pi x}, e^{\iu 2 \pi x},\ldots,e^{\iu (m-1) \pi x} \right]\T \in \C^{m\times 1}$ for $m\in \{n_c,n_r \}$, $\iu$ is the imaginary unit, and $x, y$ are trigonometric functions of the AoD and AoA at the transmitter and receiver. The explicit expressions for \cref{eq:URA response} are presented in the following discussions.

\subsubsection{BS-RIS link}

We model the BS-RIS channel as a mmWave channel, and we assume each RIS is placed at a sufficient height (e.g., on a tall
building), so there is a LOS path between the BS and the RIS.

From the coordinate system in \cref{fig:system}, let $\theta_{BR,q}$ (or $\bar{\theta}_{BR, q} $)   and  $\phi_{BR,q}$ (or $\bar{\phi}_{BR, q} $) be the elevation and azimuth AoDs %
(or AoAs) associated with $q$-th BS-RIS link, respectively. Following \cite{He2020Cascaded, chen2019channel}, we define\footnote{In this paper, in order to map the azimuth and elevation angles $(\theta, \phi)$ in 3D to the Cartesian coordinate system $(f,g,s)$, and we let $f=\cos \theta \cos \phi$, $g=\cos \theta \sin \phi$, and $s=\sin \theta$.}
\begin{align}
f_{BR,q}^{(R)} &= \cos\bar{\theta}_{BR,q} \cos \bar{\phi}_{BR,q}, ~s_{BR,q}^{(R)} = \sin \bar{\theta}_{BR,q} , \label{eq:fR1q}\\
g_{BR, q}^{(B)} &= \cos\theta_{BR,q} \sin \phi_{BR, q} ,~s_{BR,q}^{(B)} = \sin \theta_{BR,q}. \label{eq:fBrq}
\end{align}
From the \gls{OFDM} assumption, the $k$-th subcarrier of the $q$-th BS-RIS channel is then given by
\begin{align}
    \bH_{BR,k, q}
    &=   h_{BR, q}e^{-\iu 2\pi \frac{kW}{K} \tau_{BR, q}} \ba_M(f_{BR,q}^{(R)}, s_{BR,q}^{(R)}  )
  \times \ba_N\He(g_{BR, q}^{(B)},s_{BR,q}^{(B)}),   \label{eq:Channel_model_BS_RIS}
\end{align}
where $ h_{BR, q} = \alpha_{BR, q}\sqrt{\beta_{BR, q}MN}$ with $\beta_{BR, q}$ being the large scale path gain and $\alpha_{BR, q}$ being a complex-valued coefficient, $W$ is the transmission bandwidth, and $\tau_{BR, q}$ is the propagation delay of the signal from BS to the $q$-th RIS.
The quantities $\ba_M(f_{BR,q}^{(R)}, s_{BR,q}^{(R)}) \in \C^{M\times 1}$ and  $\ba_N(g_{BR, q}^{(B)},s_{BR,q}^{(B)})\in \C^{N\times 1}$ are the URA response vectors of the RIS and BS, respectively,  as defined in by \cref{eq:URA response}.

\subsubsection{RIS-UE link}
For the channel between the $q$-th RIS and the UE, we assume a LOS path. Compared to the RIS, there are more scatterers around the UE. Therefore, we model the RIS-UE link channel using the Rician fading model given by
\begin{align}
\bH_{RU,k,q} =\widebar{\bH}_{RU,k,q} + {\bZ}_{RU,k,q}, \label{eq:Channel_model_RIS_UE Z}
\end{align}
where $\widebar{\bH}_{RU,k,q}$ is the LOS path, and $\bZ_{RU,k,q}$ denotes small-scale fading\footnote{The small scale-fading includes the NLOS components of the RIS-scatterer-UE paths.} whose entries are \gls{iid} according to $\cC\cN(0,\beta_{RU,q}/{(1+K_{RU})})$ with $\beta_{RU,q}$ being the large scale path gain and $K_{RU}$ being the Rician factor. Without loss of generality, we assume that the Rician factor is constant for all the $Q$ RIS-UE links.
 The expression of $\widebar{\bH}_{RU, k,q} $ is given by
\begin{align}
  \widebar{\bH}_{RU, k,q}
  &= h_{RU, q} e^{-\iu 2\pi \frac{kW}{K}\tau_{RU, q}}\ba_D(g_{RU, q}^{(U)},  s_{RU, q}^{(U)})  \ba_M\He(f_{RU,q}^{(R)}, s_{RU,q}^{(R)}),   \label{eq:Channel_model_RIS_UE}
\end{align}
where $h_{RU, q} = \alpha_{RU, q}\sqrt{ \frac{K_{RU}}{1+K_{RU}} \beta_{RU, q}MD} $ with $\alpha_{RU, q}$ being complex-valued channel coefficient, and $\tau_{RU, q}$ is the delay.
The  $\ba_M(f_{RU,q}^{(R)},s_{RU,q}^{(R)})$ and $\ba_D(g_{RU, q}^{(U)}, s_{RU, q}^{(U)})$ are the URA response vectors of RIS and UE  defined in \cref{eq:URA response}. Based on the coordinate systems defined in \cref{fig:system,fig:rotation}, we have
\begin{small}
\begin{align}
& f_{RU,q}^{(R)} = \cos \theta_{RU,q} \cos \phi_{RU,q}, ~s_{RU,q}^{(R)} = \sin \theta_{RU,q} ,\label{AoA:RU_q 1} \\
&
\begin{bmatrix}
g_{RU, q}^{(U)}\\
s_{RU, q}^{(U)}
\end{bmatrix}=
\bO_R
\begin{bmatrix}
{f_{RU,q}^{(R)}}\\
\cos\theta_{RU,q} \sin \phi_{RU,q}\\
{s_{RU,q}^{(R)}}
\end{bmatrix}, \label{AoA:RU_q}
\end{align}
\end{small}%
where  $\theta_{RU,q}$  and  $\phi_{RU,q}$ are the elevation and azimuth AoDs associated with the RIS-UE link
and  $\bO_R =-[\bO]_{2:3,:}$(cf.\ \cref{fig:rotation}). Abusing terminology, we refer to $(f_{RU,q}^{(R)},s_{RU,q}^{(R)})$ as the AoD of the $q$-th RIS, and $(g_{RU, q}^{(U)},s_{RU, q}^{(U)})$ as the AoA of the UE on the reflection path.

\subsubsection{BS-UE link}
Similar as the RIS-UE link, we model the BS-UE link channel using the Rician fading model given by
\begin{align}
\bH_{BU,k} =\widebar{\bH}_{BU,k} + {\bZ}_{BU,k}, \label{eq:Channel_model_BS_UE}
\end{align}
where  $\widebar{\bH}_{BU,k}$ is the LOS path between BS and UE, and $\bZ_{BU,k}$ denotes small-scale fading\footnote{The small scale-fading includes the NLOS components of  the BS-scatterer-UE paths.} whose entries are \gls{iid} according to $\cC\cN(0,{\beta_{BU}}/(1+K_{BU}))$ with $\beta_{BU}$ being the large scale path gain and $K_{BU}$ being the Rician factor for BS-UE link.
The expression of $\widebar{\bH}_{BU,k}$ is given by
\begin{align}
\widebar{\bH}_{BU,k} =h_{BU} e^{-\iu2\pi \frac{kW}{K}\tau_{BU}}\ba_D(g_{BU}^{(U)}, s_{BU}^{(U)})\ba_N\He(g_{BU}^{(B)},s_{BU}^{(B)}),
\end{align}
where $h_{BU} =\sqrt{ \frac{K_{BU}}{1+K_{BU}}\beta_{BU} ND}  \alpha_{BU}$ with $\alpha_{BU}$ being complex-valued channel coefficient, and the URA response vectors of the UE and the BS, $\ba_D(g_{BU}^{(U)}, s_{BU}^{(U)})$ and $\ba_N(g_{BU}^{(B)}, s_{BU}^{(B)})$ are the URA response vectors of UE and BS defined in \cref{eq:URA response}. Based on the coordinate systems defined in \cref{fig:system} and \cref{fig:rotation}, we have
\begin{small}
\begin{align}
 g_{BU}^{(B)} &=\cos \theta_{BU}\sin \phi_{BU} , ~s_{BU}^{(B)}=\sin \theta_{BU}, \label{AoD:B_d}\\
\begin{bmatrix}
g_{BU}^{(U)}\\
s_{BU}^{(U)}
\end{bmatrix}
&=
\bO_R
\begin{bmatrix}
\cos\theta_{BU} \cos \phi_{BU}\\
 g_{BU}^{(B)}\\
s_{BU}^{(B)}
\end{bmatrix}, \label{AoD:U_d}
\end{align}
\end{small}%
where $\theta_{BU}$  and  $\phi_{BU}$ are the elevation and azimuth AoDs associated with the BS-UE link. Abusing terminology, we refer to $(g_{BU}^{(B)},s_{BU}^{(B)})$ as the AoD of the BS, and $(g_{BU}^{(U)},s_{BU}^{(U)})$ as the AoA of the UE on the LOS path.

In summary, using the channel models of the BS-RIS link in \cref{eq:Channel_model_BS_RIS}, the RIS-UE link in \cref{eq:Channel_model_RIS_UE}, and the BS-UE link in \cref{eq:Channel_model_BS_UE}, the effective channel between the BS and UE on the $k$-th subcarrier can be written as
\begin{small}
\begin{align}
\bH_k &= {\bH}_{BU,k}+\sum_{q=1}^{Q} \bH_{RU, k,q} \bTheta_q \bH_{BR, k,q}  \label{eq:final expression of Hk}\\
&= \widebar{\bH}_{BU,k}+\sum_{q=1}^{Q} \widebar{\bH}_{RU, k,q} \bTheta_q \bH_{BR, k,q} +{\bZ}_{k} \nonumber \\
&=  h_{BU} e^{-\iu 2\pi \frac{kW}{K}\tau_{BU}} \ba_D(g_{BU}^{(U)}, s_{BU}^{(U)})\ba_N\He(g_{BU}^{(B)},s_{BU}^{(B)}) 
+\sum_{q=1}^{Q} h_{R, q} e^{-\iu 2\pi \frac{kW}{K}(\tau_{BR, q} + \tau_{RU, q})} \nonumber \\
&\quad\quad\times \ba_D (g_{RU, q}^{(U)}, s_{RU, q }^{(U)})\ba_N\He(g_{BR, q}^{(B)}, s_{BR,q}^{(B)})  +{\bZ}_{k},\nonumber
\end{align}
\end{small}%
where $\bTheta_q=\diag(\btheta_q)$ with $\btheta_q = [\theta_q^{(1)}, \ldots, \theta_q^{(M)}]\T$ denoting the phase shifts of the $q$-th RIS,  $h_{R, q}=h_{BR, q}h_{RU, q}\ba_M\He(f_{RU,q}^{(R)},s_{RU,q}^{(R)})\bTheta_q \ba_M(f_{BR,q}^{(R)},s_{BR,q}^{(R)})$, and $\bZ_k = \bZ_{BU,k}+ \sum_{q=1}^{Q} \bZ_{RU, k,q} \bTheta_q \bH_{BR, k,q}$.
It is worth noting that different from the existing works \cite{wang2021joint,Lin2021Channel,alexandropoulos2022localization} where only one RIS is activated at each time, the multiple RISs work simultaneously in our framework. In addition, one can observe that the entries in $\bZ_k$ can be approximated by the \gls{iid} Gaussian distribution $\cC\cN\left(0,\frac{\beta_{BU}}{1+K_{BU}}+\frac{QM\beta_{RU}\beta_{BR}}{1+K_{RU}} \right)$.
Here, we define the channel parameters in \cref{eq:final expression of Hk} as
\begin{small}
\begin{subequations}\label{eq:bmeta}
\begin{align}
& \boldsymbol{\eta}_{B} =[\opRe\{h_{BU}\}, \opIm\{ h_{BU} \},\tau_{BU}, g_{BU}^{(U)},s_{BU}^{(U)},g_{BU}^{(B)},s_{BU}^{(B)}]\T, \\
& \boldsymbol{\eta}_{R, q} =[\opRe\{h_{R, q}\},\opIm\{h_{R, q}\}, \tau_{RU, q},g_{RU, q}^{(U)},s_{RU, q}^{(U)}]\T, \\
& \boldsymbol{\eta}=[\boldsymbol{\eta}_{B}\T,\boldsymbol{\eta}_{R,1}\T, \ldots, \boldsymbol{\eta}_{R, Q}\T] \T\in \R^{(7+5Q)\times 1}.
\end{align}
\end{subequations}
\end{small}%

To relate the channel parameters $\bmeta$ to the UE position and clock bias, we let
\begin{small}
\begin{subequations}\label{eq:bxi}
\begin{align}
& \bxi_B = [({\bp}_U^{(B)})\T,\opRe\{h_{BU}\}, \opIm\{h_{BU}\} ]\T, \\
& \bxi_{R,q} = [({\bp}_U^{(R, q)})\T, \opRe\{ h_{R, q}\}, \opIm\{h_{R, q}\}]\T, \\
& \bxi = [\bp_U\T,\Delta_\mathrm{clock}, \opRe\{h_{BU}\}, \opIm\{h_{BU}\},\opRe\{h_{R,1}\}, \opIm\{h_{R,1}\},
\ldots,\opRe\{h_{R, Q}\}, \opIm\{h_{R, Q}\}]\T ,
\end{align}
\end{subequations}
\end{small}%
where $\Delta_\mathrm{clock}$ is the clock bias between BS and UE, and ${\bp}_U^{(B)}$ and ${\bp}_U^{(R, q)}$ are given as follows
\begin{small}
\begin{subequations}\label{eq:sub clock}
\begin{align}
{\bp}_U^{(B)} &= \bp_U-c\Delta_{\text{clock}}\frac{\bp_U-\bp_B}{\| \bp_U-\bp_B \|_2},\\
{\bp}_U^{(R, q)} &= \bp_U-c\Delta_{\text{clock}}\frac{ \bp_U-\bp_{R, q}}{\|  \bp_U-\bp_{R, q} \|_2}.
\end{align}
\end{subequations}
\end{small}%
Then, we can define a function
\begin{small}
\begin{align}
F(\bxi)= \bmeta, \label{eq:F}
\end{align}
\end{small}
\!\!\!from the relations of \cref{eq:fR1q}, \cref{eq:fBrq}, \cref{AoA:RU_q 1}, \cref{AoA:RU_q}, \cref{AoD:B_d}, \cref{AoD:U_d}, and  the following equalities:
\begin{small}
\begin{align*}
\tau_{BU}  &={\| \bp_U-\bp_B \|_2}/{c}+\Delta_\mathrm{clock}, \quad
\tau_{RU, q}  = {\| \bp_U-\bp_{R, q} \|_2}/{c}+\Delta_\mathrm{clock}, \nonumber \\
\theta_{BU} &= \arcsin {\frac{z_U-z_B}{\| \bp_U-\bp_B \|_2}},\quad
\phi_{BU}  =  \arctan 2(y_U-y_B, x_U-x_B), \nonumber \\
\theta_{RU,q} &= \arcsin{\frac{z_U-z_{R, q}}{\| \bp_U-\bp_{R, q} \|_2}},\quad
\phi_{RU,q}   =  \arctan 2(y_U-y_{R, q},x_U-x_{R, q}).
\end{align*}
\end{small}

\subsection{Received Signal at the UE}\label{sect:received signal}
Suppose that the UE receives signals over $T$ time slots with $T\ge N$. From the channel model \cref{eq:final expression of Hk}, the received signal at the UE at each time $t=1,\ldots, T$ on the $k$-th subcarrier is given by
\begin{align}
  \br_k(t) &= \bH_k \bx (t)  +  \bn_k(t), \label{eq:t receive signal}
\end{align}
where $\bx(t) \in \C^{N \times 1}$ denotes the signal after precoding by the BS at time $t$, and $\bn_k(t) \in \C^{D \times 1}$ is a noise vector with entries \gls{iid} according to the complex Gaussian distribution $\dist{\cC\cN}[0, \sigma^2 ]$ and independent across time.
Specifically, the expression of $\bx(t)$ is given by $\bx(t)=\bW(t) \bs(t)$, where $\bW(t) \in \C^{N \times d}$ is the precoding matrix and $\bs(t) \in \C^{d \times 1}$ is the transmitted signal with $d$ data streams.
To satisfy the hybrid structure that the precoding matrix $\bW(t)$ must be unit modulus \cite{Heath16}, we let $\bW(t)$ be the sub-matrix of the discrete Fourier transform (DFT) matrix and $\bs(t)$ be the standard basis vector with proper normalization.

Let
$\bR_k = [\br_k(1),\ldots, \br_k(T)] \in \C^{N\times T}$,
$\bX= [\bx(1),\ldots, \bx(T)] \in \C^{N\times T}$, and
$\bN_k = [\bn_k(1),\ldots, \bn_k(T)] \in \C^{D\times T}$.
Since $T\ge N$, we can achieve $\bX\bX^H = \frac{pT}{N}\bI$ based on the discussions above, where $p=\|\bx(t)\|_2^2$ is the transmit power, for $t=1,\ldots, T$,  and $\bI$ is the identity matrix.
The compact form of the received signal in \cref{eq:t receive signal} is given by
$
\bR_k = \bH_k \bX + \bN_k.
$
Right multiplying $\bR_k $ by $\frac{N}{pT}\bX\He$, we have
\begin{align}
\frac{N}{pT}\bR_k\bX\He  =\bH_k+ \frac{N}{pT}\bN_k\bX\He. \label{eq:mid receive signal}
\end{align}
We can check that the entries in $\frac{N}{pT}\bN_k\bX \He$ are \gls{iid} Gaussian $\cC\cN(0, \sigma^2 \frac{N}{pT})$ random variables. Here, we define $\tilde{\bR}_k = \frac{N}{pT}\bR_k\bX\He$, and for convenience, denoting $\widebar{\bH}_k =\widebar{\bH}_{BU,k} +  \sum_{q=1}^{Q} \widebar{\bH}_{RU, k,q} \bTheta_q \bH_{BR,k, q}$ in \cref{eq:final expression of Hk}, we obtain
\begin{align}
\tilde{\bR}_k  = \widebar{\bH}_k + \bZ_{k} + \frac{N}{pT}\bN_k\bX\He
=\widebar{\bH}_k + \widetilde{\bN}_k, \label{eq:R observations}
\end{align}
where $\widetilde{\bN}_k ={\bZ}_{k}+\frac{N}{pT}\bN_k\bX\He$, and its entries are distributed as $\cC\cN(0, \tilde{\sigma}^2 )$ with $ \tilde{\sigma}^2=\frac{N}{pT}\sigma^2+\frac{\beta_{BU}}{1+K_{BU}} + \frac{QM\beta_{RU}\beta_{BR}}{1+K_{RU}}$.
Our objective is to infer the position of the UE and clock bias by using the observations $\{ \tilde{\bR}_k \}_{k=1}^K$ in \cref{eq:R observations}. Because directly estimating the UE position from \cref{eq:R observations} is challenging due to the nonlinearity of $\widebar{\bH}_k$  as a function of $\bxi$ in \cref{eq:bxi}, we first estimate the channel parameters $\bmeta$ in \cref{eq:bmeta}, from which the UE position and clock bias are then inferred.

\section{CRB for UE Position Estimation}\label{sec:CRB}
In this section, we derive the CRB for the UE position estimation based on the observations in \cref{eq:R observations}. We will compare the performance of our proposed method against this bound in the numerical results in \cref{sec:numerical}.

To compute the FIM $\bF_{\bmeta}$ of $\boldsymbol{\eta}$ in \cref{eq:bmeta} based on observations $\tilde{\bR}_k$, $k=0,\ldots,K-1$, in \cref{eq:R observations}, let $\bF_{\bmeta}^{(k)} \in  \R^{(7+5Q)\times (7+5Q)}$ be the FIM of $\bmeta$ based on the observations from the $k$-th subcarrier. Then, $\bF_{\bmeta}= \sum_{k=0}^{K-1} \bF^{(k)}_{\bmeta}$. Because the noise in \cref{eq:R observations} is Gaussian, we have the following
\begin{align}
\ln f(\tilde{\bR}_k| \boldsymbol{\eta}) =  -\frac{1}{ \tilde{\sigma}^2}\tr((\tilde{\bR}_k-\widebar{\bH}_k)\He (\tilde{\bR}_k-\widebar{\bH}_k)) + C,
\end{align}
where $C$ is a normalization constant. The $(i, j)$-th element of $\bF_{\bmeta}^{(k)}$ is then given by
$
[\bF^{(k)}_{\bmeta}]_{i, j}  =-\Eb[\frac{\partial^2 \ln f(\tilde{\bR}_k| \boldsymbol{\eta})}{\partial {\eta}_i \partial {\eta}_j}].
$
After simplifications, we have
$
[\bF^{(k)}_{\bmeta}]_{i, j} = \frac{2}{ \tilde{\sigma}^2}\opRe \left\{ \tr\left( \frac{\partial \widebar{\bH}_k\He}{\partial \eta_i} \frac{\partial \widebar{\bH}_k}{\partial \eta_j} \right) \right\}.
$
Section SI of the supplementary material provides detailed derivation for the terms in the FIM.

To derive the FIM for the parameter $\bxi$ in \cref{eq:bxi} from the derived $\bF_{\bmeta}$ above, we use the relation $F(\cdot)$ in \cref{eq:F}. The Jacobian matrix $\bJ =\frac{\partial \bmeta}{\partial\bxi\T} \in \R^{(7+5Q) \times (6+2Q)}$ of $F$ is given in Section SII of the supplementary material.
The FIM for $\bxi$ is then given by
\begin{align}\label{eq:Ibxi}
\bF_{\bxi} = \bJ\T \sum_{k=0}^{K-1} \bF^{(k)}_{\bmeta} \bJ \in \R^{(6+2Q) \times (6+2Q)}.
\end{align}
Accordingly, a position error bound (PEB) is as follows:
$
\sqrt{\Eb [\| \hat{\bp}_U-\bp_U\|^2_2] } \ge \tr ([\bF_{\bxi}^{-1}]_{1:3,1:3}).
$
Similarly, the clock bias error bound (CEB)  is given by:
$
\sqrt{\Eb [( \hat{\Delta}_{\text{clock}}-{\Delta}_{\text{clock}})^2] } \ge \tr ([\bF_{\bxi}^{-1}]_{4,4}).
$

\section{Estimation of Channel Parameters}\label{sec:estimation}

In this section, we formulate optimization problems to estimate the AoDs $(g_{BU}^{(B)},s_{BU}^{(B)})$ of the BS, and propagation delays $\tau_{BU}$ and $\{\tau_{RU, q}\}_{q=1}^Q$ along the LOS BS-UE path and the reflection paths from each RIS to the UE, respectively. We also estimate the AoAs $(g_{BU}^{(U)},s_{BU}^{(U)})$ and $(g_{RU, q}^{(U)},s_{RU, q}^{(U)})$ at the UE along the LOS path and reflection paths, respectively.

Because the noise $\widetilde{\bN}$ in \cref{eq:R observations} is Gaussian, the \gls{MLE} of $\bmeta$ of \cref{eq:bmeta} is given by the following:
\begin{align} \label{original eta problem}
\min_{\bmeta}
\sum_{k=0}^{K-1}\| \tilde{\bR}_k - \widebar{\bH}_k \|_{\text{F}}^2.
\end{align}
However, directly solving the above problem is challenging because it is nonlinear and non-convex in $\bmeta$. However, we note that the rank of  $\widebar{\bH}_k$ in \cref{eq:R observations} is  $Q+1$. We can leverage this low-rank property to estimate the channel parameters.

\subsection{Estimation of AoD \txp{$(g_{BU}^{(B)},s_{BU}^{(B)})$}{g\_Bd, v\_Bd}}

Recall that the AoD $(g_{BU}^{(B)},s_{BU}^{(B)})$ is for the BS-UE link given in \cref{AoD:B_d}. Here, we discuss the estimation of $g_{BU}^{(B)}$, and the estimation of $s_{BU}^{(B)}$ is done similarly.
According to \cref{eq:final expression of Hk}, we reshape $\{ \tilde{\bR}_k \}_{k=0}^{K-1}$ over the dimensions of $\tilde\ba_{N_r}(g_{BU}^{(B)})$ and $\{\tilde\ba_{N_r}(g_{BR, q}^{(B)})\}_{q=1}^Q$ as
\begin{align}
 {\bR}_{B}
 &=[\tilde\ba_{N_r}(g_{BR,1}^{(B)}),\ldots,\tilde\ba_{N_r}(g_{BR, Q}^{(B)}),\tilde\ba_{N_r}(g_{BU}^{(B)}) ]
 \bQ_{B}
 +{\bN}_{B} \in \C^{N_r \times N_c DK}, \label{eq:expression RB}
\end{align}
where $\bQ_B \in \C^{(Q+1) \times  N_c DK }$  is a coefficient matrix. For convenience, we let $\bA_B =[\bA_{BR},\tilde\ba_{N_r}(g_{BU}^{(B)})] \in \C^{N_r \times (Q+1)}$ with $\bA_{BR} =[ \tilde\ba_{N_r}(g_{BR,1}^{(B)}),\ldots,\tilde\ba_{N_r}(g_{BR, Q}^{(B)})]$.

It is worth noting that the column space of $ {\bR}_{B}$ in \cref{eq:expression RB} is spanned by the columns in $\bA_B$, which are parameterized by $\{ g_{BR, q}^{(B)}\}_{q=1}^Q$ and  $ g_{BU}^{(B)}$.
Since $g_{BR, q}^{(B)}$, for all $q$ in \cref{eq:fBrq} is known a priori as the position of the $q$-th RIS is known,\footnote{The BS controls the RISs and knows the position of each RIS. We assume this position information can be transmitted to the UE by BS.} we only need to estimate $g_{BU}^{(B)}$ from \cref{eq:expression RB} based on the column space of $\bR_B$ by solving
\begin{small}
\begin{align}
\min_{g_{BU}^{(B)}}  \lA{\bR}_{B} - \bA_B \bQ_{B}  \rA_{\text{F}}^2, \
\st  \ \bQ_{B}  = (\bA_B\He \bA_B)^{-1}\bA_B\He {\bR}_{B}. \label{eq:problem equ}
\end{align}
\end{small}%
We assume that $g_{BR, q}^{(B)}$ is distinct for each $q$. Meanwhile, we recall that $N_c>Q+1$ in \cref{sect:channel model}. Thus, $\bA_{B}\He\bA_{B}$ is invertible, and we have the following result.

\begin{Lemma}
The problem in \cref{eq:problem equ} is equivalent to
\begin{align}
\max_{g_{BU}^{(B)}} \|\widebar{\ba}\He_{N_r}(g_{BU}^{(B)}){\bR}_{B} \|_2^2,  \label{eq:transform problem}
\end{align}%
where $\widebar{\ba}_{N_r}(g_{BU}^{(B)})=\frac{\tilde\ba_{N_r}(g_{BU}^{(B)}) -\bP_R \tilde\ba_{N_r}(g_{BU}^{(B)})}{\| \tilde\ba_{N_r}(g_{BU}^{(B)}) -\bP_R \tilde\ba_{N_r}(g_{BU}^{(B)})\|_2} $ and $\bP_R =\bA_{BR} (\bA_{BR}\He\bA_{BR})^{-1}\bA_{BR}\He$.
\end{Lemma}
\begin{proof}

Note $\bP_R \tilde\ba_{N_r}(g_{BU}^{(B)})$ is the orthogonal projection onto the column space of $\bA_{BR}$, and $\widebar{\ba}_{N_r}(g_{BU}^{(B)})$ is the residual vector of projection with normalization.
Therefore, $[\bA_{BR},\tilde\ba_{N_r}(g_{BU}^{(B)})]$ spans the same subspace as $[\bA_{BR},\widebar{\ba}_{N_r}(g_{BU}^{(B)})]$.
For convenience, we define $\widebar{\bA}_{BR}$ as the Gram–Schmidt orthogonalization of columns in ${\bA}_{BR}$.
We have the equivalence in subspaces: $\Col([\bA_{BR},\tilde\ba_{N_r}(g_{BU}^{(B)})])\!=\! \Col([\widebar{\bA}_{BR},\tilde\ba_{N_r}(g_{BU}^{(B)}) ])
\!=\! \Col([\widebar{\bA}_{BR},\widebar{\ba}_{N_r}(g_{BU}^{(B)}) ])$.

Moreover, by defining $\widebar{\bA}_B \! =\! [\widebar{\bA}_{BR},\widebar{\ba}_{N_r}(g_{BU}^{(B)}) ]$, one can check that $\widebar{\bA}_B\He \widebar{\bA}_B \! =\!\bI$.
Therefore, from the equivalence in subspaces, the residual of $\bR_B$ \gls{wrt} $\Col(\bA_B)$ is the same as $\Col(\widebar{\bA}_B)$.
The objective in \cref{eq:problem equ} is rewritten as
\begin{small}
\begin{align}
 \lA {\bR}_{B} - \widebar{\bA}_B \widebar{\bQ}_{B}  \rA_F^2
&= \tr({\bR}_{B}\He {\bR}_{B} -  {\bR}_{B}\He \widebar{\bA}_B \widebar{\bQ}_{B} - \widebar{\bQ}_{B}\He \widebar{\bA}_B\He  {\bR}_{B} +\widebar{\bQ}_{B}\He  \widebar{\bA}_B\He  \widebar{\bA}_B \widebar{\bQ}_{B}) \nonumber\\
&= \tr({\bR}_{B}\He {\bR}_{B} -  {\bR}_{B}\He \widebar{\bA}_B\widebar{\bQ}_{B}) \nonumber \\
&= \tr({\bR}_{B}\He {\bR}_{B} -  {\bR}_{B}\He \widebar{\bA}_B \widebar{\bA}_B\He {\bR}_{B} ), \label{eq:obj equivalent}
\end{align}
\end{small}
\!\!\!where $\widebar{\bQ}_{B}=(\widebar{\bA}_B\He \widebar{\bA}_B)^{-1}\widebar{\bA}_B\He {\bR}_{B}$, and  the last inequality comes from $ \widebar{\bA}_B\He  \widebar{\bA}_B  = \bI$.
Therefore, we have
\begin{small}
\begin{align*}
\argmin_{g_{BU}^{(B)}}\tr({\bR}_{B}\He {\bR}_{B} -  {\bR}_{B}\He \widebar{\bA}_B \widebar{\bA}_B\He {\bR}_{B})
=\argmax_{g_{BU}^{(B)}} \|  \widebar{\bA}_B\He {\bR}_{B} \|_{\text{F}}^2
=\argmax_{g_{BU}^{(B)}}  \|\widebar{\ba}\He_{N_r}(g_{BU}^{(B)}){\bR}_{B} \|_2^2,
\end{align*}
\end{small}%
which is exactly the problem provided in \cref{eq:transform problem}.
This concludes the proof.
\end{proof}

The variable of optimization $g_{BU}^{(B)}$ in problem \cref{eq:transform problem} is scalar, and various standard optimization techniques can be applied to find the optimal solution \cite{nocedal1999numerical}.
Suppose $\hat{g}_{BU}^{(B)}$ is the optimal solution found.
How to distinguish the LOS and the reflection paths is now of interest. Let
\begin{small}
\begin{align}\label{distinguish LOS path}
\hat{\bQ}_B = \argmin_{\bQ_{B}} \|{\bR}_{B} - [\bA_{BR},\tilde\ba_{N_r}(\hat{g}_{BU}^{(B)}) ] \bQ_{B}  \|_{\text{F}}^2,
\end{align}
\end{small}%
where $\hat{\bQ}_B = [ [\hat{\bQ}_B]_{1,:},\ldots,[\hat{\bQ}_B]_{Q,:},[\hat{\bQ}_B]_{Q+1,:}]$.
Note that the values of  $\|[\hat{\bQ}_B]_{q,:}\|_2^2$, $q=1,\dots,Q$ and $\|[\hat{\bQ}_B]_{Q+1,:}\|_2^2$ are related to the energy of the reflection and LOS paths, respectively.
We sort the paths according to these values. We save the estimated sorting of the path energies as $\bkappa = [\kappa_1, \kappa_2,\ldots,\kappa_{Q+1}] \in \R^{Q+1}$, which matches the path energies with each specific path. To be more precise, we can obtain a table as shown in \cref{tab:PathsOrder}. This path order is utilized to distinguish the channel parameters associated with each path in the following subsections.

\begin{table}[!htb]
  \footnotesize
  \centering
  \caption{Path Order Information}\label{tab:PathsOrder}
  \begin{tabular}{|c|c|}\hline
    \textbf{Path}        &  \textbf{Energy index}   \\ \hline
      Reflection path of RIS $1$                 & $\kappa_1$  \\ \hline
      \vdots &      \vdots  \\ \hline
       Reflection path of RIS $Q$                 & $\kappa_Q$  \\ \hline
              LOS path  & $\kappa_{Q+1}$  \\ \hline
  \end{tabular}
\end{table}

\subsection{Estimation of \txp{$\tau_{BU}$}{tau-d} and \txp{$\{\tau_{RU, q}\}_{q=1}^Q$}{tau-r2,q}}\label{subsec:est_delays}

We define
$\bb(\tau) = [1,e^{-\iu2\pi \frac{W }{K} \tau},\ldots, e^{-\iu2\pi \frac{(K-1)W }{K} \tau}]\T$
and reshape $\{ \tilde{\bR}_k \}_{k=0}^{K-1}$ over the dimensions of $\bb(\tau_{BU})$ and $\{\bb(\tau_{BR,q}+\tau_{RU,q})\}_{q=1}^Q$ to obtain
\begin{align}
\bR_{G} & = [\bb(\tau_{BU}),\bb(\tau_{BR,1}+\tau_{RU,1}),\ldots,\bb(\tau_{BR,Q}+\tau_{RU,Q})] 
 \times \bQ_{G}
+ \bN_{G} \in  \C^{K \times  DN }, \label{eq:expression RH}
\end{align}
where $\bQ_{G} \in \C^{(Q+1) \times DN}$ and $\bN_{G} \in \C^{K \times  DN }$.
Similar to \cref{eq:expression RB},
the column space of  $\bR_{G}$ is spanned by $\bb(\tau_{BU})$ and $\{\bb(\tau_{BR, q}+\tau_{RU, q})\}_{q=1}^Q$.
We use the multiple signal classification (MUSIC) method \cite{Schmidt1986M} to estimate delays from the observations in \cref{eq:expression RH} as follows.

Letting
$\bB_{G} = [\bb(\tau_{BU}), \bb(\tau_{BR,1}+\tau_{RU,1}),\ldots,\bb(\tau_{BR,Q}+\tau_{RU,Q})] $, the covariance of \cref{eq:expression RH} is
\begin{align}
\bC_{G}  =\bB_{G}  \bQ_{G} \bQ_{G} \He \bB_{G}\He +\tilde{ \sigma}^2 \bI. \label{eq:true delay cov}
\end{align}
Intuitively, when the noise level is low, the covariance matrix $\bC_{G}$ in \cref{eq:true delay cov} can be approximated by the covariance of the signal part, i.e., $\bB_{G}  \bQ_{G} \bQ_{G} \He\bB_{G}\He$. This is the underlying methodology of MUSIC.
The covariance matrix in \cref{eq:true delay cov} can be estimated by using the sample correlation matrix
$\hat{\bC}_{G} =\bR_{G} \bR_{G}\He$ in \cref{eq:expression RH}.
Let $[\bw_1, \ldots, \bw_{K}]$ be the eigenvectors of $\hat{\bC}_{G}$, where $\bw_i$ corresponds to the $i$th largest eigenvalue.
Recall that $K>Q+1$ in \cref{sect:channel model}, we can denote $ \bW_{H}^c=[\bw_{Q+2},\ldots, \bw_{K}]$. Then, the estimation of the delays $\tau_{BU}, \{\tau_{RU, q}\}_{q=1}^Q$ is achieved by:
\begin{align} \label{eq:est delay music}
\text{find $Q+1$ peaks of}~{1}/{\| \bb\He( \tau)
 \bW_{H}^c \|_2^2} \text{~with~} \tau \le K/W.
\end{align}

\begin{Remark}\label{remark:distinguish}
Suppose the estimated delays are $\{\hat{\tau}_i\}_{i=1}^{Q+1}$.
A heuristic way to distinguish the delay for the LOS path versus those for the reflection paths is to use the minimum delay estimated. However, this approach can only distinguish the LOS path. When there are multiple RISs leading to multiple reflection paths, this delay-based method can not distinguish these reflection paths
since the order of delay values is unknown a priori.
To address this, we use \cref{distinguish LOS path} instead to assign the estimated delays in \cref{eq:est delay music} to the LOS and reflection paths. Specifically, after estimating $\{\hat{\tau}_i\}_{i=1}^{Q+1}$, we denote $\hat{\bB}_G=[\bb(\hat{\tau}_1),\ldots,\bb(\hat{\tau}_{Q+1})]$. Similar as \eqref{distinguish LOS path}, we define
$
\hat{\bQ}_{G}
=  \argmin_{\bQ_{G}}\| \bR_{G} - \hat{\bB}_{G}\bQ_{G}\|_F^2=(\hat{\bB}_{G}\He\hat{\bB}_{G})^{-1}\hat{\bB}_{G}\He \bR_{G}.
$
Then, we use the path ordering $\bkappa$ from the sorting of \cref{distinguish LOS path} and $\{\| [\hat{\bQ}_{G}]_{i,:} \|^2_2\}_{i=1}^{Q+1}$ to assign the estimated delays to path indices, and the matched estimated delays are $\{\hat\tau_{BU},\hat{\tau}_{RU,1},\ldots,\hat{\tau}_{RU,Q}\}$.
\end{Remark}
\subsection{Estimation of AoAs \txp{$(g_{BU}^{(U)},g_{RU, q}^{(U)})$}{g\_Ur, g\_Urq} and \txp{($s_{BU}^{(U)}, s_{RU, q}^{(U)})$}{s\_Ud, s\_Ud}} \label{sect:AoAs}

For the same reason, we present only the method to estimate $g_{RU, q}^{(U)}$ and $g_{BU}^{(U)}$. The same approach can be applied to the estimation of $s_{RU, q}^{(U)}$ and $s_{BU}^{(U)}$.
We reshape $\{ \tilde{\bR}_k \}_{k=0}^{K-1}$ over the dimension of $\tilde\ba_{D_r}(g_{BU}^{(U)})$ and $\{\tilde\ba_{D_r}(g_{RU, q}^{(U)})\}_{q=1}^Q$ as ${\bR}_{U} \in \C^{D_r \times D_c NK}$,
\begin{align}
{\bR}_{U} &=[\tilde\ba_{D_r}(g_{BU}^{(U)}),\tilde\ba_{D_r}(g_{RU,1}^{(U)}),\ldots,\tilde\ba_{D_r}(g_{RU, Q}^{(U)}) ] \bQ_{U}  +\widetilde{\bN}_{U},
\end{align}
where $\bQ_{U}\in \C^{(Q+1) \times D_c N K}$ and $\widetilde{\bN}_{U} \in \C^{D_r \times D_c NK }$. Note that the signal part of the column space of ${\bR}_{U}$ is spanned by $[\tilde\ba_{D_r}(g_{BU}^{(U)}) ,\tilde\ba_{D_r}(g_{RU,1}^{(U)}),\ldots,\tilde\ba_{D_r}(g_{RU, Q}^{(U)})]$.
Similar to the manipulations in \cref{subsec:est_delays}, we utilize the MUSIC method in the following,
\begin{align} \label{eq:est AoA music}
\text{find $Q+1$ peaks of}~{1}/{\| \tilde\ba_{D_r}\He( g)
 \bW_{A}^c \|_2^2},
\end{align}
 where  $\bW_{A}^c$ is defined similarly as $ \bW_{H}^c$ in \cref{eq:est delay music}, to obtain the estimation of $\{ g_{RU, q}^{(U)}\}_{q=1}^Q$ and $g_{BU}^{(U)}$ as $\{\hat{g}_{RU, q}^{(U)}\}_{q=1}^Q$ and $\hat{g}_{BU}^{(U)}$.
Then, the path matching technique in \cref{remark:distinguish} can also be employed to distinguish paths.

\subsection{Estimation of \txp{$h_{BU}$}{hd} and \txp{$\{h_{R, q}\}_{q=1}^Q$}{hq}}

Using the estimates $(\hat\tau_{BU}, \hat{g}_{BU}^{(U)},\hat{s}_{BU}^{(U)}, \hat{g}_{BU}^{(B)}, \hat{s}_{BU}^{(B)})$ and  $\{\hat{\tau}_{RU,q}, \hat{g}_{RU, q}^{(U)},\hat{s}_{RU, q}^{(U)}\}_{q=1}^Q$ in the previous subsections, we solve the following problem to estimate $h_{BU}$ and $h_{r,q}$:
\begin{small}
\begin{align}
&\widehat{\bh}=\argmin_{\bh} \sum_{k=0}^{K-1}\| \tilde{\bR}_k - \widebar{\bH}_k \|_{\text{F}}^2  \nonumber \\
&\st  (\tau_{BU},{g}_{BU}^{(U)},{s}_{BU}^{(U)}, {g}_{BU}^{(B)},{s}_{BU}^{(B)}) \nonumber= (\hat\tau_{BU},\hat{g}_{BU}^{(U)},\hat{s}_{BU}^{(U)}, \hat{g}_{BU}^{(B)}, \hat{s}_{BU}^{(B)}), \nonumber \\
& \quad \quad  ({\tau}_{RU,q},{g}_{RU, q}^{(U)},{s}_{RU, q}^{(U)}) = (\hat{\tau}_{RU,q},\hat{g}_{RU, q}^{(U)},\hat{s}_{RU, q}^{(U)}), \label{eq:estimate gain problem}
\end{align}
\end{small}%
where $\widehat{\bh}$ is the estimate of
$\bh=[{h_{BU},{h}_{r,1},\ldots,{h}_{r, Q}}]\T$.
From \cref{eq:R observations}, we vectorize the matrix as
$
{\br}_s =\vect([ \tilde{\bR}_0,\ldots,\tilde{\bR}_{K-1}])$,
$\hat{\br}_B= \vect( [ \widebar{\bH}_{BU,0},\ldots,\widebar{\bH}_{BU, K-1}])$,
 and $\hat{\br}_{R, q} = \vect([ \widebar{\bH}_{RU, 0,q} \bTheta_q \bH_{BR, 0,q},\ldots, \widebar{\bH}_{RU, K-1,q} \bTheta_q \bH_{BR, K-1,q}]).
$
Then, the formulated problem in \cref{eq:estimate gain problem} can be expressed as
$
\widehat{\bh}=\argmin_{\bh} \| \br_s - \hat{\bR}
\bh \|_2^2,
$
where $\hat{\bR}= [\hat{\br}_{B},\hat{\br}_{R,1},\ldots,\hat{\br}_{R, Q}]$.
The solution is $\widehat{\bh}=(\hat{\bR}\He \hat{\bR})^{-1} \hat{\bR}\He\br_s$.

\section{UE Position Estimation} \label{sec:fusion}
In this section, we present a fusion method to infer the UE position from the estimated channel parameters.

\subsection{Fusion via Weighted Least Squares} \label{sec:linear fusion}
For convenience, we denote the unit directional vectors of the LOS and reflection paths as
\begin{align*}
\br^{(B)} = \frac{ \bp_U-\bp_B}{\| \bp_U-\bp_B \|_2},\
\br^{(R, q)} = \frac{ \bp_U-\bp_{R, q}}{\| \bp_U-\bp_{R, q} \|_2},
\end{align*}
respectively.
Then, from the relations in \cref{eq:sub clock}, we have
\begin{align}
{\bp}_U& = {\bp_B + c \br^{(B)}\tau_{BU}}- c \br^{(B)}\Delta_\mathrm{clock}, \label{eq:re d}\\
{\bp}_U& = {\bp_{R, q} + c \br^{(R, q)}\tau_{RU,q}}- c \br^{(R, q)}\Delta_\mathrm{clock} .\label{eq:re rq}
\end{align}
Here, we denote the estimator of ${\br}^{(B)}$ as $\hat{\br}^{(B)}$ and that of $\tau_{BU}$ as $\hat{\tau}_{BU}$, where the detailed estimation procedure is presented in \cref{sect: cal pU}.
We assume that the estimate $\hat{\bp}_U^{(B)}=\bp_B+c\hat{\br}^{(B)}\hat\tau_{BU}$ is unbiased such that $\Eb[ \hat{\bp}_{U}^{(B)}]= {\bp}_{U}^{(B)}$.
Similarly, for the $q$-th reflection path, we denote the  estimations of ${\br}^{(R, q)}$ and $\tau_{RU, q}$ as $\hat{\br}^{(R, q)}$ and $\hat{\tau}_{RU, q}$, respectively.
We also let $\hat{\bp}_U^{(R, q)}=\bp_{R, q}+c\hat{\tau}_{RU, q}\hat{\br}^{(R, q)}$ with $\Eb[ \hat{\bp}_{U}^{(R, q)}]= {\bp}_{U}^{(R, q)}$.

\begin{Proposition} \label{pro:crb bound}
The error covariance matrix of $\hat{\bp}_{U}^{(B)}$ given by
\begin{align} \label{eq:cov_direct}
 \bC_{{\bp}_{U}}^{(B)}=\Eb[(\hat{\bp}_{U}^{(B)} -{\bp}_{U}^{(B)}) (\hat{\bp}_{U}^{(B)}- {\bp}_{U}^{(B)})\T]
 \end{align}
satisfies the following bound:
\begin{align}
\bC_{{\bp}_{U}}^{(B)} &\succeq  \left[ ~ \widebar{\bC}_{\bxi_B}\right]_{1:3,1:3} \succeq  \left[ \left(\bJ_B\T \bF_{\bmeta_B}  \bJ_B \right)^{-1} \right]_{1:3,1:3}, \label{eq:crb direct}
\end{align}
where
$\widebar{\bC}_{\bxi_B}=(\bJ_B\T \widebar{\bC}_{{\bmeta}_B}^{-1}  \bJ_B )^{-1}$ with $\widebar{\bC}_{{\bmeta}_B} = \left[ \bF_{\bmeta} ^{-1}\right]_{1:7,1:7}, \bJ_B  = \frac{\partial \bmeta_B }{\partial \bxi_B\T}\in \R^{7 \times 5}$, and $\bF_{\bmeta_B} \in \C^{7\times 7}$ is the FIM of ${\bmeta_B}$.
Similarly, the error covariance matrix of $\hat{\bp}_{U}^{(R, q)}$ given by
\begin{align}
\bC_{{\bp}_{U}}^{(R, q)}= \Eb[(\hat{\bp}_{U}^{(R, q)} -{\bp}_{U}^{(R, q)}) (\hat{\bp}_{U}^{(R, q)}- {\bp}_{U}^{(R, q)})\T]
\label{eq:cov_reflect}
\end{align}
satisfies the following bound:
\begin{align}
\bC_{{\bp}_{U}}^{(R, q)} &\succeq  \left[ \widebar{\bC}_{\bxi_{R, q}}\right]_{1:3,1:3} \succeq  \left[ \left(\bJ_{R, q}\T \bF_{\bmeta_{R, q}}  \bJ_{R, q} \right)^{-1} \right]_{1:3,1:3}, \label{eq:crb reflect}
\end{align}
where
$ \widebar{\bC}_{\bxi_{R, q}}=(\bJ_{R, q}\T \widebar{\bC}_{{\bmeta}_{R, q}}^{-1}  \bJ_{R, q} )^{-1}$ with
 $\widebar{\bC}_{{\bmeta}_{R, q}} = [ \bF_{\bmeta} ^{-1}]_{5q+3:5q+7,5q+3:5q+7}$,
$\bJ_{R, q}  = \frac{\partial \bmeta_{R, q}}{\partial \bxi_{R, q}\T}\in \R^{5 \times 5}$, and $\bF_{\bmeta_{R, q}} \in \C^{5\times 5}$ is the FIM of ${\bmeta}_{R, q}$.
\end{Proposition}
\begin{proof}
See Section SIII of the supplementary material.
\end{proof}

\cref{pro:crb bound} shows the error covariance matrices of $\hat{\bp}_{U}^{(B)}$ and $\{\hat{\bp}_{U}^{(R, q)}\}_{q=1}^Q$ in \cref{eq:cov_direct} and \cref{eq:cov_reflect}, respectively.
The following lemma presents the proposed fusion method based on these error covariance matrices.

\begin{Lemma}
Assume that the estimations of $\hat{\bp}_{U}^{(B)}$ and $\{\hat{\bp}_{U}^{(R, q)}\}_{q=1}^Q$ are based on independent measurements. Then, according to \cref{eq:re d} and \cref{eq:re rq}, the UE position and clock bias can be obtained by the following \gls{WLS} problem:
\begin{align}
 \min_{\bp_U,\Delta_\mathrm{clock}} & \Big\{(\hat{\bp}_U^{(B)}-(\bp_U + c\Delta_\mathrm{clock} {\br}^{(B)}))\T \left( \bC_{\bp_U}^{(B)}\right)^{-1}   (\hat{\bp}_U^{(B)}-(\bp_U + c\Delta_\mathrm{clock} {\br}^{(B)}))\nonumber \\
  &   \ + \sum_{q=1}^Q (\hat{\bp}_U^{(R, q)} \!\!-\!\! (\bp_U \!+\! c\Delta_\mathrm{clock} {\br}^{(R, q)}))\T \left( \bC_{\bp_U}^{(R, q)}\right)^{-1}   (\hat{\bp}_U^{(R, q)}-(\bp_U + c\Delta_\mathrm{clock} {\br}^{(R, q)})) \Big\}. \label{eq:linear combination}
\end{align}
\begin{proof}
Given the definitions in \cref{eq:re d} and \cref{eq:re rq}, this is a standard formulation of \gls{WLS} problem. Thus we omit the proof here.
\end{proof}
\end{Lemma}

\begin{Remark}\label{remark:equivalent LS}
When the LOS path or some reflection paths have low SNR, the related covariance matrices associated with these paths will have a high magnitude. Therefore,  the proposed \gls{WLS} method will adaptively rely less on the paths with low SNR to make the final decision about the UE position.
\end{Remark}

\begin{Remark} \label{remark:bound of Cpu}
Denoting $\bF_{\hat{\bmeta}_B} = \evalat*{\bF_{{\bmeta}_B}}_{{{\bmeta}_B}=\hat{\bmeta}_B}$, $\bF_{\hat{\bmeta}_{R, q}} = \evalat*{\bF_{{\bmeta}_{R, q}}}_{{{\bmeta}_{R, q}}=\hat{\bmeta}_{R, q}}$,  $\tilde{\bJ}_B =  \evalat*{{\bJ}_B }_{\bxi_B=\hat{\bxi}_B}$, and $\tilde{\bJ}_{R, q}=  \evalat*{{\bJ}_{R, q}}_{\bxi_{R, q}=\hat{\bxi}_{R, q}}$, we have the following bounds for the error covariance matrices,
\begin{align}
{\bC}_{{\bp}_{U}}^{(B)} &\overset{(a)}\succeq {\left[ \left(\bJ_B\T \bF_{{\bmeta}_B}  \bJ_B \right)^{-1} \right]_{1:3,1:3}}
\overset{(b)}\approx {\left[ \left(\tilde{\bJ}_B\T \bF_{\hat{\bmeta}_B}  \tilde{\bJ}_B \right)^{-1} \right]_{1:3,1:3}} \triangleq \tilde{\bC}_{{\bp}_{U}}^{(B)}, \label{eq:def_Cd}\\
{\bC}_{{\bp}_{U}}^{(R, q)}&\overset{(c)}\succeq {\left[ \left({\bJ}_{R, q}\T \bF_{{\bmeta}_{R, q}}  {\bJ}_{R, q} \right)^{-1} \right]_{1:3,1:3}} 
 \overset{(d)}\approx {\left[ \left(\tilde{\bJ}_{R, q}\T \bF_{\hat{\bmeta}_{R, q}}  \tilde{\bJ}_{R, q} \right)^{-1} \right]_{1:3,1:3}}\triangleq \tilde{\bC}_{{\bp}_{U}}^{(R, q)},\label{eq:def_Cr}
\end{align}
where  $(a)$ and $(c)$ are from \cref{pro:crb bound}, and the approximations in $(b)$ and $(d)$ hold when $\bxi_B \approx \hat{\bxi}_B$ and $\bxi_{R, q} \approx \hat{\bxi}_{R, q}$, where $\hat{\bxi}_B$ and $\hat{\bxi}_{R, q}$ are calculated in \cref{sect: cal pU}.
In the following, we denote the bounds on \gls{RHS} of \cref{eq:def_Cd} and \cref{eq:def_Cr} as $\tilde{\bC}_{{\bp}_{U}}^{(B)}$ and $\tilde{\bC}_{{\bp}_{U}}^{(R, q)}$, respectively.
\end{Remark}

When the exact error covariances in \cref{eq:linear combination} are not available, we can employ the lower bounds in \cref{eq:def_Cd} and \cref{eq:def_Cr} and formulate the problem in \cref{eq:linear combination} as\footnote{When the rotation matrix $\bO$ can not be obtained from an IMU at the UE, the problem in \cref{eq:linear combination bound} can optimize $\bp_U,\Delta_\mathrm{clock}$ and $\bO$ jointly, where $\bO$ is determined by the Euler angles in 3D.}
\begin{small}
\begin{align} \label{eq:linear combination bound}
 \min_{\bp_U,\Delta_\mathrm{clock}} & \Big\{ (\hat{\bp}_U^{(B)}-(\bp_U + c\Delta_\mathrm{clock} {\br}^{(B)}))\T \left( \tilde{\bC}_{\bp_U}^{(B)}\right)^{-1} 
 \times (\hat{\bp}_U^{(B)}-(\bp_U + c\Delta_\mathrm{clock} {\br}^{(B)})) \nonumber \\
  &  \quad +\sum_{q=1}^Q (\hat{\bp}_U^{(R, q)}-(\bp_U + c\Delta_\mathrm{clock} {\br}^{(R, q)}))\T \left( \tilde{\bC}_{\bp_U}^{(R, q)}\right)^{-1}
  (\hat{\bp}_U^{(R, q)}-(\bp_U + c\Delta_\mathrm{clock} {\br}^{(R, q)})) \Big\}.
\end{align}%
\end{small}%
To solve \cref{eq:linear combination bound} efficiently,  we can further make the approximations ${\br}^{(B)}\approx \hat{\br}^{(B)}$ and ${\br}^{(R, q)}\approx \hat{\br}^{(R, q)}$. A closed-form solution for \cref{eq:linear combination bound} with fixed $\Delta_\mathrm{clock}$ can be shown to be
\begin{small}
\begin{align}
\hat{\bp}_{U}(\Delta_\mathrm{clock}) &= \tilde{\bC}_{\bp_U} \bigg(  (\tilde{\bC}^{(B)}_{\bp_U})^{-1} (\hat{\bp}_{U}^{(B)}-c\Delta_\mathrm{clock} \hat{\br}^{(B)}) 
 \sum_{q=1}^{Q} (\tilde{\bC}^{(R, q)}_{\bp_U})^{-1}(\hat{\bp}_{U}^{(R, q)}-c\Delta_\mathrm{clock} \hat{\br}^{(R, q)})\bigg),\label{eq:wls solution}
\end{align}
\end{small}%
where $\tilde{\bC}_{\bp_U} = ( (\tilde{\bC}_{{\bp}_{U}}^{(B)} )^{-1}+  \sum_{q=1}^{Q} (\tilde{\bC}_{{\bp}_{U}}^{(R, q)})^{-1} )^{-1}$.
Then, the problem in \cref{eq:linear combination bound} can be solved efficiently by performing a one-dimensional iterative method to search for the optimal $\Delta_\text{clock}$.
A coarse search and iterative update procedure \cite{nocedal1999numerical} can be utilized to find the solution.
Specifically, during the coarse search, we evaluate the objective value at discrete points of $\Delta_{\text{clock}}$ and find the minimal point as the initial point. Then from the initial point, we iteratively update $\Delta_{\text{clock}}$ using the gradient-descent method, until a convergence criterion is reached. By using a sufficient number of discrete points in the coarse search, one is likely to find a solution close to the global optimum.

\subsection{Asymptotic MLE} \label{sect:asy MLE}
Though the lower bounds for the covariance matrices are utilized in \cref{eq:linear combination bound}, we now show that the proposed \gls{WLS} in \cref{eq:linear combination bound} is approximately the \gls{MLE} in the asymptotic regime of a large sample size.
We first introduce the extended invariance principle (ExIP) \cite{stoica1989reparametrization}, which is asymptotically equivalent to the \gls{MLE}.
Then we show that the \gls{WLS} method is approximately the optimal solution for ExIP.

\begin{Theorem}[ExIP theorem\cite{stoica1989reparametrization}]\label{thm:ExIP}
Suppose the loss function for estimating the parameters $\bxi $ is given by $L(\by; \bxi)$, where $\by\in \R^{Z\times 1}$ are observations. Suppose there exists a function $\bmeta = F(\bxi)$ with loss function
$L(\by; \bmeta) = L(\by; F(\bxi) )=L(\by; \bxi)$.
The estimation of $\bxi$ and $\bmeta$ are given by
$\hat{\bxi} = \argmin L(\by; \bxi)$ and $\hat{\bmeta} =\argmin L(\by; \bmeta)$.
If
$
\lim_{Z \rightarrow \infty}\hat{\bmeta} = \lim_{Z \rightarrow \infty}F(\hat{\bxi}),
$
then
\begin{align} \label{eq:ExIP form}
 \breve{\bxi} = \argmin_{\bxi} [\hat{\bmeta}-F(\bxi)]\T\bW [\hat{\bmeta}-F(\bxi)]
\end{align}
is asymptotically equivalent to $\hat{\bxi}$ as $Z\to\infty$, where
$
\bW = \evalat*{\Eb[\frac{\partial^2 L(\by; \bmeta)}{\partial \bmeta \bmeta\T}]}_{\bmeta = \hat{\bmeta}}.
$
\end{Theorem}

In \cref{eq:F}, the UE position parameters are related to the channel parameters $\bmeta$ via a function $F(\cdot)$. We can apply the ExIP approach to obtain the UE position estimate from $\hat{\bmeta}$. Specifically, from the estimator $\hat{\bmeta}$ in \cref{sec:estimation}, applying \cref{thm:ExIP}, we can solve the following \gls{WLS} problem
\begin{align}
\hat{\bxi} = \argmin_{\bxi} [\hat{\bmeta}-F(\bxi)]\T \bF_{\hat{\bmeta}}  [\hat{\bmeta}-F(\bxi)], \label{eq:UE position estimation}
\end{align}
where the weight matrix $\bW$ is
$
\bF_{\hat{\bmeta}}=\evalat*{\Eb[\frac{\partial^2 L(\{\tilde{\bR}_k\}_{k=1}^K; \bmeta)}{\partial \bmeta \bmeta\T}]}_{\bmeta = \hat{\bmeta}},
$
and the loss function $L(\{\tilde{\bR}_k\}_{k=1}^K; \bmeta) = \sum_{k=1}^{K}\| \tilde{\bR}_k - \widebar{\bH}_k \|_F^2$ (see \cref{original eta problem}).
Note that the inference model in \cite{wang2021joint} using the LS method is equivalent to letting $\bW = \bI$ in \cref{eq:ExIP form}.

Though the gradient-based method can be utilized to find the optimum in \cref{eq:UE position estimation}, the high nonlinearity of the function $F(\cdot)$ makes the problem in \cref{eq:UE position estimation} sensitive to the initialization. Instead, as we discussed in \cref{sec:linear fusion}, the expression in \cref{eq:wls solution} has a closed-form solution with fixed $\Delta_\mathrm{clock}$, and we can solve \cref{eq:linear combination bound} efficiently by performing a one-dimensional iterative search for $\Delta_\mathrm{clock}$.
In what follows, we will show that \cref{eq:linear combination bound} is the approximate solution to the problem \cref{eq:UE position estimation}.

Recall the definitions of
$\bxi_B  $ and $ \bxi_{R, q}$ in \cref{eq:bxi}, and the definitions of $\bmeta$, $\boldsymbol{\eta}_B$ and $\boldsymbol{\eta}_{R, q}$ in \cref{eq:bmeta}. We let $F_B$ and $F_{R, q}$, $q=1,\ldots, Q$, be functions such that $\bmeta_B=F_B(\bxi_{B})$ and $\bmeta_{R, q}=F_{R, q}(\bxi_{R, q})$.
Given the estimation results $\hat{\bmeta}$, we obtain $\hat{\bxi}_{B}$ from $\hat{\bmeta}_{B}$ and $\hat{\bxi}_{R,q}$ from $\hat{\bmeta}_{R,q}$ (see \cref{sect: cal pU}).
Therefore, the objective function in \cref{eq:UE position estimation} is approximated as
\begin{small}
\begin{align}
(\hat{\bmeta}-F(\bxi))\T\bF_{\hat{\bmeta}} (\hat{\bmeta}-F(\bxi))
  &\overset{(a)}{\approx}\!
  \begin{bmatrix}
   F_B(\hat{\bxi}_B)-F_B(\bxi_B) \\
  F_{r,1}(\hat{\bxi}_{R,1})-F_{R,1}(\bxi_{R,1}) \\
  \vdots \\
  F_{r, Q}(\hat{\bxi}_{R, Q})-F_{R, Q}(\bxi_{R, Q}) \\
 \end{bmatrix}\T
\! \! \bF_{\hat{\bmeta}} \!\!
  \begin{bmatrix}
   F_B(\hat{\bxi}_B)-F_B(\bxi_B) \\
  F_{r,1}(\hat{\bxi}_{R,1})-F_{R,1}(\bxi_{R,1}) \\
  \vdots \\
  F_{r, Q}(\hat{\bxi}_{R, Q})-F_{R, Q}(\bxi_{R, Q}) \\
 \end{bmatrix}\nonumber \\
  &\overset{(b)}{\approx}\!
   \begin{bmatrix}
   \tilde{\bJ}_B(\hat{\bxi}_B-\bxi_B) \\
 \tilde{\bJ}_{R,1}(\hat{\bxi}_{R,1}-\bxi_{R,1}) \\
  \vdots \\
  \tilde{\bJ}_{R, Q}(\hat{\bxi}_{R, Q}-\bxi_{R, Q}) \\
 \end{bmatrix}\T
\bF_{\hat{\bmeta}}
  \begin{bmatrix}
   \tilde{\bJ}_B(\hat{\bxi}_B-\bxi_B) \\
  \tilde{\bJ}_{R,1}(\hat{\bxi}_{R,1}-\bxi_{R,1}) \\
  \vdots \\
  \tilde{\bJ}_{R, Q}(\hat{\bxi}_{R, Q}-\bxi_{R, Q}) %
 \end{bmatrix},\label{eq:partial decom ExIP} %
\end{align}%
\end{small}%
where the approximation $(a)$ is from the fact that $F_B(\hat{\bxi}_B)\approx \hat{\bmeta}_B$ in \cref{sect: cal pU}, and the approximation in $(b)$ is due to the Taylor series expansion with $\tilde{\bJ}_B$ and $\tilde{\bJ}_{R, q}$ defined in \cref{remark:bound of Cpu}.
Based on \cref{eq:partial decom ExIP}, we can approximately formulate the ExIP problem in \cref{eq:UE position estimation} as the following:
\begin{align}
\min_{\bxi} \ \cref{eq:partial decom ExIP},  \    \st \ \cref{eq:bxi}.   \label{eq:approx_formulation}
\end{align}
Therefore, the solution in \cref{eq:approx_formulation} is the approximate solution of \cref{eq:UE position estimation}, which is asymptotically \gls{MLE} in the large-sample regime.
Moreover, the following proposition shows the \gls{WLS} problem in \cref{eq:linear combination bound} is equivalent to the solution in \cref{eq:approx_formulation} when the paths are independent.

\begin{Proposition} \label{lem:equivalence}
Suppose the paths are independent, in other words, $\bF_{\hat{\bmeta}}$ has the following form
\begin{small}
\begin{align}\label{eq:diag F}
\bF_{\hat{\bmeta}} =
\begin{bmatrix}
  \bF_{\hat{\bmeta}_B} & \boldsymbol{0}& \cdots & \boldsymbol{0}\\
   \boldsymbol{0} & \bF_{\hat{\bmeta}_{R,1}} & \cdots & \boldsymbol{0} \\
   \vdots & \vdots& \ddots& \vdots\\
      \boldsymbol{0} &    \boldsymbol{0} & \cdots & \bF_{\hat{\bmeta}_{R, Q}}
\end{bmatrix}.
\end{align}
\end{small}%
Then, the approximated ExIP problem in \cref{eq:approx_formulation} can be written as
\begin{small}
\begin{subequations} \label{eq:approx_formulation diag}
\begin{align}
\min_{\bxi} &  \ (\hat{\bxi}_B-\bxi_B)\T \tilde{\bJ}_B\T   \bF_{\hat{\bmeta}_B}\tilde{\bJ}_B (\hat{\bxi}_B-\bxi_B) +\sum_{q=1}^Q (\hat{\bxi}_{R, q}-\bxi_{R, q})\T \tilde{\bJ}_{R, q}\T   \bF_{\hat{\bmeta}_{R, q}}\tilde{\bJ}_{R, q} (\hat{\bxi}_{R, q}-\bxi_{R, q})
 \\
 \st & \ \cref{eq:bxi}.
\end{align}
\end{subequations}
\end{small}%
Moreover, the problem in \cref{eq:approx_formulation diag} is equivalent to the \gls{WLS} problem in \cref{eq:linear combination bound}.
\end{Proposition}
\begin{proof}
See Section SIV of the supplementary material.
\end{proof}

\begin{Remark}
Based on \cref{lem:equivalence}, we have shown that the \gls{WLS} problem in \cref{eq:linear combination bound} is approximately the optimal solution of the ExIP method (up to estimation errors in $\hat{\bmeta}$, $\hat{\bxi}_B$, $\hat{\bxi}_{R,q}$, $q=1,\ldots,Q$) when the paths are independent.
Therefore, it is approximately equivalent to the MLE in the large-sample region.
Compared with the original ExIP problem in \cref{eq:UE position estimation}, our estimation of the UE position in \cref{eq:linear combination bound} is expressed more explicitly, which can be solved efficiently.
In addition, the simulations in \cref{sec:numerical} also verify our estimation of the UE position in \cref{eq:linear combination bound} is close to the PEB.
\end{Remark}

\subsection{Estimation of \txp{$\hat{\bp}_{U}^{(B)}$}{p\_Ud} and \txp{$\{\hat{\bp}_{U}^{(R, q)}\}_{q=1}^Q$}{p\_Uq}}  \label{sect: cal pU}

\subsubsection{Inferring \txp{$\hat{\bp}_{U}^{(B)} (\hat{\bxi}_B)$}{p\_Ud} from \txp{$\hat{\bmeta}_{d}$}{bmeta\_d}}

We first focus on the LOS path and discuss how to obtain the refined channel parameters associated with the LOS path. Define ${\boldf} = [{g}_{BU}^{(U)},{s}_{BU}^{(U)},{g}_{BU}^{(B)},{s}_{BU}^{(B)}]\T$,
and $\bz = [f_{BU}^{(B)},{g}_{BU}^{(B)},{s}_{BU}^{(B)}]\T$ with $f_{BU}^{(B)}=\cos\theta_{BU}^{(B)} \cos \phi_{BU}^{(B)}$.
From \cref{AoD:B_d} and \cref{AoD:U_d}, we have the relation
${\boldf}=
[
  \bO_R;
  \boldsymbol{0}, ~\bI
] \bz= \tilde{\bA}\bz,
$
where we denote $ \tilde{\bA}=[
      \bO_R;
      \boldsymbol{0}, ~\bI
]  \in \R^{4 \times 3}$ for convenience.
To estimate $\bz$ from $\hat{\boldf}$, we employ the \gls{WLS} method by solving
\begin{small}
\begin{align} \label{eq:direct Pu}
\min_{\bz} (\hat{\boldf} - \tilde{\bA} \bz)\T \widebar{\bC}_{\hat{\boldf}}^{-1}  (\hat{\boldf} - \tilde{\bA} \bz),\ \st  ~\|  \bz\|_2^2 = 1,
\end{align}
\end{small}%
where
$\widebar{\bC}_{\hat{\boldf}} =[\bF_{\hat{\bmeta}}^{-1}]_{4:7,4:7}$.
If we ignore the constraint on $\bz$, i.e, $\|  \bz\|_2^2 = 1$, the solution is given by
$(\tilde{\bA} \T \widebar{\bC}_{\hat{\boldf}}^{-1} \tilde{\bA})^{-1} \tilde{\bA} \T \widebar{\bC}_{\hat{\boldf}}^{-1} \hat{\boldf}$.
Then, we project this solution to the feasible region of the problem in \cref{eq:direct Pu}.
We denote the final solution of \cref{eq:direct Pu} after projection as $ \tilde{\bz}= [\tilde{f}_{BU}^{(B)},\tilde{g}_{BU}^{(B)},\tilde{s}_{BU}^{(B)}]\T$.

For convenience, we define ${d}_{BU}=c\tau_{BU}$
Then, the estimators $(\hat{d}_{BU},\hat{\theta}_{BU},\hat{\phi}_{BU})$ are given by
$
\hat{d}_{BU} = c\hat\tau_{BU},
\hat{\theta}_{BU} = \arcsin \tilde{s}_{BU}^{(B)}$, and $
\hat{\phi}_{BU} = \arctan 2 (\tilde{g}_{BU}^{(B)},\tilde{f}_{BU}^{(B)}).
$
Furthermore, the estimator $\hat{\bp}_{U}^{(B)}=[\hat{x}_{U}^{(B)},\hat{y}_{U}^{(B)},\hat{z}_{U}^{(B)}]\T$ is given by
\begin{small}
\begin{align}
\begin{cases}
\hat{z}_{U}^{(B)}-z_B= \hat{d}_{BU}   \sin \hat{\theta}_{BU} =  \hat{d}_{BU}   \tilde{s}_{B, d}^{(B)}, \nonumber\\
\hat{x}_{U}^{(B)}-x_B= \hat{d}_{BU} \cos \hat{\theta}_{BU}  \cos \hat{\phi}_{BU}   = \hat{d}_{BU}  \tilde{f}_{B, d}^{(B)}, \nonumber\\
\hat{y}_{U}^{(B)}-y_B = \hat{d}_{BU} \cos \hat{\theta}_{BU}  \sin \hat{\phi}_{BU} =  \hat{d}_{BU} \tilde{g}_{B, d}^{(B)}.
\end{cases}
\end{align}
\end{small}%
The estimation of directional vector is given by $\hat{\br}^{(B)}=[\tilde{f}_{BU}^{(B)},\tilde{g}_{BU}^{(B)},\tilde{s}_{BU}^{(B)}]\T$. Here, the estimation of  $(\opRe\{h_{BU}\}, \opIm\{h_{BU}\} )$ in $\hat{\bxi}_B$ is equivalent to that in $\hat{\bmeta}_B$. Recall the definition of $F_B(\cdot)$ in \cref{eq:partial decom ExIP}, since $\hat{\boldf}$ is not strictly equal to $\tilde{\bA} \tilde{\bz}$ when solving \cref{eq:direct Pu}, one has $F_B(\hat{\bxi}_B)\approx \hat{\bmeta}_B$.

\subsubsection{Inferring \txp{$\hat{\bp}_{U}^{(R, q)}  (\hat{\bxi}_{r,q})$}{p\_Uq} from \txp{$\hat{\bmeta}_{R, q}$}{bmeta\_r}}\label{sect: cal pUr}
For the $q$-th reflection path, we define $g_{RU,q}^{(R)}  = \cos \theta_{RU,q} \sin \phi_{RU,q}$ and  ${f}_{RU, q}^{(U)} = -\sqrt{1-({g}_{RU, q}^{(U)})^2-({s}_{RU, q}^{(U)})^2}$.
Based on the relations in \cref{AoA:RU_q} and rotation matrix $\bO$, we have
\begin{align}
\begin{bmatrix}
f_{RU,q}^{(R)}\\
g_{RU,q}^{(R)}\\
s_{RU,q}^{(R)}
\end{bmatrix} =
\begin{bmatrix}
\cos \theta_{RU,q} \cos \phi_{RU,q}\\
\cos \theta_{RU,q} \sin \phi_{RU,q}\\
\sin \theta_{RU,q}
\end{bmatrix}
=
-\bO^{-1}\begin{bmatrix}
f_{RU, q}^{(U)}\\
g_{RU, q}^{(U)}\\
s_{RU, q}^{(U)}
\end{bmatrix}.
\end{align}
Therefore,
$$[\hat{f}_{RU,q}^{(R)},\hat{g}_{RU,q}^{(R)},\hat{s}_{RU,q}^{(R)}]\T
=-\bO^{-1}[\hat{f}_{RU, q}^{(U)},\hat{g}_{RU, q}^{(U)},\hat{s}_{RU, q}^{(U)}]\T.$$
Similarly, let ${d}_{RU,q} =  c{\tau}_{RU,q}$, then the estimations of $\{\hat{d}_{RU,q},\hat{\theta}_{RU,q} ,\hat{\phi}_{RU,q} \}_{q=1}^Q$ are given by
$\hat{d}_{RU,q} =  c\hat{\tau}_{RU,q}$, $\hat{\theta}_{RU,q} = \arcsin \hat{s}_{RU,q}^{(R)}$, and $\hat{\phi}_{RU,q}  = \arctan 2(\hat{g}_{RU,q}^{(R)}, \hat{f}_{RU,q}^{(R)}).$
The estimation $\hat{\bp}_{U}^{(R, q)}=[\hat{x}_{U}^{(R, q)},\hat{y}_{U}^{(R, q)},\hat{z}_{U}^{(R, q)}]\T$ is given by
\begin{small}
\begin{align}
\begin{cases}
\hat{z}_{U}^{(R, q)} - {z}_{R, q}= \hat{d}_{RU,q} \sin \hat{\theta}_{RU,q}  = \hat{d}_{2,q}  \hat{s}_{RU,q}^{(R)},\nonumber\\
\hat{x}_{U}^{(R, q)}-{x}_{R, q}= \hat{d}_{RU,q} \cos \hat{\theta}_{RU,q} \cos \hat{\phi}_{RU,q}   =  \hat{d}_{RU,q} \hat{f}_{RU,q}^{(R)}, \nonumber\\
\hat{y}_{U}^{(R, q)}-{y}_{R, q} = \hat{d}_{RU,q}\cos \hat{\theta}_{RU,q} \sin \hat{\phi}_{RU,q}= \hat{d}_{RU,q} \hat{g}_{RU,q}^{(R)}.   \nonumber
\end{cases}
\end{align}
\end{small}%
The estimation of directional vector is $\hat{\br}^{(R, q)}=[\hat{f}_{RU,q}^{(R)},\hat{g}_{RU,q}^{(R)},\hat{s}_{RU,q}^{(R)}]\T$. The estimation of  $(\opRe\{h_{R,q}\}, \opIm\{h_{R,q}\} )$ in $\hat{\bxi}_{R,q}$ is equivalent to that  in $\hat{\bmeta}_{R,q}$. In addition, from the  definition of $F_{R,q}(\cdot)$ in \cref{eq:partial decom ExIP}, one has $F_{R,q}(\hat{\bxi}_{R,q})= \hat{\bmeta}_{R,q}$.

\section{Discussions} \label{sec:discussions}
In this section, we propose methods to optimize the phase shifts of a RIS to position a UE. We also discuss the extension of our proposed framework to the multi-BS and multi-UE scenarios.

\subsection{Signal Overhead and Computational Complexity} \label{sec:complexity}
The multiple RISs provide us with more reflection paths.
In our approach, all RISs are used simultaneously and scheduling is not required.
Therefore, different from the existing work \cite{wang2021joint}, where the signal overhead is linearly proportional to the number of RISs,
the signal overhead of the proposed method is independent of the number of RISs. Recall that
the signal overhead requirement is $T\ge N$ in \cref{sect:received signal}, and it means that the number of signal transmissions is required to be larger than the number of antennas at the BS.

We next analyze the computational complexity of the proposed method. The first step of the proposed method consists of estimating the AoD, delays, AoAs, and path gains.
For the AoD estimation, we suppose the number of iterations for solving the problem in \cref{eq:transform problem} is $n_{iter}$, whose value relies on the desired error tolerance \cite{nocedal1999numerical}.\footnote{For convenience, we use the notation $n_{iter}$ to stand for the number of iterations for the optimization problem in the paper.} Since each function evaluation in \cref{eq:transform problem} has complexity of $\mathcal{O}(NDK)$, thus, the computational complexity of estimating AoD is $\mathcal{O}(n_{iter} NDK)$.
Similarly, the computational complexities for estimating the delays in  \cref{eq:est delay music}  and AoAs in \cref{eq:est AoA music} are given by $\mathcal{O}(n_{iter} Q  ND K)$ and $\mathcal{O}(n_{iter} Q  ND K)$, respectively. Meanwhile, the complexity to estimate the path gains in \eqref{eq:estimate gain problem} is $\mathcal{O}(Q^2  ND K)$. When the number of RISs is small such that $Q\ll  n_{iter}$, the overall computational complexity for the first step is given by $\mathcal{O}(n_{iter} Q  ND K)$.
For the second step of inferring the UE position from the channel parameters, the computational complexity is dominated by solving \cref{eq:linear combination bound}, which can be approximated as a one-dimensional optimization. Since the complexity of evaluating the function in \cref{eq:linear combination bound} is $\mathcal{O}(Q)$, the complexity for the second step is given by $\mathcal{O}(n_{iter} Q)$. Therefore, after combing the two steps, the overall computational complexity of the proposed positioning method is $\mathcal{O}(n_{iter}Q  ND K+n_{iter} Q) = \mathcal{O}(n_{iter}Q  ND K)$.

In the scenario of the multiple RISs, one can check that the proposed method has the same computational complexity as the AoA and AoD fine search in \cite{wang2021joint}. However, apart from the fine search in the existing work \cite{wang2021joint, Key2021SISO}, a two-dimensional search is also required to initialize the AoA or AoD, which may have high complexity, i.e., $\mathcal{O}(Z_{1}Z_2)$, with $Z_1$ and $Z_2$ being the two-dimensional quantization levels.
As discussed above, the proposed method is different from the existing work, where it only performs the one-dimensional search in the first or second step.
Therefore, due to the efficiency of one-dimensional search, the proposed positioning method is more potential to be applied in the real-time system.

\subsection{Design of RIS Phase Shifts} \label{sec:phase shift}
The design of phase shifts of each RIS is an important topic for RIS-aided positioning \cite{Ji2020large,elzanaty2021reconfigurable,fascista2022ris}. However, directly minimizing the PEB involves high complexity. Moreover,  it requires the exact position of the UE, which is unknown a priori. In the following, we provide an SVD-based design for the phase shifts of a RIS, which aims to serve multiple UEs in a certain range. Moreover, the proposed SVD-based design can be readily extended to the scenario of multiple BSs in \cref{section multi UEs BSs}.

Specifically, in our work, the phase shifts of a RIS are designed to serve the UEs with elevation angles in the range $[\theta_l, \theta_u]$ and azimuth angles in $[\phi_l, \phi_u]$. For example, the LOS between the BS and UEs within this region of interest may be blocked with high probability. Therefore, the RIS phase shifts are designed to aid these UEs.

 Recall that the gain of the reflection path from the $q$-th RIS is proportional to $| \ba_M\He(f_{RU,q}^{(R)},s_{RU,q}^{(R)}) \bTheta_q \allowbreak \ba_M(f_{BR,q}^{(R)},s_{BR,q}^{(R)})| $. Inspired by \cite{Ji2020large,elzanaty2021reconfigurable}, which maximizes the gain of the reflection path, and because the quantities $f_{BR,q}^{(R)}$ and $s_{BR,q}^{(R)}$ are known a priori, we propose an unbiased design $\bTheta_q$ in the following.
For convenience, we combine $ \bTheta_q  \ba_M(f_{BR,q}^{(R)},s_{BR,q}^{(R)})$ as one variable $ \tilde{\btheta}_q =[\tilde{\theta}_q^{(1)},\ldots, \tilde{\theta}_q^{(M)}] \T\in \C^{M \times 1}$. If the UEs are uniformly distributed  in the  elevation range $[\theta_{q,l}, \theta_{q,u}]$ and azimuth range $[\phi_{q,l}, \phi_{q,u}]$, we consider the following optimization problem:
\begin{align}

\max_{ \tilde{\btheta}_q }&\ \Eb[ | (\tilde\ba_{M_r}(f) \otimes  \tilde\ba_{M_c}(s) )\He \tilde{\btheta}_q  |^2]  \nonumber\\
\st &\ f = \cos \theta \cos \phi, ~s = \sin\theta,  \nonumber\\
& \qquad \quad \text{for}~ \theta \sim U[{\theta}_{q, l},{\theta}_{q, u}] , ~\phi \sim U[{\phi}_{q, l},{\phi}_{q, u} ],  \nonumber \\
& \ |\tilde{\theta}_q^{(m)}|=1, \forall m. \label{eq: phase expectation}
\end{align}
The expectation in \cref{eq: phase expectation} is a unit-modulus optimization over high dimensions (if $M$ is large), which is numerically expensive to compute. Instead, to perform approximation, we define a matrix
 $\bD_A \in \C^{M\times Z}$ with its columns having the form of $ \tilde\ba_{M_r}(f) \otimes  \tilde\ba_{M_c}(s) $, where  $(f, s)$ is chosen in a discretized range. Then, one has
$
\Eb[ | (\tilde\ba_{M}(f) \otimes  \tilde\ba_{M}(s) )\He \tilde{\btheta}_q  |^2]\approx 1/Z \| \bD_A\He \tilde{\btheta}_{q}   \|_2^2.
$
Then, we can reformulate the  problem in \cref{eq: phase expectation} as
\begin{align}
\hat{\tilde{\btheta}}_q = \argmax_{\tilde{\btheta}_{q} }\| \bD_A\He \tilde{\btheta}_{q} \|_2^2, \ \st |\tilde{\theta}_q^{(m)}|=1, \forall m. \label{eq: reform RIS phase}
\end{align}
If there is no constraint for $\hat{\tilde{\btheta}}_q $, the solution is the dominant left singular vector of $\bD_A$.
In order to satisfy the constraint imposed on $\tilde{\btheta}_{q}$, we let $\hat{\tilde{\btheta}}_q $ be the complex angle of the dominant left singular vector of $\bD_A$.
The design of the phase shifts of the $q$-th RIS, $\bTheta_q$, is then given by  $\diag(\btheta_q)$ with
$
\angle\btheta_q = -\angle\ba_M(f_{BR,q}^{(R)},s_{BR,q}^{(R)}) + \angle \hat{\tilde{\btheta}}_q.
$

It is worth noting that the proposed design of RIS is fixed over time.
When the RIS profile can be designed over time, the effective channel in \cref{eq:final expression of Hk} is expressed as
$\bH_{k}(t) = {\bH}_{BU,k}+\sum_{q=1}^{Q} \bH_{RU, k,q} \bTheta_q (t) \bH_{BR, k,q}$, and the received signal is $\br_k(t) = \bH_k(t) \bx (t)  +  \bn_k(t)$.
The design of the dynamic RIS profile has been studied in recent literature \cite{keykhosravi2022ris,fascista2022ris}.
In \cite{keykhosravi2022ris}, the RIS profile is randomly designed over part of the time slots and is concentrated to the UE position for the rest of the time slots. On the other hand, \cite{fascista2022ris} proposes a combination of directional and derivative beams for the RIS profile.

When there is only a single antenna at the BS and UE \cite{keykhosravi2022ris,fascista2022ris}, it is necessary to change the RIS profile over time to take advantage of its resolution to estimate the AoA/AoD.
However, when there are multiple antennas at the UE and BS, fixing the RIS profile to act as a strong reflector can result in accurate AoA/AoD estimates at the UE/BS, at the expense of not utilizing the resolution capability of the RIS.
The decision to change the RIS profile over time is a trade-off, which depends on the number of antennas at the UE and BS. Our investigations have shown that in cases with sufficient antennas at the UE and BS, a fixed RIS profile can result in slightly more accurate results than a changing profile, especially in low SNR environments.
Given this trade-off, in future work, it is of interest to investigate the design of the RIS profile in MIMO scenarios and to apply deep learning techniques such as deep reinforcement learning \cite{sutton2018reinforcement} to the RIS profile design.

\subsection{Extension to Multiple UEs and Multiple BSs} \label{section multi UEs BSs}
Suppose that there are $n_U$ UEs, $n_B$ BSs and $Q$ RISs. Similar to the single-UE case, in which the UE receives the signals reflected by $Q$ RISs simultaneously, the multiple UEs receive signals reflected by $Q$ RISs. In other words, these $Q$ RISs serve multiple UEs simultaneously.
By letting all BSs transmit orthogonal signals, each UE can distinguish each BS's signal and estimate its position using the approach presented in the previous sections.
The UE $l$ estimates its position through the received signal of the $i$th BS as $\hat{\bp}_{U, l}^{(i)}$. Suppose the corresponding covariance matrix is $\bC_{{\bp}_{U, l}^{(i)}} \in \R^{3\times 3}$.
In particular, the relative positions of the UEs are estimated through inter-UE measurements and message exchanges \cite{wymeersch2009cooperative, xu2015distributed, conti2012network}. Here, we let the relative position between the $l$th UE and $l'$th UE be denoted by ${\bGamma}_{l,l'} = \bp_{U,l'}- \bp_{U,l}$, and its estimation  be given by $\hat{\bGamma}_{l,l'}$ with the covariance matrix ${\bC}_{{\bGamma}_{l,l'}}$. In particular, if $l=l'$, one has ${\bC}_{{\bGamma}_{l,l}}=\boldsymbol{0}$.

The position of the $l$th UE is obtained through the following \gls{WLS} formulation described in \cref{sec:linear fusion}:
\begin{align*}
  \min_{\bp_{U,l}} \sum_{i=1}^{n_B} \sum_{l'=1}^{n_U} &(\bp_{U,l}-\hat{\bp}_{U, l'}^{(i)}+\hat{\bGamma}_{l,l'})\T  ({\bC}_{\bp_{U, l'}^{(i)}}^{-1} +{\bC}_{{\bGamma}_{l,l'}}^{-1} )  (\bp_{U,l}-\hat{\bp}_{U, l'}^{(i)}+\hat{\bGamma}_{l,l'}).
\end{align*}
The closed-form solution is given by
$
\hat{\bp}_{U,l}=\sum_{i=1}^{n_B} \sum_{l'=1}^{n_U} \bA_{i, l'} (\hat{\bp}_{U, l'}^{(i)}-\hat{\bGamma}_{l,l'})$,
where $\bA_{i, l'}\in \R^{3 \times 3}$ is the combining matrix given by
$\bA_{i, l'}=(\sum_{i=1}^{n_B} \sum_{l=1}^{n_U} ({\bC}_{\bp_{U, l'}^{(i)}}^{-1} +{\bC}_{{\bGamma}_{l,l'}}^{-1} ))^{-1} ({\bC}_{\bp_{U, l'}^{(i)}}^{-1} +{\bC}_{{\bGamma}_{l,l'}}^{-1})$.
When the error covariance $\bC_{{\bp}_{U, l'}^{(i)}}$ is not available, we can employ the CRB as an alternative, which can still achieve near-optimal performance as discussed in \cref{sec:linear fusion,sect:asy MLE}.

When there are multiple BSs, the SVD-based design of phase shifts of RIS can also be applied. For the $q$-th RIS, it is assumed to serve the BSs and UEs in a certain range. Specifically, the UEs are uniformly distributed in the elevation range $[\theta^{(U)}_{q,l}, \theta^{(U)}_{q,u}]$ and azimuth range $[\phi^{(U)}_{q,l}, \phi^{(U)}_{q,u}]$, and the BSs are uniformly distributed in the elevation range $[\theta^{(B)}_{q,l}, \theta^{(B)}_{q,u}]$ and azimuth range $[\phi^{(B)}_{q,l}, \phi^{(B)}_{q,u}]$. We propose to solve the following problem:
\begin{small}
\begin{align}
\max_{ \bTheta_q  }&\ \Eb[ | \ba_M\He(f_1,s_1) \bTheta_q  \ba_M(f_2,s_2)|], \nonumber \\
\st &\ f_1= \cos\theta \cos \phi, ~s_1 = \sin\theta,  ~\text{for}~ \theta \sim U[{\theta}^{(U)}_{q, l},{\theta}^{(U)}_{q, u}] , ~\phi \sim U[{\phi}^{(U)}_{q, l},{\phi}^{(U)}_{q, u}], \nonumber \\
&\ f_2= \cos\theta \cos \phi, ~s_2 = \sin\theta, ~\text{for}~ \theta \sim U[{\theta}^{(B)}_{q, l},{\theta}^{(B)}_{q, u}] , ~\phi \sim U[{\phi}^{(B)}_{q, l},{\phi}^{(B)}_{q, u}],\nonumber \\
& \ \bTheta_q = \diag(\btheta_q), ~ |\theta_q^{(m)}| = 1, \forall m. \label{multiRISobj}
\end{align}
\end{small}%
By discretizing the continuous range, we let the columns of $\bA_1$ have the expressions of $\ba_M (f_1,s_1) $, and columns of $\bA_2$ have the expressions of $\ba_M (f_2,s_2) $. Then,
the above problem in \cref{multiRISobj} becomes
\begin{align*}
\begin{aligned}
\max_{ \bTheta_q  }& \ \|  \bA_1\He \bTheta_q  \bA_2\|_F^2, \nonumber \\
 \st & \ \bTheta_q = \diag(\btheta_q), ~ |\theta_q^{(m)}| = 1, \forall m,
 \end{aligned}
\end{align*}
or equivalently,
\begin{align}
\max_{ \bTheta_q  }  &\ \|  (\bA_1\He \otimes \bA_2 \T) \vect(\bTheta_q) \|_2^2, \nonumber \\
 \ \st &\ \bTheta_q = \diag(\btheta_q), ~ |\theta_q^{(m)}| = 1, \forall m. \label{eq:multi_BS_RIS}
\end{align}
It is worth noting that when there is no constraint on $\bTheta_q$, the problem can be solved through SVD. However, there are two constraints for $\bTheta_q$: 1) it is a diagonal matrix, and 2) the diagonal entries are unit-modulus. To handle this, we define a sub-matrix $\bA_S$ of $\bA_1\He \otimes \bA_2\T $, which collects the corresponding columns indexed by the non-zero entries of  $\vect(\bTheta_q)$. Then the problem in \cref{eq:multi_BS_RIS} becomes
\begin{align}
\max_{ \btheta_q  } \|  \bA_S \btheta_q \|_2^2,\ \st  |\theta_q^{(m)}| = 1, \forall m.
\end{align}
Similarly, the problem can be solved by using the proposed SVD-based method in \cref{sec:phase shift}.

\section{Numerical Results}\label{sec:numerical}
In this section, we numerically evaluate the proposed RIS-aided positioning method in a mmWave setting. We compare the achieved UE positioning accuracy to the PEB under varying noise levels. Numerical experiments are also conducted to provide insights into the phase shifts of a RIS on the UE positioning accuracy. Finally, we present numerical results for the multi-UE and multi-BS scenarios.

In the numerical experiments, we use the root-mean-square error, RMSE $=\sqrt{\Eb[\|\bp_U-\hat{\bp}_U\|_2^2]}$, to measure the positioning accuracy, where $\hat{\bp}_U$ is the estimated UE position.
Using similar settings as those in \cite{Wen2021Pos,zhang2017performance,3gpp.25.996,samimi201628}, we set our experiment parameters as shown in \cref{tab:simulation parameter}, except for the parameter that we are varying.
By varying the noise level $\sigma$, the SNR is defined as the ratio of the power of the received signal to that of noise in \cref{eq:t receive signal}.
Note that when the power of the received signal changes, the received SNR will change accordingly, even with a fixed noise level $\sigma$. For a fair comparison, we let the SNR value in the following simulations correspond to the noise level $\sigma$.

\begin{table}[t]
  \footnotesize
  \centering
  \caption{Simulation Parameters}
  \begin{tabular}{|c|c|}\hline
    \textbf{Parameter}                        & \textbf{Value}      \\ \hline
    Number of BS antennas                     & $N=10\times 10$             \\ \hline
    Number of UE antennas                     & $D= 8 \times 8$              \\ \hline
    RIS size                                  & $M=20 \times 20$             \\ \hline
    Transmission bandwidth                    & $W = 100$ MHz \\ \hline
    Carrier frequency                         & $f_c=30$ GHz  \\ \hline
    Number of OFDM subcarriers                & $K=64$              \\ \hline
    Rician factor                             & $K_{BU}= K_{RU}=10$        \\ \hline
    Number of time slots                      & $T=200$             \\  \hline
    Path loss exponent of the LOS path     & $L_{BU}=4.5$           \\ \hline
    Path loss exponent of the reflection path & $L_R=2$             \\ \hline
    SNR                              & $10$ dB             \\ \hline
  \end{tabular}
  \label{tab:simulation parameter}
\end{table}

\subsection{Accuracy of UE Positioning and Clock Bias} \label{sect:sim Accuracy}

In this simulation, we evaluate the accuracy of UE positioning and clock bias of the proposed RIS-aided positioning method with a single BS and UE. All position coordinates are measured in meters with the BS at the origin as illustrated in \cref{fig:system}.
The UE position is $\bp_U=[50,10,-20]\T$ and two RISs are located at position $\bp_{R,1}=[30,-5,-2]\T, \bp_{R,2}=[16,20,31]\T$. We let the clock bias $\Delta_\mathrm{clock} = 50 ~\text{ns}$, and evaluate the positioning accuracy of the following methods:
\begin{itemize}
  \item The proposed positioning method that distinguishes the LOS path based on the path energy is labeled as the ``Proposed (energy)'' (cf.\ \cref{distinguish LOS path}).
  \item The proposed positioning method that distinguishes the LOS path based on the estimated delay is labeled as the ``Proposed (delay)'' (cf.\ \cref{eq:est delay music}).
  \item The LS positioning method in \cite{wang2021joint}.
   \item The SISO positioning method in \cite{Key2021SISO}.
\end{itemize}
In this experiment, a single RIS is used for SISO \cite{Key2021SISO} for a fair comparison,  since it does not consider multiple RISs. For LS \cite{wang2021joint} and the proposed methods, we consider the case of the multiple RISs.

\begin{figure}[!htbp]
\centering
\begin{minipage}[t]{0.42\textwidth}
\centering
\includegraphics[width=0.95\textwidth]{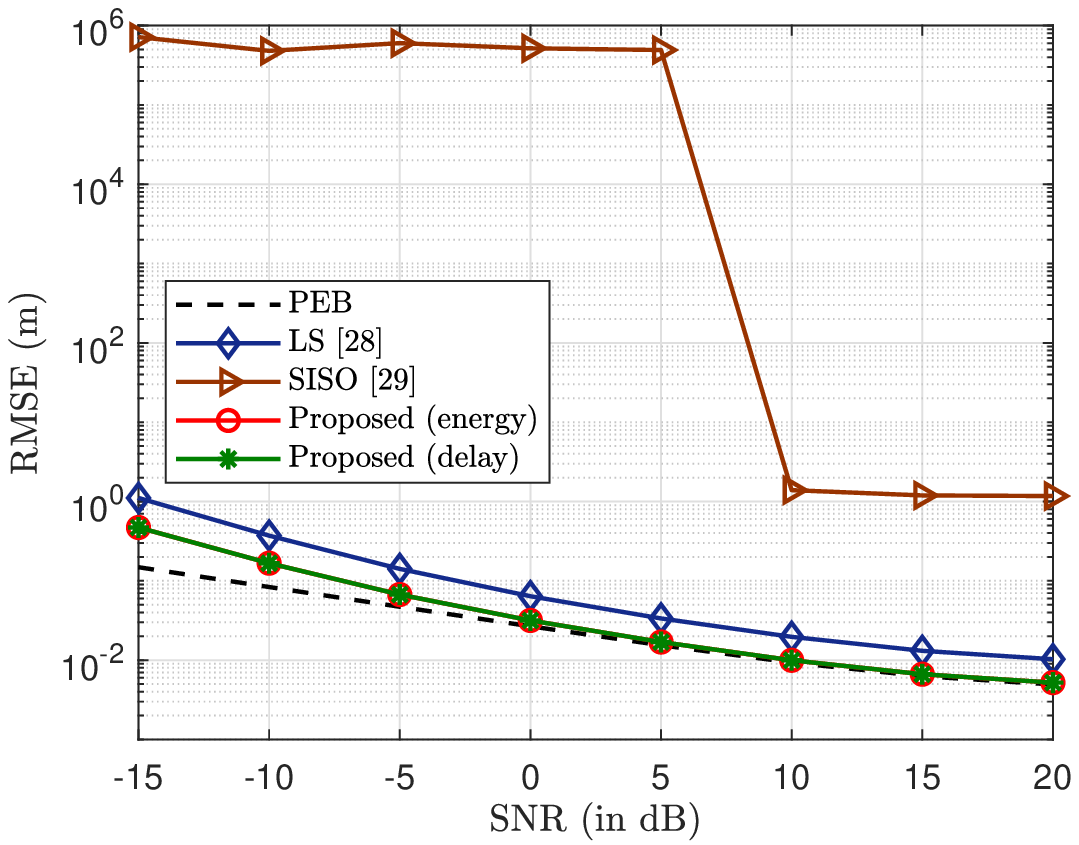}
\caption{RMSE of UE position versus SNR {(single RIS)}.  }  \label{fig:fig position error One}
\end{minipage}
\hspace{1.0cm}
\begin{minipage}[t]{0.42\textwidth}
\centering
\includegraphics[width=0.95\textwidth]{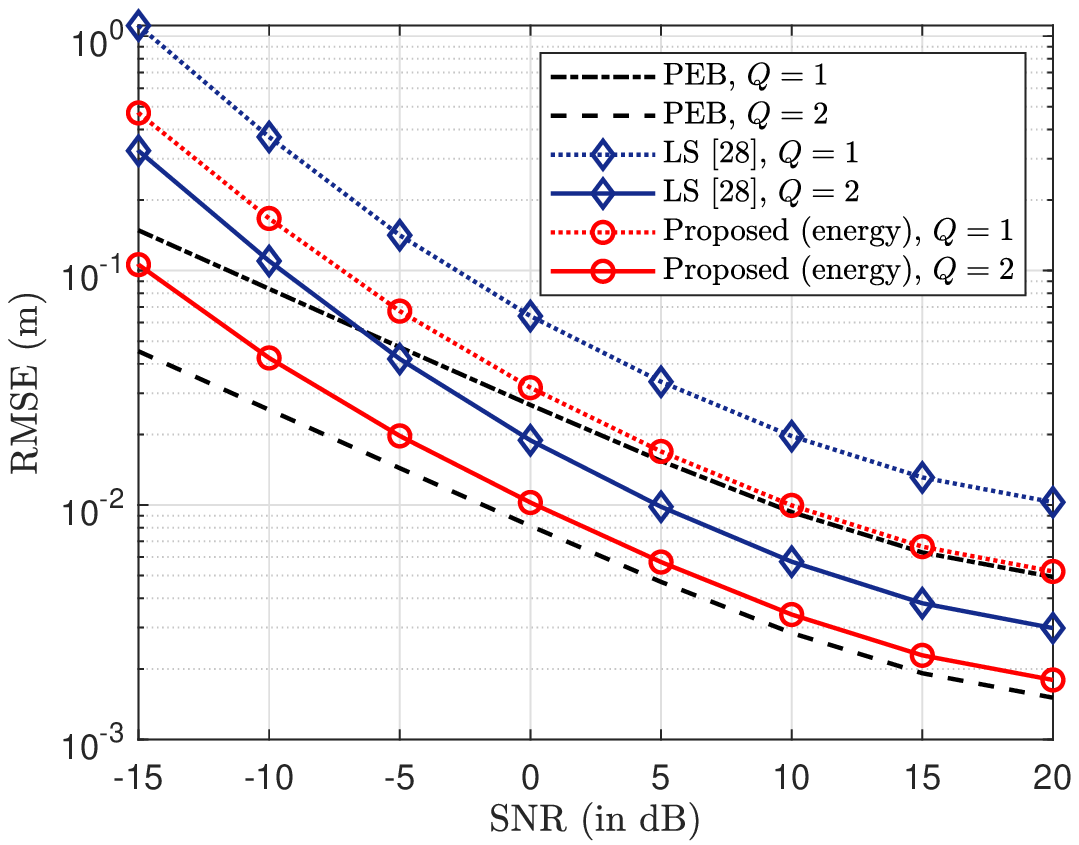}
\caption{RMSE of UE position versus SNR {(multiple RISs)}.  }  \label{fig:fig position error Two}
\end{minipage}
\end{figure}

In \cref{fig:fig position error One}, we first illustrate the position accuracy of the benchmarks in the scenario of a single RIS.
We observe from \cref{fig:fig position error One} that both our proposed methods outperform the other two benchmarks under varying SNRs. The result illustrates that distinguishing the paths based on path energy provides approximately equal performance compared with the delay-based approach.
Furthermore, compared to the LS positioning method in \cite{wang2021joint}, which places equal weight on each path, the proposed methods incorporate the accuracy of the estimated parameters in making the positioning inference. This allows the accuracy of the proposed methods to approach the PEB.
Additionally, compared to  the SISO positioning method presented in \cite{Key2021SISO}, the proposed methods in this work have the advantage of being able to estimate the AoA and AoD using both multiple-antenna UE and BS, resulting in improved positioning accuracy as demonstrated in  \cref{fig:fig position error One}.

In \cref{fig:fig position error Two}, we illustrate the position accuracy in the scenario of multiple RISs. We observe from \cref{fig:fig position error Two} that using multiple RISs can improve the positioning accuracy of the benchmarks compared to a single RIS, which is due to the availability of additional reflection paths.
Similar to the scenario of a single RIS, due to the incorporation of the estimated parameters' accuracies, the proposed methods outperform the LS positioning method \cite{wang2021joint}  and approach the PEB under varying SNRs.

\begin{figure}[!htbp]
\centering
\begin{minipage}[t]{0.32\textwidth}
\centering
\includegraphics[width=0.95\textwidth]{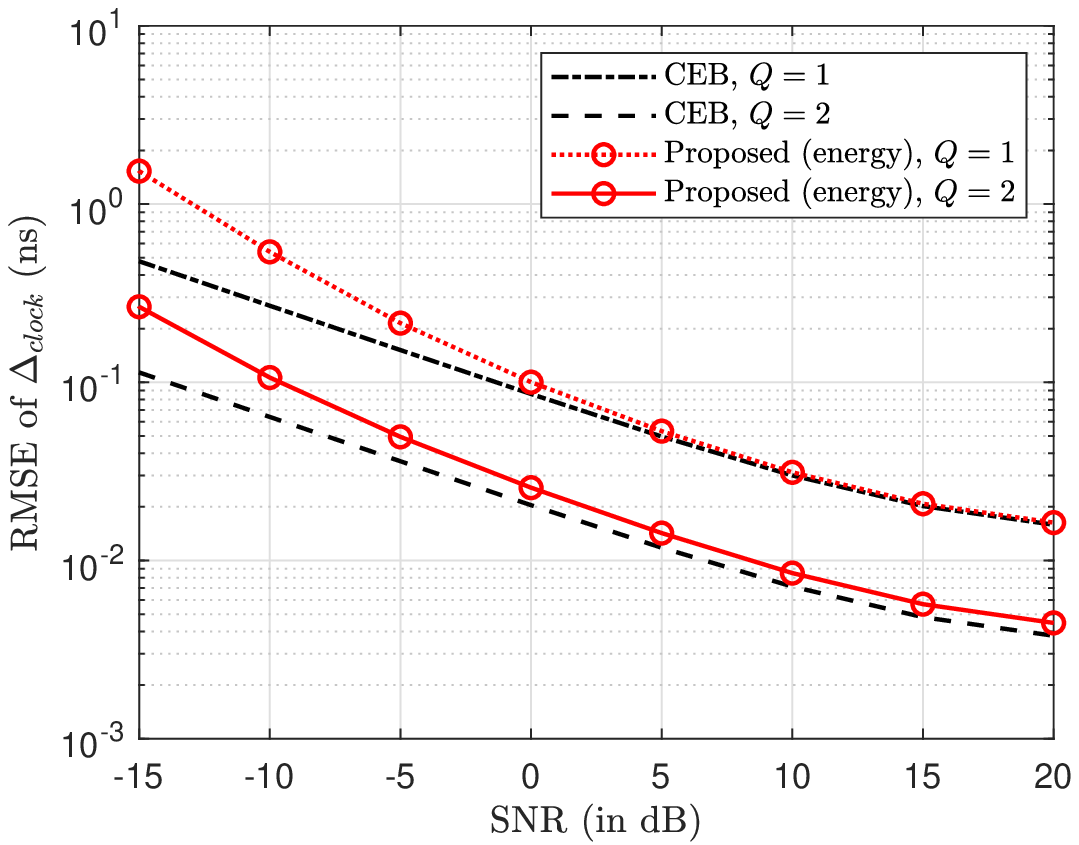}
\caption{RMSE of clock bias versus SNR.  }  \label{fig:fig clock error}
\end{minipage}
\begin{minipage}[t]{0.32\textwidth}
\centering
\includegraphics[width=0.95\textwidth]{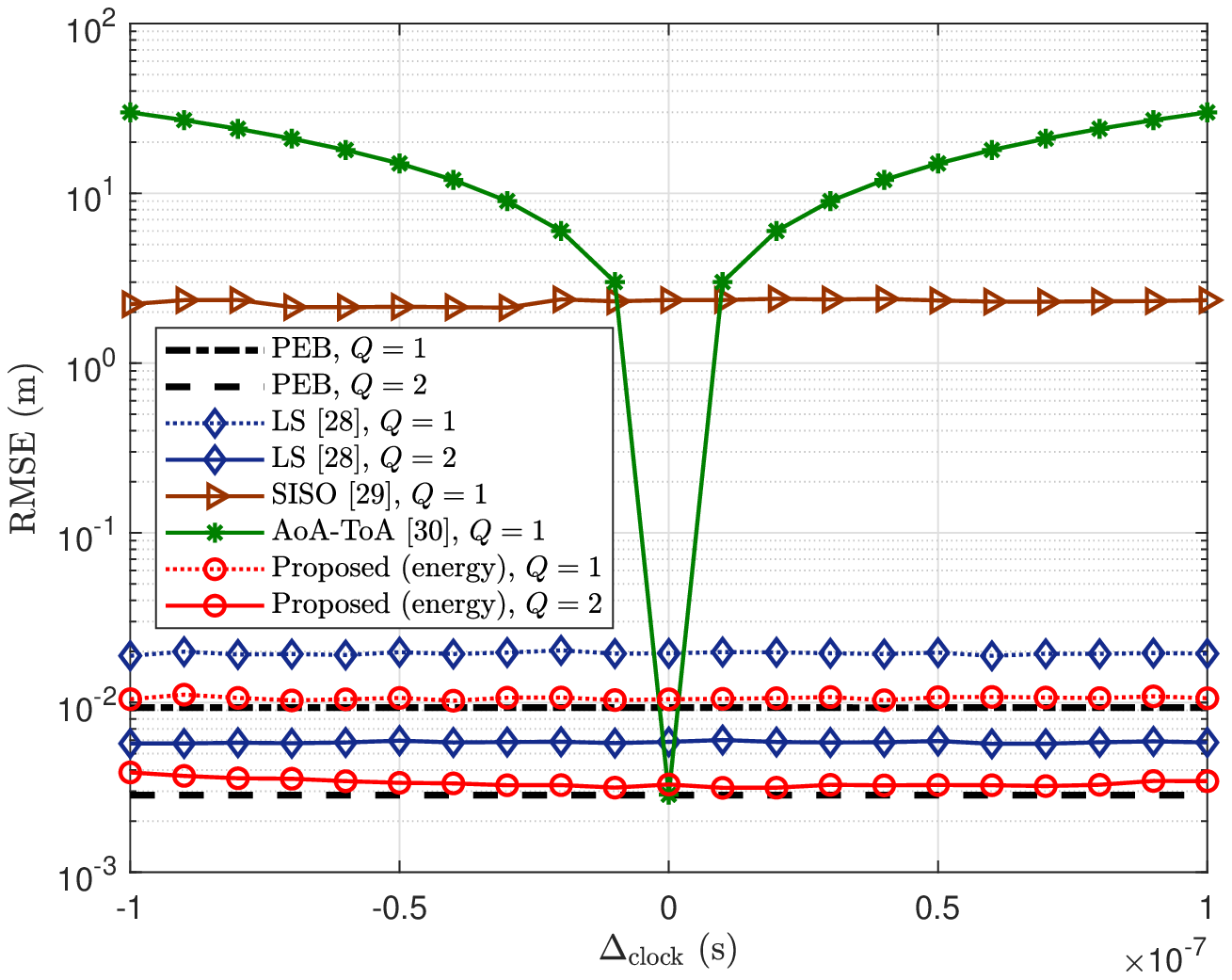}
\caption{RMSE of UE position versus clock bias. } \label{fig:different_clock_bias}
\end{minipage}
\begin{minipage}[t]{0.33\textwidth}
\centering
\includegraphics[width=0.95\textwidth]{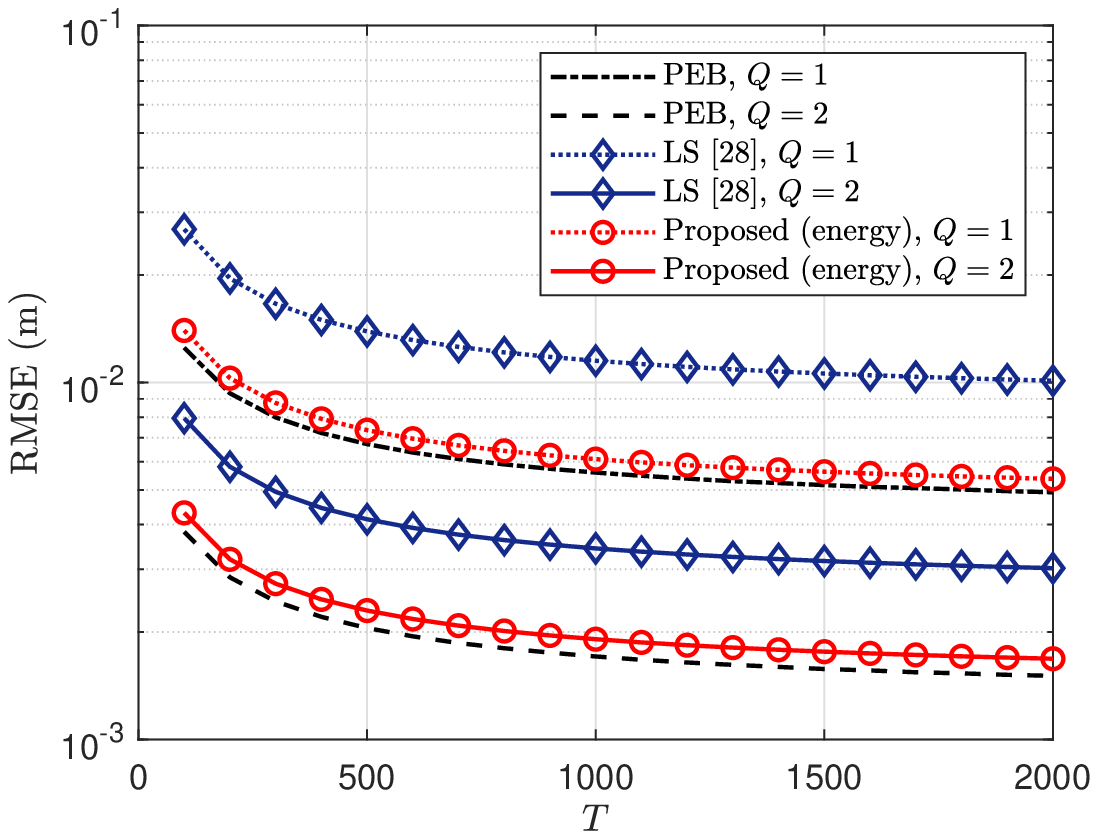}
\caption{RMSE of UE position versus the number of samples. } \label{fig:Num_Samples}
\end{minipage}
\end{figure}

In \cref{fig:fig clock error}, we evaluate the accuracy of the clock bias estimation of the benchmarks. Since the LS positioning method in \cite{wang2021joint} does not estimate the clock bias, we do not compare it in \cref{fig:fig clock error}. Similar to the RMSE of UE position, the proposed method achieves a near-optimal estimation of the clock bias. Meanwhile, the multiple RISs can also improve the estimation accuracy of the clock bias.

In \cref{fig:different_clock_bias},  we evaluate the positioning accuracy of the RIS-aided positioning methods with different values of clock bias $\Delta_{\text{clock}}$.
We vary $\Delta_{\text{clock}}$ in the range of $[-10/f_s, 10/f_s]$, and the other parameters are the same as \cref{tab:simulation parameter}.
As seen from \cref{fig:different_clock_bias}, because the proposed method estimates the clock bias and the UE position, the positioning accuracy of the proposed method is approximately the same for different clock biases, which is consistent with the theoretical PEB.
However, for the AoA-ToA positioning method in \cite{Lin2021Channel}, since it does not consider the estimation of clock bias between BS and UE, the value of clock bias affects the positioning accuracy significantly.
Since LS \cite{wang2021joint} uses only AoA, it is not affected by clock bias, as seen in the figure.
For the SISO method \cite{Key2021SISO}, the positioning accuracy is also constant with different biases. However, its accuracy is much worse than the proposed method due to the single antenna at the BS and UE.
 Overall, the proposed method outperforms the LS and SISO methods.

\subsection{Number of Samples}\label{sect:NumSam}
In \cref{fig:Num_Samples}, we compare the positioning accuracy of the proposed method under different numbers of samples $T$. We observe that when more samples are utilized, the proposed methods and LS can achieve more accurate UE positioning results.
This is because, for the MIMO setting, more samples can lead to a smaller equivalent noise level in \cref{eq:R observations}.

\subsection{Number of RIS elements and RISs}\label{sect:RIS sim}
\begin{figure}[!htbp]
\centering
\begin{minipage}[t]{0.42\textwidth}
\centering
\includegraphics[width=0.95\textwidth]{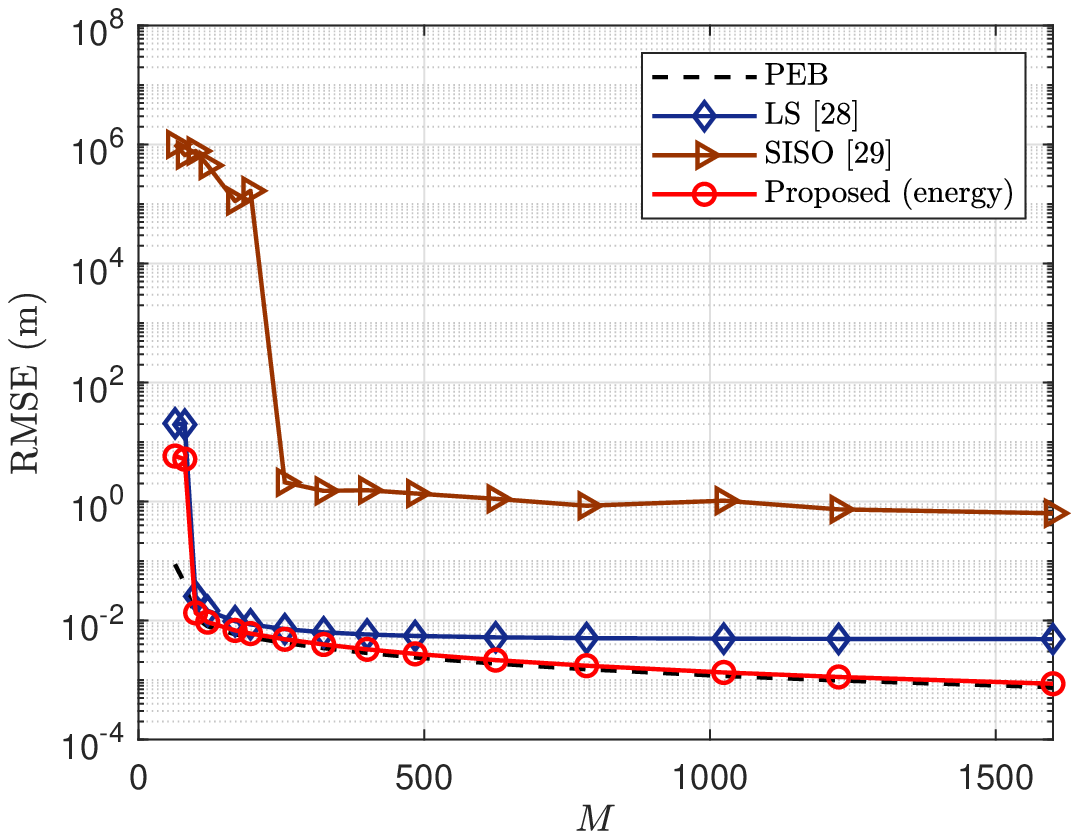}
\caption{RMSE of UE position versus the number of RIS elements (single RIS). }  \label{fig:fig RIS_element}
\end{minipage}
\hspace{1.0cm}
\begin{minipage}[t]{0.42\textwidth}
\centering
\includegraphics[width=0.95\textwidth]{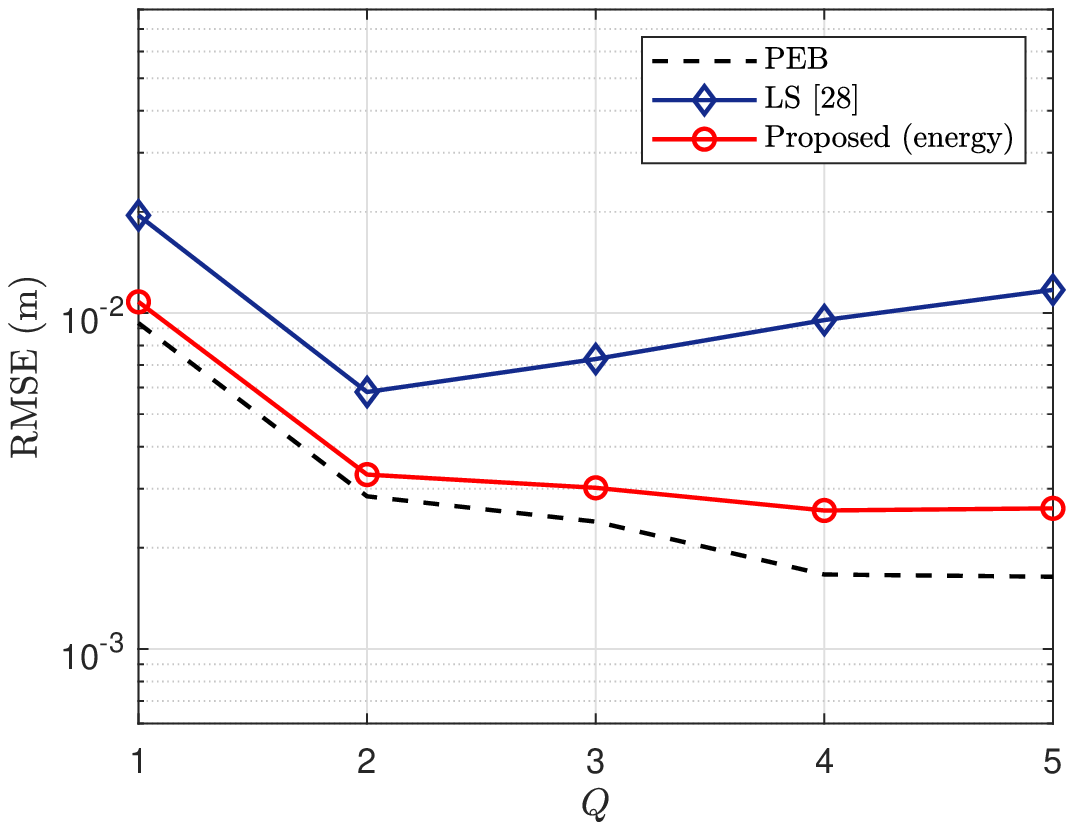}
\caption{RMSE of UE position versus the number of RISs. }  \label{fig:two_RIS}
\end{minipage}
\end{figure}

In \cref{fig:fig RIS_element}, we evaluate the UE positioning accuracy with a varying number of elements $M$ in the RIS URA.
The other simulation parameters are the same as \cref{tab:simulation parameter}. It can be observed that when the RIS has more elements, more accurate positioning results can be achieved for all methods. This is because more RIS elements lead to an increase in the power of the reflection path, which benefits the positioning accuracy.

Interestingly, we observe that the positioning errors of all the methods experience a sharp decrease with $M$. This happens because the positioning methods require sufficient power in the reflection path to achieve a valid estimation of the UE position.
For our proposed method and LS \cite{wang2021joint}, the power of the reflection path is proportional to $M^2$. However, for the SISO method \cite{Key2021SISO}, the power of the reflection path is proportional to $M$.
Therefore, our proposed method and LS \cite{wang2021joint} require
fewer RIS elements before the sharp decrease in RMSE in \cref{fig:fig RIS_element}.


In \cref{fig:two_RIS},  we evaluate the positioning accuracy while varying the number of RISs, i.e., $Q=1,2,\ldots,5$. The other simulation parameters are as in \cref{tab:simulation parameter}.
We observe that when multiple RISs are utilized for positioning, more accurate positioning results can be achieved for the proposed method.
 This is because more RISs can provide more reliable reflection paths to achieve positioning. However, LS \cite{wang2021joint} treats different paths equally, which does not optimally infer the UE position. This explains why its performance does not improve monotonically as the number of RIS increases.
 Interestingly, as we observe from \cref{fig:two_RIS}, with an increasing number of RISs, the performance gap between the PEB and the positioning RMSE of the proposed method increases. This is because more RISs can lead to increased interference for the reflection signals from different RISs, which is not mitigated in either the proposed method or the LS method. Advanced signal processing techniques may be used to reduce interference and will be considered in future research.

\subsection{LOS Path Gains}
\begin{figure}[!htbp]
\centering
\begin{minipage}[t]{0.42\textwidth}
\centering
\includegraphics[width=0.95\textwidth]{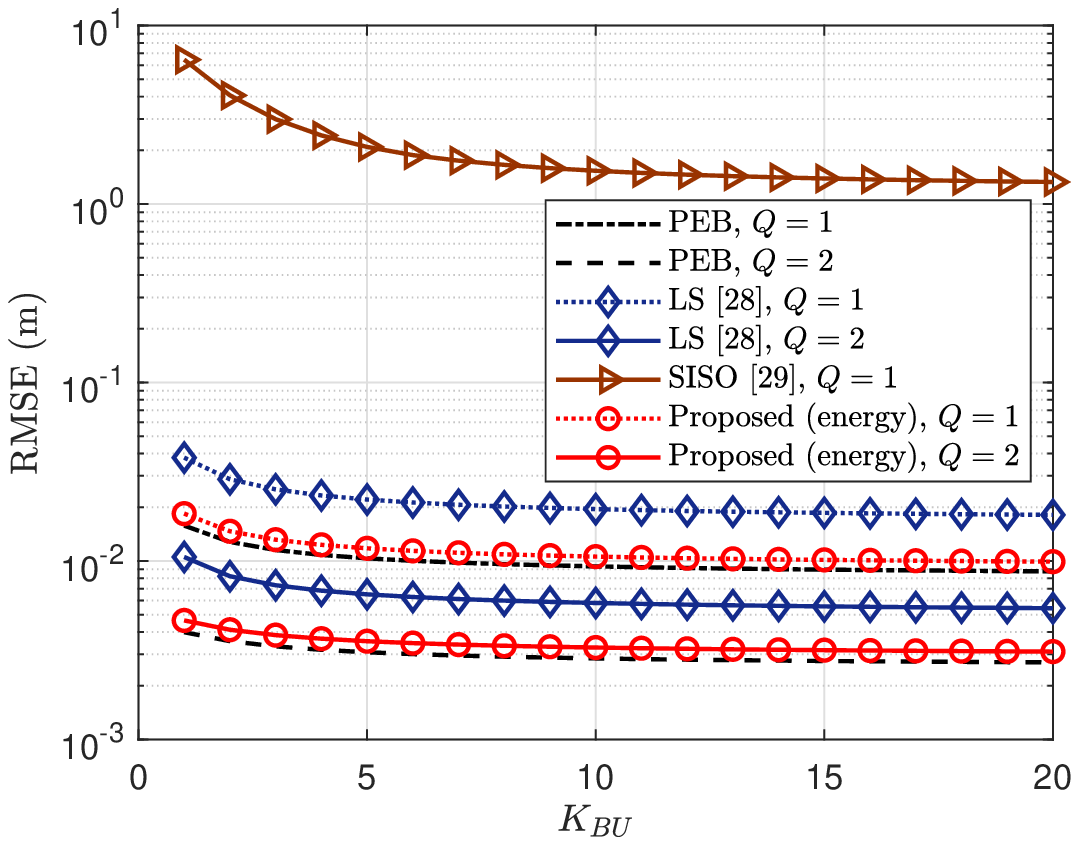}
\caption{RMSE of UE position versus Rician factor of LOS path. } \label{fig:direct_fading_K_BU}
\end{minipage}
\hspace{1.0cm}
\begin{minipage}[t]{0.42\textwidth}
\centering
\includegraphics[width=0.95\textwidth]{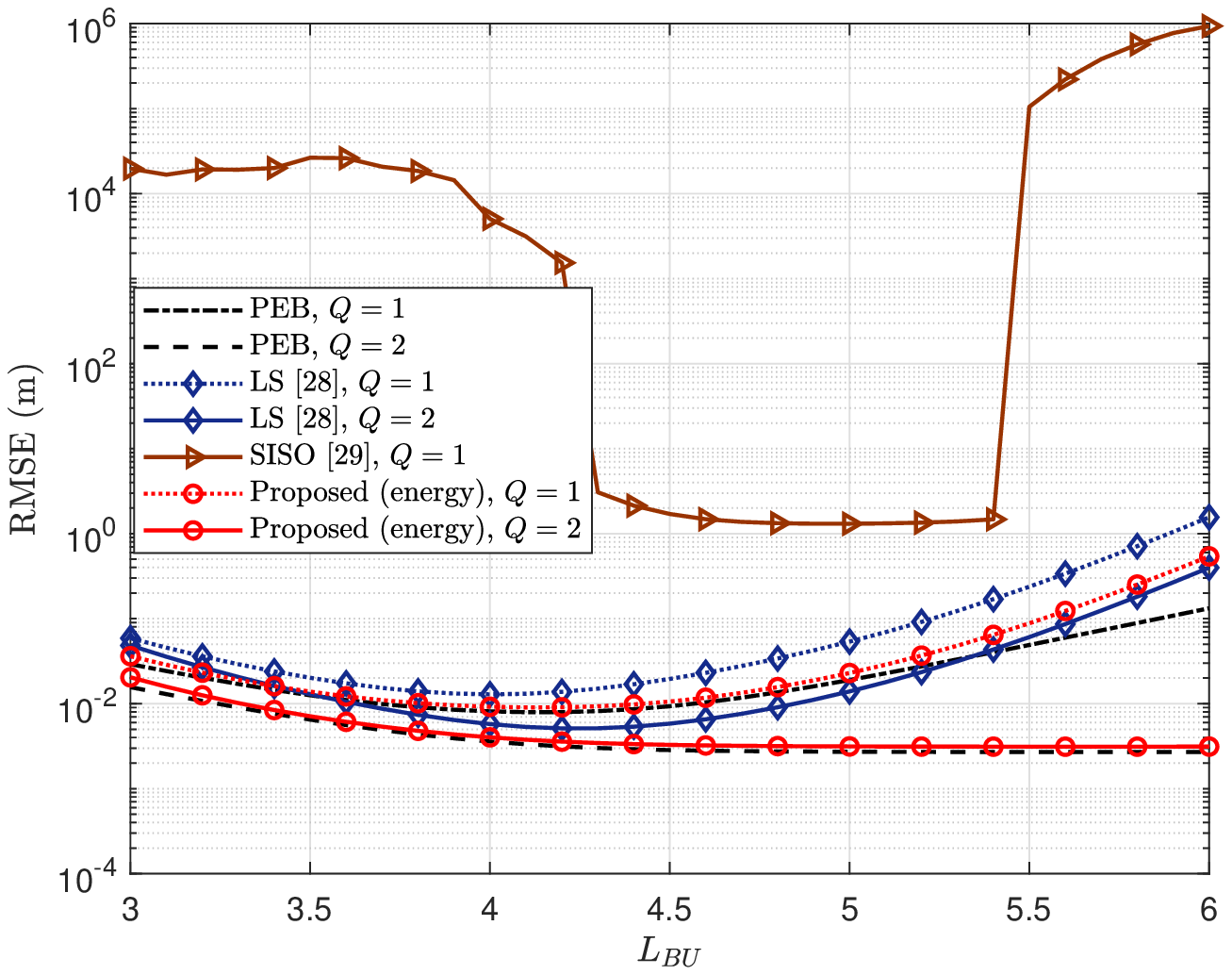}
\caption{RMSE of UE position versus LOS path loss. } \label{fig:direct_fading}
\end{minipage}
\end{figure}

In \cref{fig:direct_fading_K_BU}, we evaluate the positioning accuracy with different Rician factor $K_{BU}$ of the LOS path between the BS and UE in \cref{eq:Channel_model_BS_UE}. A larger $K_{BU}$ means a higher power of the signal from the LOS path.
With increasing $K_{BU}$, all the methods achieve a more accurate positioning result. Since the proposed method adaptively relies on the LOS and reflection paths, it is very close to the PEB for different Rician factors $K_{BU}$.
We can see from \cref{fig:direct_fading_K_BU} that with the aid of the multiple RISs, the proposed method's accuracy is close to the PEB even when $K_{BU}$ is small.
This is because the reflection paths established by the RISs make the positioning task less dependent on the LOS path for the proposed method.


In \cref{fig:direct_fading},  we evaluate the positioning accuracy of the RIS-aided positioning methods with different path gains on the BS-UE link given by $\beta_{BU}={\lambda_c^2}/(16\pi^2 \| \bp_U-\bp_B\|_2^{L_{BU}})$, where $\lambda_c$ is the wavelength and $L_{BU}$ is the path loss exponent \cite{goldsmith2005wireless}.
We observe from \cref{fig:direct_fading} that for the scenario of single RIS ($Q=1$), the value of $L_{BU}$ has a trade-off effect on the estimation accuracy of the benchmarks.
Because increasing  $L_{BU}$ not only leads to the decrease of the energy of the LOS path but also leads to a smaller effective noise $\tilde{\sigma}$ in \cref{eq:R observations}.
Since these two factors have two opposite effects on the positioning accuracy, the positioning errors of benchmarks first decrease and then increase as illustrated in \cref{fig:direct_fading}.
In particular, for the SISO method \cite{Key2021SISO}, the performance gap between the proposed method and LS \cite{wang2021joint} is large because of the employment of a single antenna.
When there are multiple RISs $(Q=2)$, the proposed method
monotonically decreases with $L_{BU}$. Because multiple RISs make the proposed method rely less on the LOS path, and the smaller effective noise dominates among the two effects we mentioned above.
We can also observe the LS positioning method in \cite{wang2021joint} has a large gap with the PEB, especially when $L_{BU}$ is large. Because the inference method in \cite{wang2021joint} does not take the accuracy of the estimated channel parameters into account, and large $L_{BU}$ makes the difference between the accuracy of parameters very large.
Overall, the proposed positioning method is closer to the PEB with different LOS path gains in \cref{fig:direct_fading}.
This verifies that the proposed method adaptively relies less on those paths with low SNR to infer the UE position.

\subsection{RIS Phase Shifts}
\begin{figure}[!htbp]
\centering
\begin{minipage}[t]{0.42\textwidth}
\centering
\includegraphics[width=0.95\textwidth]{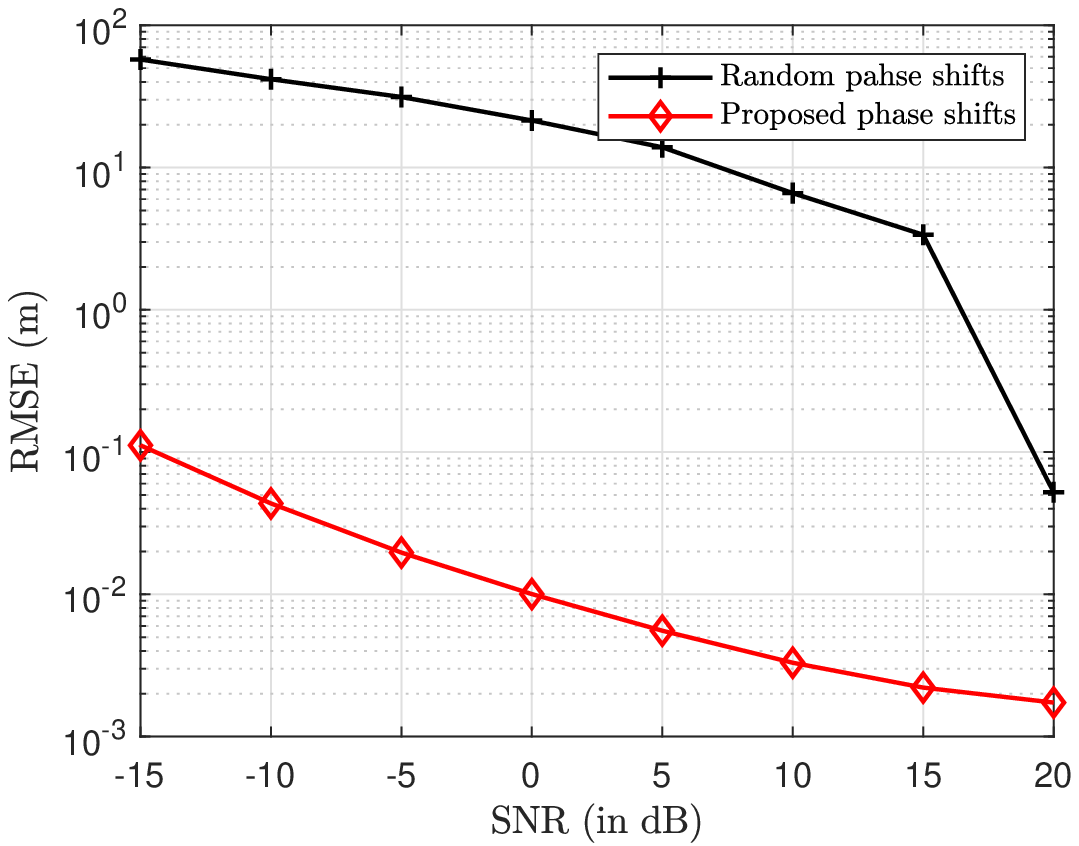}
\caption{RMSE of UE position with different phase shifts (multiple RISs). } \label{fig:random}
\end{minipage}
\hspace{1.0cm}
\begin{minipage}[t]{0.42\textwidth}
\centering
\includegraphics[width=0.95\textwidth]{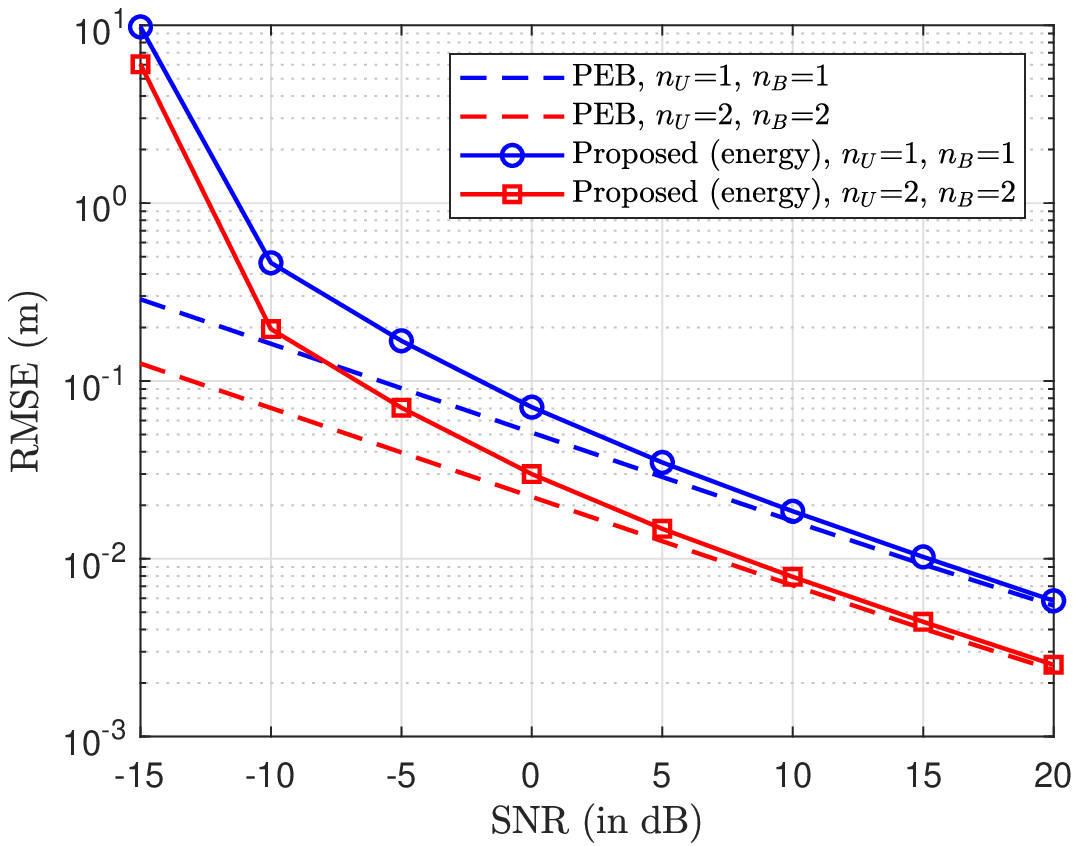}
\caption{RMSE of UE position versus SNR with multiple BSs and multiple UEs (multiple RISs).} \label{fig:multi_BS}
\end{minipage}
\end{figure}

In \cref{fig:random}, we evaluate the proposed design of phase shifts in \cref{sec:phase shift}. Specifically, for the proposed method, we configure the phase shifts of two RISs $(Q=2)$ as the design in \cref{sec:phase shift}. For the design of random phase shifts, we let $\angle\theta_q^{(m)}\sim U[0,2\pi], \forall q, m$. The other simulation parameters are the same as \cref{tab:simulation parameter}.
One can find from \cref{fig:random} that by using the proposed phase shift design, more accurate positioning can be achieved, which verifies the effectiveness of the proposed phase shift design.

\subsection{Multi-UE and Multi-BS}
In \cref{fig:multi_BS}, we compare the positioning accuracy of the proposed method when there are multiple BSs and multiple UEs with the scenario of a single BS and single UE.
The positions of the two BSs are at $[0,0,0]\T$ and $[0,10,0]\T$. The positions of the two UEs are at $[50,10,20]\T$ and $[52,10,20]\T$. Two RISs are aiding the positioning task. The other simulation parameters are the same as in \cref{tab:simulation parameter}.
In \cref{fig:multi_BS}, we only plot the position error of UE at $[50,10,20]\T$, and ${\bC}_{{\bGamma}_l}=\boldsymbol{0}$ for simplicity.
We see from \cref{fig:multi_BS} that by using the techniques in \cref{sec:discussions}, the proposed positioning method in this scenario also achieves performance close to the PEB. It verifies that positioning accuracy can be further improved when more than one BS and UE can cooperate and exchange information.

\section{Conclusion}\label{sec:conclusions}
In this paper, we have developed a RIS-aided positioning framework. The framework first estimates the RIS-aided channel parameters from received signals and then uses these estimates to infer the UE position and clock bias. The proposed fusion of the estimates from the LOS and reflection paths is achieved via the ExIP framework to approximate the \gls{MLE} asymptotically when the estimates are independent.
The advantage of our approach is computational tractability, making it amenable to real-time implementation, as compared to the direct estimation of the UE position from the received signals.
The proposed RIS-aided positioning method can be readily extended to the multi-UE and multi-BS scenarios. Finally, the numerical results illustrate that the positioning accuracy of the proposed method is close to the PEB, which verifies that the proposed method is the approximate MLE.

\bibliographystyle{IEEEtran}

\bibliography{IEEEabrv, Conference_mmWave_CS}

\end{document}



\title{Supplementary Material for the Paper:
“Approximate Maximum-Likelihood RIS-Aided Positioning”}

\maketitle
\IEEEpeerreviewmaketitle

In this document, we present the calculation to obtain the Fisher information matrix (FIM), and we also show the detailed proofs of Proposition 1 and Proposition 2.

\section{CRB Derivation}\label{supp:CRB}
Taking the derivatives of $\widebar{\bH}_k$ \gls{wrt} the parameters of the LoS path, we have
\begin{align*}
\frac{\partial \widebar{\bH}_k}{\partial \tau_{BU}}&= -\iu2\pi\frac{kW}{K} \widebar{\bH}_{BU,k}, \\
\frac{\partial \widebar{\bH}_k}{\partial\opRe\{ h_{BU} \}}&= \widebar{\bH}_{BU,k}/ { h_{BU}},~\frac{\partial \widebar{\bH}_k}{\partial \opIm\{h_{BU}\}} = \iu \widebar{\bH}_{BU,k}/ { h_{BU}} , \nonumber \\
\frac{\partial \widebar{\bH}_k}{\partial g_{BU}^{(U)}} &=  h_{BU,k}\frac{\partial \ba_D(g_{BU}^{(U)}, s_{BU}^{(U)})}{\partial g_{BU}^{(U)}}\ba_N\He(g_{BU}^{(B)},s_{BU}^{(B)}) , \\
\frac{\partial \widebar{\bH}_k}{\partial s_{BU}^{(U)}} &= h_{BU,k} \frac{\partial \ba_D(g_{BU}^{(U)}, s_{BU}^{(U)})}{\partial s_{BU}^{(U)}}\ba_N\He(g_{BU}^{(B)},s_{BU}^{(B)}) , \\
\frac{\partial \widebar{\bH}_k}{\partial g_{BU}^{(B)}}&=  h_{BU,k} \ba_D(g_{BU}^{(U)}, s_{BU}^{(U)})\frac{\partial \ba_N\He(g_{BU}^{(B)},s_{BU}^{(B)}) }{\partial g_{BU}^{(B)}} , \\
\frac{\partial \widebar{\bH}_k}{\partial s_{BU}^{(B)}}&=h_{BU,k}
\ba_D(g_{BU}^{(U)}, s_{BU}^{(U)})\frac{\partial \ba_N\He(g_{BU}^{(B)},s_{BU}^{(B)}) }{\partial s_{BU}^{(B)}}, \nonumber
\end{align*}
where $h_{BU,k}=h_{BU} e^{-\iu2\pi \frac{kW}{K}\tau_{BU}}$. Taking the derivative \gls{wrt} the parameters of the reflection path, we obtain
\begin{align*}
\frac{\partial \widebar{\bH}_{k}}{\partial \tau_{RU, q}}&= -\iu2\pi\frac{kW}{K} \widebar{\bH}_{R, k,q}, \\
\frac{\partial \widebar{\bH}_{k}}{\partial \opRe\{h_{R, q}\}}&= \widebar{\bH}_{R, k,q}/ { h_{R, q}},~\frac{\partial \widebar{\bH}_{k}}{\partial\opIm\{h_{R, q}\}} = \iu \widebar{\bH}_{R, k,q}/ { h_{R, q}} , \nonumber \\
\frac{\partial \widebar{\bH}_{k}}{\partial g_{RU, q}^{(U)}} &=h_{R,q,k}    \frac{\partial \ba_D(g_{RU, q}^{(U)}, s_{RU, q }^{(U)})}{\partial g_{RU, q}^{(U)}}\ba_N\He(g_{BR, q}^{(B)},s_{BR, q}^{(B)}),\\
\frac{\partial \widebar{\bH}_{k}}{\partial s_{RU, q}^{(U)}}&= h_{R,q,k}  \frac{\partial \ba_D(g_{RU, q}^{(U)}, s_{RU, q}^{(U)})}{\partial s_{RU, q}^{(U)}}\ba_N\He(g_{BR, q}^{(B)},s_{BR, q}^{(B)}),\nonumber
\end{align*}
where $\widebar{\bH}_{R, k,q} = \widebar{\bH}_{RU, k,q} \bTheta_q \bH_{BR, k,q}$ and $h_{R,q,k} =h_{R, q} e^{-\iu 2\pi \frac{kW}{K}(\tau_{BR, q} + \tau_{RU, q})}$.
We also have
\begin{align*}
\frac{\partial \ba_n(x, y)}{\partial x} &=\left(\iu\pi \tilde\ba_{n_r}(x) \circ \bE_r\right) \otimes   \tilde\ba_{n_c}(y),\\
\frac{\partial \ba_n(x, y)}{\partial y}&=\tilde\ba_{n_r}(x) \otimes \left(\iu\pi \tilde\ba_{n_c}(y) \circ \bE_c\right),
\end{align*}
where $n\in \{N,D \}$,  $\bE_r=[0,1,\ldots, n_r]\T$, and $\bE_c=[0,1,\ldots, n_c]\T$.

\section{Jacobian Matrix Derivation} \label{supp:jaco}
In this section, we derive the Jacobian matrix $\bJ \in \R^{(7+5Q) \times (5+2Q)}$ used in (22).
Recall that $\boldsymbol{\eta}$ in (14) and $\bxi$ in (15), we write $\bJ$ in as
$
  \bJ=\begin{bmatrix}
        \bar{\bJ}_B\T,
        \bar{\bJ}_{R,1}\T,
         ,\ldots,
       \bar{\bJ}_{R, Q}\T
       \end{bmatrix}\T,
$
where $\bar{\bJ}_B=\frac{\partial \bmeta_B}{\partial\bxi\T}\in \R^{7\times (6+2Q)}$ and $\bar{\bJ}_{R, q}=\frac{\partial \bmeta_{R,q}}{\partial\bxi\T}\in \R^{5\times (6+2Q)}$.
We first derive the Jacobian matrix of the LoS path $\bar{\bJ}_B$, whose entries can be obtained through
\begin{align*}
\frac{\partial \tau_{BU}}{\partial \bp_U\T}&=\frac{\bp_U\T}{c \| \bp_U-\bp_B \|_2},\
\frac{\partial \tau_{BU}}{\partial \Delta_\mathrm{clock}}=1,
\frac{\partial\opRe\{ h_{BU} \}}{\partial\opRe\{ h_{BU} \}}=1,\ \frac{\partial \opIm\{h_{BU}\}}{\partial \opIm\{h_{BU}\}}=1,\\
\frac{\partial x_B}{\partial \bp_U\T}&=\frac{\partial x_B}{\partial \theta_{BU}}\frac{\theta_{BU}}{\partial \bp_U\T} + \frac{\partial x_B}{\phi_{BU}}\frac{\partial \phi_{BU}}{\partial \bp_U\T}. \nonumber
\end{align*}
with $x_B$ denoting any entry in $\bmeta_B$.
We have
\begin{align*}
\frac{\partial \theta_{BU}}{\partial \bp_U\T}&=\frac{1}{ \| \bp_U-\bp_B \|_2^3 {\left(1-\frac{(z_U)^2}{\| \bp_U-\bp_B \|_2^2}\right)^{\frac{1}{2}}}}\left[x_Uz_U, y_Uz_U, -x_U^2-y_U^2\right],\\
\frac{\partial \phi_{BU}}{\partial  \bp_U\T}&= \left[ -\frac{y_U}{x_U^2+y_U^2},\frac{x_U}{x_U^2+y_U^2} ,0 \right], \\
\frac{\partial g_{BU}^{(B)}}{\partial \theta_{BU}}&=- \sin \theta_{BU} \sin\phi_{BU},
~\frac{\partial g_{BU}^{(B)}}{\partial \phi_{BU}}=  \cos \theta_{BU} \cos\phi_{BU},\\
\frac{\partial s_{BU}^{(B)}}{\partial \theta_{BU}}&= \cos \theta_{BU},~\frac{\partial s_{BU}^{(B)}}{\partial \phi_{BU}}= 0,\\
\frac{\partial g_{BU}^{(U)}}{\partial \theta_{BU}}&= - [\bO_R]_{1,1} \sin \theta_{BU}  \cos \phi_{BU}-  [\bO_R]_{1,2}  \sin \theta_{BU}  \sin \phi_{BU} +  [\bO_R]_{1,3}  \cos \theta_{BU}, \\
\frac{\partial g_{BU}^{(U)}}{\partial \phi_{BU}}&= -[\bO_R]_{1,1} \sin \phi_{BU}\cos \theta_{BU}
 +  [\bO_R]_{1,2} \cos \phi_{BU}\cos \theta_{BU},\\
\frac{\partial s_{BU}^{(U)}}{\partial \theta_{BU}}&= -[\bO_R]_{2,1}  \sin  \theta_{BU}\cos \phi_{BU} - [\bO_R]_{2,2}  \sin \theta_{BU}  \sin \phi_{BU} +   [\bO_R]_{2,3}  \cos \theta_{BU},\\
\frac{\partial s_{BU}^{(U)}}{\partial \phi_{BU}}&=  -[\bO_R]_{2,1} \sin \phi_{BU}\cos \theta_{BU}  +  [\bO_R]_{2,2} \cos \phi_{BU} \cos \theta_{BU}.
\end{align*}

We next derive the Jacobian matrix of the reflection path $\bar{\bJ}_{R, q}$, whose entries are can be obtained through
\begin{align*}
\frac{\partial \tau_{RU, q}}{\partial \bp_U\T}&=\frac{\bp_U\T-\bp_{R, q}\T}{c \| \bp_U -\bp_{R, q} \|_2},\
\frac{\partial \tau_{RU, q}}{\partial \Delta_\mathrm{clock}}=1,\
\frac{\partial \opRe\{h_{R, q}\}}{\partial \opRe\{h_{R, q}\}}=1, \ \frac{\partial\opIm\{h_{R, q}\}}{\partial\opIm\{h_{R, q}\}}=1,\\
\frac{\partial x_R}{\partial \bp_U\T}&=\frac{\partial x_R}{\partial \theta_{RU,q}}\frac{ \theta_{RU,q}}{\partial \bp_U\T} + \frac{\partial x_R}{ \phi_{RU,q}}\frac{\partial  \phi_{RU,q}}{\partial \bp_U\T},\nonumber
\end{align*}
with $x_R$ denoting any entry in $\bmeta_{R,q}$.
Here, we denote $\tilde{\bp}_{U, q} = \bp_U - \bp_{R, q}  = \left[ \tilde{x}_{U, q},\tilde{y}_{U, q},\tilde{z}_{U, q} \right]\T$.
Therefore,
\begin{align*}
\frac{\partial \theta_{RU,q}}{\partial  \bp_U\T}&=\frac{ \left[{\tilde{x}_{U, q}\tilde{z}_{U, q}}, {\tilde{y}_{U, q}\tilde{z}_{U, q}},
-{\tilde{x}_{U, q}^2-\tilde{y}_{U, q}^2}\right]}{ \|\tilde{\bp}_{U, q} \|_2^3 {\left(1-\frac{\tilde{z}_{U, r}^2}{\| \tilde{\bp}_{U, q} \|_2^2}\right)^{\frac{1}{2}}}},\\
\frac{\partial \phi_{RU,q}}{\partial  \bp_U\T}&= \Big[ -\frac{\tilde{y}_{U, q}}{\tilde{x}_{U, q}^2+\tilde{y}_{U, q}^2}, \frac{\tilde{x}_{U, q}}{\tilde{x}_{U, q}^2+\tilde{y}_{U, q}^2} ,0 \Big],\\
\frac{\partial g_{RU, q}^{(U)}}{\partial \theta_{RU,q}}&=-[\bO_R]_{1,1}  \sin \theta_{RU,q}\cos \phi_{RU,q} -     [\bO_R]_{1,2}  \sin \theta_{RU,q}\sin \phi_{RU,q} + [\bO_R]_{1,3}  \cos \theta_{RU,q}, \\
\frac{\partial g_{RU, q}^{(U)}}{\partial \phi_{RU,q}}&=-[\bO_R]_{1,1} \cos \theta_{RU,q} \sin \phi_{RU,q}+   [\bO_R]_{1,2} \cos \phi_{RU,q} \cos \theta_{RU,q}, \\
\frac{\partial s_{RU, q}^{(U)}}{\partial \theta_{RU,q}}&=   -[\bO_R]_{2,1}  \sin \theta_{RU,q}\cos \phi_{RU,q}  -     [\bO_R]_{2,2}  \sin \theta_{RU,q}\sin \phi_{RU,q} +[\bO_R]_{2,3}  \cos \theta_{RU,q}, \\
\frac{\partial s_{RU, q}^{(U)}}{\partial \phi_{RU,q}}&= -[\bO_R]_{2,1} \cos\theta_{RU,q} \sin \phi_{RU,q} +   [\bO_R]_{2,2} \cos \phi_{RU,q} \cos \theta_{RU,q}.
\end{align*}

\section{Proof of  Proposition 1 }\label{supp:single position}

When we utilize only the BS-UE link to estimate ${\bxi_B}$, the CRB is given by
$
\widebar{\bC}_{\bxi_B}=\left( \bJ_B\T  \widebar{\bC}_{{\bmeta}_B}^{-1} \bJ_B \right)^{-1},
$
where $\widebar{\bC}_{{\bmeta}_B} = \left[ \bF_{\bmeta} ^{-1}\right]_{1:7,1:7}$, and $ \bJ_B  = \frac{\partial \bmeta_B }{\partial \bxi_B\T}\in \R^{7 \times 5}$. Thus, the error covariance matrix satisfies the following:
\begin{small}
\begin{align}
\bC_{{\bp}_{U}}^{(B)} &\succeq   \left[ \widebar{\bC}_{\bxi_B}\right]_{1:3,1:3}. \label{eq: crb direct p1}
\end{align}
\end{small}
\!\!\!Since
$
\bF_{\bmeta_B}^{-1} \succeq \left[ \bF_{\bmeta} ^{-1}\right]_{1:7,1:7}
$, the following equation holds,
\begin{small}
\begin{align}
\widebar{\bC}_{\bxi_B}=\left( \bJ_B\T \left ( \left[ \bF_{\bmeta} ^{-1}\right]_{1:7,1:7}\right)^{-1} \bJ_B \right)^{-1}
\succeq
\left( \bJ_B\T \bF_{\bmeta_B}   \bJ_B \right)^{-1}. \label{eq: crb direct p2}
\end{align}
\end{small}
\!\!\!Therefore, combining \cref{eq: crb direct p1} and \cref{eq: crb direct p2}, we have
$
\bC_{{\bp}_{U}}^{(B)} \succeq  \left[ \widebar{\bC}_{\bxi_B}\right]_{1:3,1:3} \succeq  \left[ \left(\bJ_B\T \bF_{\bmeta_B}  \bJ_B \right)^{-1} \right]_{1:3,1:3}.
$
This concludes the proof in (38).
Similarly, when only the $q$-th RIS link is utilized to estimate $\bxi_{R, q}$, we can also obtain
$
\bC_{{\bp}_{U}}^{(R, q)} \succeq  \left[ \widebar{\bC}_{\bxi_{R, q}}\right]_{1:3,1:3} \succeq  [ (\bJ_{R, q}\T \bF_{\bmeta_{R, q}}  \bJ_{R, q} )^{-1} ]_{1:3,1:3}.
$
This concludes the proof for (40).

\section{Proof of Proposition 2}\label{supp:equivalence}
It is worth noting that we can prove the equivalence by showing that the optimality condition  for (21) is the same as the one for
(44). For convenience, we denote the objective function in (51) as $J$. The corresponding optimality condition for (51) is
\begin{small}
\begin{align*}
\frac{\partial J}{\partial \bxi}=\left( \frac{\partial \bxi_B}{\partial \bxi} \right)\T\frac{\partial J}{\partial \bxi_B}+ \sum_{q=1}^{Q} \left( \frac{\partial \bxi_{R, q}}{\partial \bxi}\right)\T \frac{\partial J}{\partial \bxi_{R, q}} = \boldsymbol{0}.
\end{align*}
\end{small}
\!\!\!Here, we denote $\bxi=[  \bgamma_0\T,  \bgamma_B\T,  \bgamma_{R,1}\T,\ldots,\bgamma_{R, Q}\T]\T $ where
 $ \bgamma_0 = [\bp_U^T, \Delta_\mathrm{clock}]\T,
  \bgamma_B = [\opRe\{h_{BU}\}, \opIm\{h_{BU}\} ]\T,
  \bgamma_{R, q} = [\opRe\{h_{R, q}\}, \opIm\{h_{R, q}\} ]\T, q=1,\ldots, Q$.
Then, we separate the derivative above as three parts, i.e., $\frac{\partial J}{\partial \bgamma_B}, \frac{\partial J}{\partial \bgamma_{R, q}}$, and $\frac{\partial J}{\partial \bgamma_0}$.
First, the expression of $\frac{\partial J}{\partial \bgamma_B} $ is given by
\begin{small}
\begin{align}
\frac{\partial J}{\partial \bgamma_B} &=\left( \frac{\partial \bxi_B }{\partial \bgamma_B \T} \right)\T\frac{\partial J}{\partial \bxi_B} \nonumber\\
&=2 \begin{bmatrix}
 \mathbf{0}_{3\times 2}\\
  \mathbf{I}_2
\end{bmatrix}\T
\tilde{\bJ}_B\T   \bF_{\hat{\bmeta}_B}\tilde{\bJ}_B  (\bxi_B-\hat{\bxi}_B) \nonumber\\
& =  [ \tilde{\bJ}_B\T   \bF_{\hat{\bmeta}_B}\tilde{\bJ}_B ]_{4:5,1:3}({\bp}_U^{(B)}-\hat{\bp}_U^{(B)})+ [ \tilde{\bJ}_B\T   \bF_{\hat{\bmeta}_B}\tilde{\bJ}_B ]_{4:5,4:5}(\bgamma_B-\hat{\bgamma}_B).\label{eq:obj deriva d}
\end{align}
\end{small}%
Similarly, we have the expression of $\frac{\partial J}{\partial \bgamma_{R, q}} $ in the following
\begin{small}
\begin{align}
\frac{\partial J}{\partial \bgamma_{R, q}}
& =  [ \tilde{\bJ}_{R, q}\T   \bF_{\hat{\bmeta}_{R, q}}\tilde{\bJ}_{R, q} ]_{4:5,1:3}({\bp}_U^{(R, q)}-\hat{\bp}_U^{(R, q)})+ [ \tilde{\bJ}_{R, q}\T   \bF_{\hat{\bmeta}_{R, q}}\tilde{\bJ}_{R, q} ]_{4:5,4:5}(\bgamma_{R, q}-\hat{\bgamma}_{R, q}). \label{eq:obj deriva rq}
\end{align}
\end{small}

From the optimality condition of (51), the first-order derivatives in \cref{eq:obj deriva d} and \cref{eq:obj deriva rq} should be zero. Equivalently, the following two formulas hold,
\begin{small}
\begin{align}
(\bgamma_B-\hat{\bgamma}_B) &= -\left(  [ \tilde{\bJ}_B\T   \bF_{\hat{\bmeta}_B}\tilde{\bJ}_B ]_{4:5,4:5}\right)^{-1}[ \tilde{\bJ}_B\T   \bF_{\hat{\bmeta}_B}\tilde{\bJ}_B ]_{4:5,1:3}({\bp}_U^{(B)}-\hat{\bp}_U^{(B)}) ,\label{eq:partial equality} \\
 (\bgamma_{R, q}-\hat{\bgamma}_{R, q})&=-\left(  [ \tilde{\bJ}_{R, q}\T   \bF_{\hat{\bmeta}_{R, q}}\tilde{\bJ}_{R, q} ]_{4:5,4:5} \right)^{-1}[ \tilde{\bJ}_{R, q}\T   \bF_{\hat{\bmeta}_{R, q}}\tilde{\bJ}_{R, q} ]_{4:5,1:3}({\bp}_U^{(R, q)}-\hat{\bp}_U^{(R, q)}).\label{eq:partial equality 2}
\end{align}
\end{small}
\!\!\!In addition, the expression of $\frac{\partial J}{\partial \bgamma_0}$ is given by
\begin{small}
\begin{align}
\frac{\partial J}{\partial \bgamma_0} &=
\left( \frac{\partial \bp_U^{(B)}}{\partial \bgamma_0\T} \right)\T \left(\frac{\partial \bxi_B}{\partial (\bp_U^{(B)})\T} \right)\T
\frac{\partial J}{\partial \bxi_B} +
\sum_{q=1}^{Q}
\left( \frac{\partial \bp_U^{(R, q)}}{\partial \bgamma_0\T} \right)\T \left(\frac{\partial \bxi_{R, q}}{\partial (\bp_U^{(R, q)})\T} \right)\T
\frac{\partial J}{\partial \bxi_{R, q}}  \nonumber\\
&=2 \left( \frac{\partial \bp_U^{(B)}}{\partial \bgamma_0\T} \right)\T \begin{bmatrix}
  \mathbf{I}_3\\
  \mathbf{0}_{2\times 3}
\end{bmatrix}\T
\tilde{\bJ}_B\T   \bF_{\hat{\bmeta}_B}\tilde{\bJ}_B  (\bxi_B-\hat{\bxi}_B)
+2 \sum_{q=1}^{Q}\left( \frac{\partial \bp_U^{(R, q)}}{\partial \bgamma_0\T} \right)\T \begin{bmatrix}
  \mathbf{I}_3\\
  \mathbf{0}_{2\times 3}
\end{bmatrix}\T
\tilde{\bJ}_{R, q}  \T   \bF_{\hat{\bmeta}_{R, q}  }\tilde{\bJ}_{R, q}  (\bxi_{R, q}  -\hat{\bxi}_{R, q}  )
\nonumber\\
& =2 \left( \frac{\partial \bp_U^{(B)}}{\partial \bgamma_0\T} \right)\T   \left([ \tilde{\bJ}_B\T   \bF_{\hat{\bmeta}_B}\tilde{\bJ}_B ]_{1:3,1:3}
({\bp}_U^{(B)}-\hat{\bp}_U^{(B)})
+ [ \tilde{\bJ}_B\T   \bF_{\hat{\bmeta}_B}\tilde{\bJ}_B ]_{1:3,4:5}(\bgamma_B-\hat{\bgamma}_B)\right)  \nonumber\\
&~~~+
2 \sum_{q=1}^{Q}\left( \frac{\partial \bp_U^{(R, q)}}{\partial \bgamma_0\T} \right)\T \left([ \tilde{\bJ}_{R, q} \T   \bF_{\hat{\bmeta}_{R, q} }\tilde{\bJ}_{R, q} ]_{1:3,1:3}({\bp}_U^{(R, q)}-\hat{\bp}_U^{(R, q)})
+ [ \tilde{\bJ}_{R, q} \T   \bF_{\hat{\bmeta}_{R, q} }\tilde{\bJ}_{R, q} ]_{1:3,4:5}(\bgamma_{R, q} -\hat{\bgamma}_{R, q}  \right). \label{eq:sum opt}
\end{align}
\end{small}
Substituting  \cref{eq:partial equality} into the first part of \cref{eq:sum opt} yields
\begin{small}
\begin{align*}
&  [ \tilde{\bJ}_B\T   \bF_{\hat{\bmeta}_B}\tilde{\bJ}_B ]_{1:3,1:3}
({\bp}_U^{(B)}-\hat{\bp}_U^{(B)})
+ [ \tilde{\bJ}_B\T   \bF_{\hat{\bmeta}_B}\tilde{\bJ}_B ]_{1:3,4:5}(\bgamma_d-\hat{\bgamma}_B)\\
&=[ \tilde{\bJ}_B\T   \bF_{\hat{\bmeta}_B}\tilde{\bJ}_B ]_{1:3,1:3}
({\bp}_U^{(B)}-\hat{\bp}_U^{(B)})
- [ \tilde{\bJ}_B\T   \bF_{\hat{\bmeta}_B}\tilde{\bJ}_B ]_{1:3,4:5} \left(\left(  [ \tilde{\bJ}_B\T   \bF_{\hat{\bmeta}_B}\tilde{\bJ}_B ]_{4:5,4:5}\right)^{-1}[ \tilde{\bJ}_B\T   \bF_{\hat{\bmeta}_B}\tilde{\bJ}_B ]_{4:5,1:3}({\bp}_U^{(B)}-\hat{\bp}_U^{(B)})\right)  \\
&\overset{(a)}{=}  \left(  \left[ \left({\bJ}_{R, q}\T \bF_{{\bmeta}_{R, q}}  {\bJ}_{R, q} \right)^{-1} \right]_{1:3,1:3}\right)^{-1} ({\bp}_U^{(B)}-\hat{\bp}_U^{(B)})\\
&\overset{(b)}{=}  \left(\tilde{\bC}_{{\bp}_{U}}^{(B)}\right)^{-1} ({\bp}_U^{(B)}-\hat{\bp}_U^{(B)}),
\end{align*}
\end{small}
\!\!\! \!where $(a)$ is from the Schur complement, and ${(b)}$ is from (42).
Similarly, substituting  \cref{eq:partial equality 2} into the second part of \cref{eq:sum opt} yields the following,
\begin{small}
\begin{align*}
[ \tilde{\bJ}_{R, q} \T   \bF_{\hat{\bmeta}_{R, q} }\tilde{\bJ}_{R, q} ]_{1:3,1:3}({\bp}_U^{(R, q)}-\hat{\bp}_U^{(R, q)})
+ [ \tilde{\bJ}_{R, q} \T   \bF_{\hat{\bmeta}_{R, q} }\tilde{\bJ}_{R, q} ]_{1:3,4:5}(\bgamma_{R, q} -\hat{\bgamma}_{R, q})
=\left(\tilde{\bC}_{{\bp}_{U}}^{(R, q)}\right)^{-1} ({\bp}_U^{(R, q)}-\hat{\bp}_U^{(R, q)}).
\end{align*}
\end{small}
\!\!\!Therefore, the expression of $\frac{\partial J}{\partial \bgamma_0}$ in \cref{eq:sum opt} is simplified as the following
\begin{small}
\begin{align*}
\frac{\partial J}{\partial \bgamma_0} &= 2 \left( \frac{\partial \bp_U^{(R, q)}}{\partial \bgamma_0\T} \right)\T \left(\tilde{\bC}_{{\bp}_{U}}^{(B)}\right)^{-1} ({\bp}_U^{(B)}-\hat{\bp}_U^{(B)})+ 2 \sum_{q=1}^{Q}\left( \frac{\partial \bp_U^{(R, q)}}{\partial \bgamma_0\T} \right)\T \left(\tilde{\bC}_{{\bp}_{U}}^{(R, q)}\right)^{-1} ({\bp}_U^{(R, q)}-\hat{\bp}_U^{(R, q)}),
\end{align*}
\end{small}
\!\!\!which is exactly the derivative of the objective function in (44). Therefore, the optimality conditions for (44) and (51) are the same, which implies these two problems are equivalent. This concludes the proof.